\documentclass[prb,twocolumn,longbibliography]{revtex4-1}
\usepackage{graphicx}
\usepackage{bm} % bold math
\usepackage{wrapfig}
\usepackage{color}
\usepackage{natbib}
\usepackage{hyperref}

\usepackage{amssymb} % use this package to enable \nrightarrow command
\usepackage{amsmath} % use this package to enable \xrightarrow command
\usepackage{geometry}
\begin{document}
\relpenalty=9999
\binoppenalty=9999

\title{Thermopower and thermal conductance of a superconducting quantum point contact}

\author{Sergey S. Pershoguba and Leonid I. Glazman}

\affiliation{Department of Physics, Yale University, New Haven, CT 06520, USA}

\begin{abstract}
We find the charge and heat currents caused by a temperature difference applied to a superconducting point contact or to a quantum point contact between a superconducting and normal conductors. The results are formulated in terms of the properties of the electron scattering matrix of the quantum point contact in its normal state, and are valid at any  transmission coefficient. In the low-transmission limit, the  theory provides reliable results, setting the limits for the use of the popular method of tunneling Hamiltonian. 
\end{abstract}

\maketitle
%\tableofcontents
%\newpage

%%%%%%%%%%%%%%%%%%%%%%%%%%%%%%%%%%%%%%%%%%%%%%%%%%%%%%%%%%%%%%%%%%%%%%%%%%%%%%%%%%%%%%
\section{Introduction}
%%%%%%%%%%%%%%%%%%%%%%%%%%%%%%%%%%%%%%%%%%%%%%%%%%%%%%%%%%%%%%%%%%%%%%%%%%%%%%%%%%%%%%
Superconductivity changes drastically the spectrum of low-energy electron excitations. Their energy distribution and dynamics define the equilibrium thermal properties of a superconductor, as well as charge and entropy transport caused by a temperature gradient.

In a bulk superconductor, observation of the electronic component of the entropy transport at low temperatures is masked by a bigger phonon component~[\onlinecite{Mendelssohn1953},\onlinecite{Bardeen1959}]. The conventional manifestation of thermopower for a normal-state conductor is an electric potential build-up in an open circuit. That does not happen in a superconductor because of the shunting effect of the supercurrent [\onlinecite{Ginzburg1944},\onlinecite{GinzburgRMP2004}]. Due to it, a temperature gradient applied to an inhomogeneous superconducting ring creates a persistent current in the ring. Its value, inferred from the magnetic flux associated with the current, serves as a proxy for thermopower. Such measurement scheme turned out to be prone to errors caused by spurious Meissner currents [\onlinecite{VanHarlingenPRB1980},\onlinecite{Shellye-Matrozova-Petrashov2016}]. Alternatively, one may infer the thermopower from the measurements of the charge imbalance near the ends of a superconductor in an open-circuit geometry [\onlinecite{Mamin-Clarke-VanHarlingen-PRB1984}]. This inference, however, involves assumptions regarding the inelastic electron scattering leading to the charge imbalance relaxation.

Charge and entropy current responses to a temperature difference applied to a weak link depend, in addition, on the difference between the superconducting order parameter phases in the leads [\onlinecite{Maki1966,GuttmanPRB1997a,GuttmanPRB1997b,GiazottoAPL2012,SaulsPRL2003,SaulsPRB2004}]. This phase dependence was experimentally demonstrated [\onlinecite{Giazotto-Nature2012}] and used to control the heat current. Theoretical consideration of Ref. [\onlinecite{GurevichEPJB2006}] also favors including  a superconducting weak link in a ring geometry designed to measure the thermopower. The downside of using weak links for studying thermopower is the temperature dependence of the {\it equilibrium} dissipationless (Josephson) current [\onlinecite{Josephson1962}] which should be discriminated from the specific for thermopower dissipative current component associated with the lack of particle-hole symmetry.

The existing theory of thermally-induced charge and entropy currents through weak links employs the tunneling Hamiltonian approximation in considering superconductor-insulator-superconductor~(SIS) junctions [\onlinecite{GuttmanPRB1997a},\onlinecite{GuttmanPRB1997b},\onlinecite{SmithPRB1980}] or more complex structures [\onlinecite{HwangPRB2016},\onlinecite{TrochaPRB2017}]. Other approaches use semiclassical description of diffusive [\onlinecite{Bezuglyi2003},\onlinecite{Golubov2005}] or ballistic [\onlinecite{SaulsPRL2003},\onlinecite{SaulsPRB2004}] weak links or junctions between a normal-state material and superconductor (NS junction). There are certain limitations of these approximations. Due to the singularity in the quasiparticle density of states, the lowest-order tunneling Hamiltonian formalism leads to divergent results for  charge [\onlinecite{SmithPRB1980}] and heat [\onlinecite{Maki1966}] current; some qualitative considerations are customarily used to cut off the divergence. Furthermore, the tunneling Hamiltonian makes it difficult to correctly account for the absence of particle-hole symmetry in tunneling of electrons with energies, respectively, below and above the Fermi level; that leads to unreliable results for thermopower [\onlinecite{GuttmanPRB1997a}]. The semiclassical approximation, while adequately describing junctions of arbitrary transmission, nominally requires the junction width to exceed the Fermi wavelength, {\sl i.e.}, the approximation assumes a large number of electron modes propagating through the junction. The limitations of the existing theory makes its results hardly applicable to single- or a few-channel quantum point contacts of arbitrary transmission. These kinds of contacts are currently studied in several different experimental settings. These include proximized semiconductor quantum wires [\onlinecite{Goffman2017},\onlinecite{Mourik2012}], atomic point contacts [\onlinecite{DellaRocca2007},\onlinecite{Bretheau2013}], and trapped cold atoms [\onlinecite{Stadler2012,BrantutScience2013,HusmannScience2015,Husmann-PNAS2018}]. 

The scattering formalism for thermoelectric effects in contacts between normal-state conductors is well-known [\onlinecite{LesovikUFN2011}]. In this work, we develop a scattering theory for an evaluation of the charge and heat currents generated by a temperature difference applied to a superconducting quantum point contact. In obtaining concrete results, which are valid at any transmission, we assume the length of a single-mode contact short compared to the superconducting coherence length. 

Scattering theory allows us to find the dependence of thermal conductance on the transmission coefficient $\tau$ in the entire interval $1\geq \tau>0$. The small-$\tau$ limit of our result elucidates the correct regularization of the perturbative in $\tau$ expressions. 

To evaluate the charge current, we account for the violation of particle-hole symmetry in the scattering matrix. In the course of calculation presented in Sec.~\ref{sec:particle_current_SXS}, we highlight the discrepancy 
between the perturbative-in-$\tau$ results of Refs.~[\onlinecite{SmithPRB1980}] and [\onlinecite{GuttmanPRB1997a}], respectively. The root of the inconsistency is in the use~[\onlinecite{GuttmanPRB1997a}] of the tunneling Hamiltonian which is poorly suited for the accounting of the finite thickness $d$ of the tunneling barrier. Inadequate accounting for a finite value of $d$ yields an error in the
evaluation of a response which relies on a particle-hole symmetry violation. We demonstrate this, and correct the error by performing  expansion of the particle current in powers of $d$ in Appendix~\ref{sec:bdg_appendix}.

The scattering theory also allows us to single out, at any $\tau$, the dissipative charge current response to the applied temperature bias and to clarify the role of Andreev levels and of inelastic electron scattering in the full current response. Furthermore, by considering the thermopower of an NSN junction (relevant for the cold-atoms realization [\onlinecite{Husmann-PNAS2018}]) we demonstrate that it is determined by the thermopower of the NS boundaries rather than by the thermopower of the point contact.

The paper is organized as follows. In Sec. \ref{sec:2} we present the general result for the scattering matrix of Bogoliubov quasiparticles, valid in the absence of particle-hole symmetry. General expressions for the energy and charge currents generated by a difference in temperatures of the quasiparticles impinging on the junction are derived in Sec. \ref{sec:3}. These expressions are simplified for the case of weak particle-hole asymmetry in Sec. \ref{sec:sj}, where we also analyze the limit $\tau\ll 1$. In Secs. \ref{sec:nsj} and \ref{sec:nsnj}, we apply the general theory of entropy and particle currents driven by temperature bias to NS and NSN junctions, respectively.

The developed theory is applicable to electron transport in superconducting nanostructures, and to transport of neutral cold fermions in spatially-restricted clouds [\onlinecite{Husmann-PNAS2018}]. Therefore we will make no distinction between the references to charge and particle currents. We retain the absolute value $e$ of electron charge in the final results; for cold-atom applications, one may replace $e\to 1$.

%%%%%%%%%%%%%%%%%%%%%%%%%%%%%%%%%%%%%%%%%%%%%%%%%%%%%%%%%%%%%%%%%%%%%%%%%%%%%
\begin{figure}
(a) \includegraphics[width=0.9\linewidth]{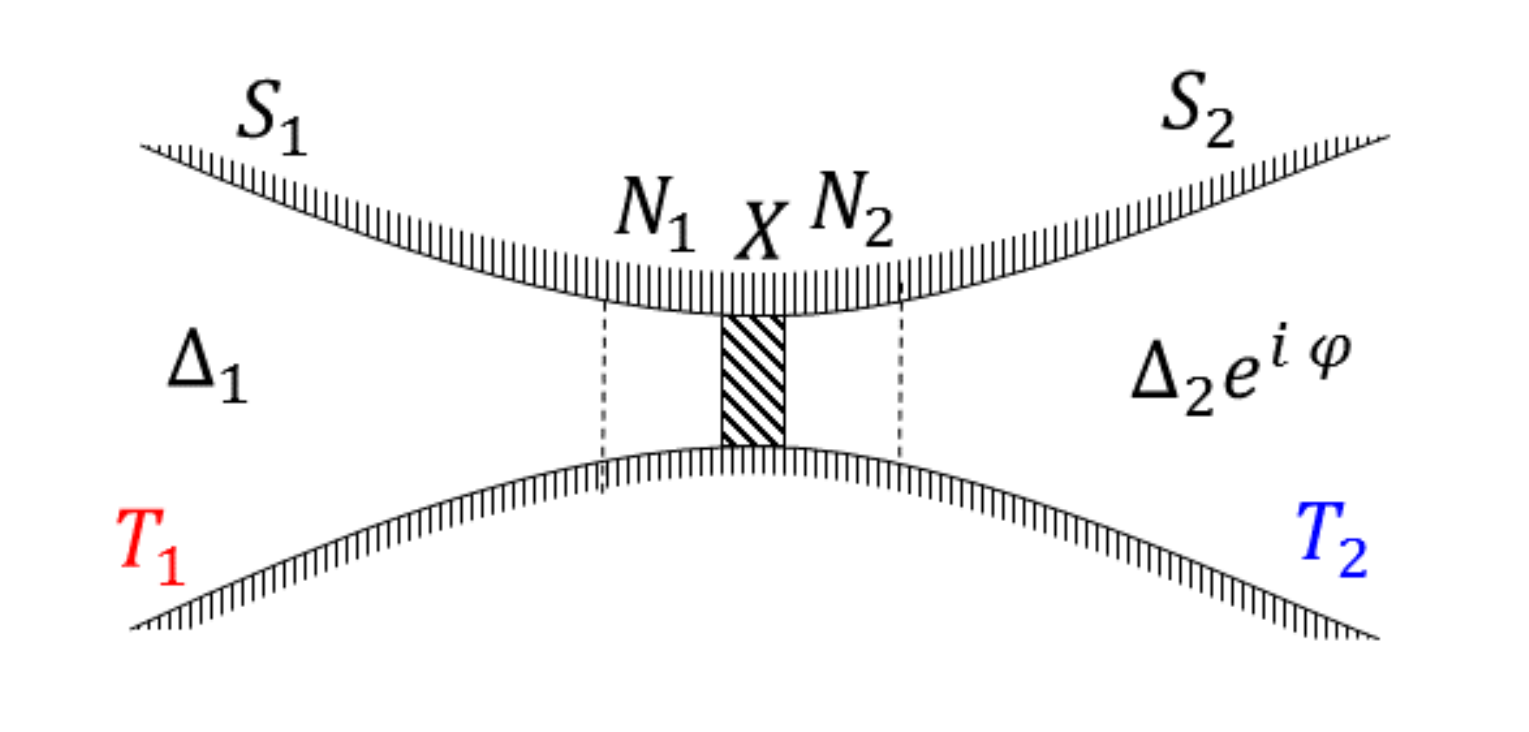}  \\ \vspace{.1cm} (b) \includegraphics[width=0.9\linewidth]{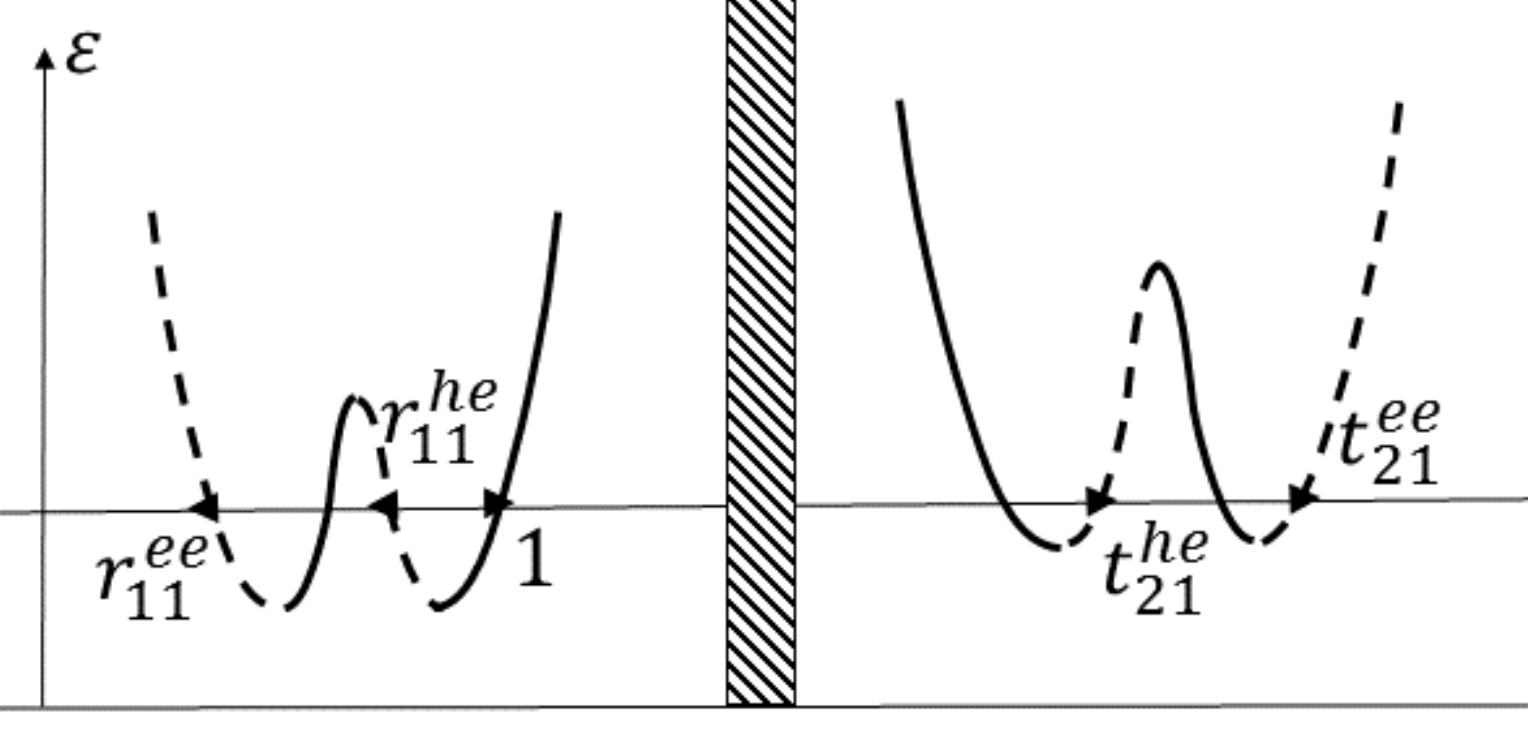} \\ \vspace{.5cm} (c) \includegraphics[width=0.9\linewidth]{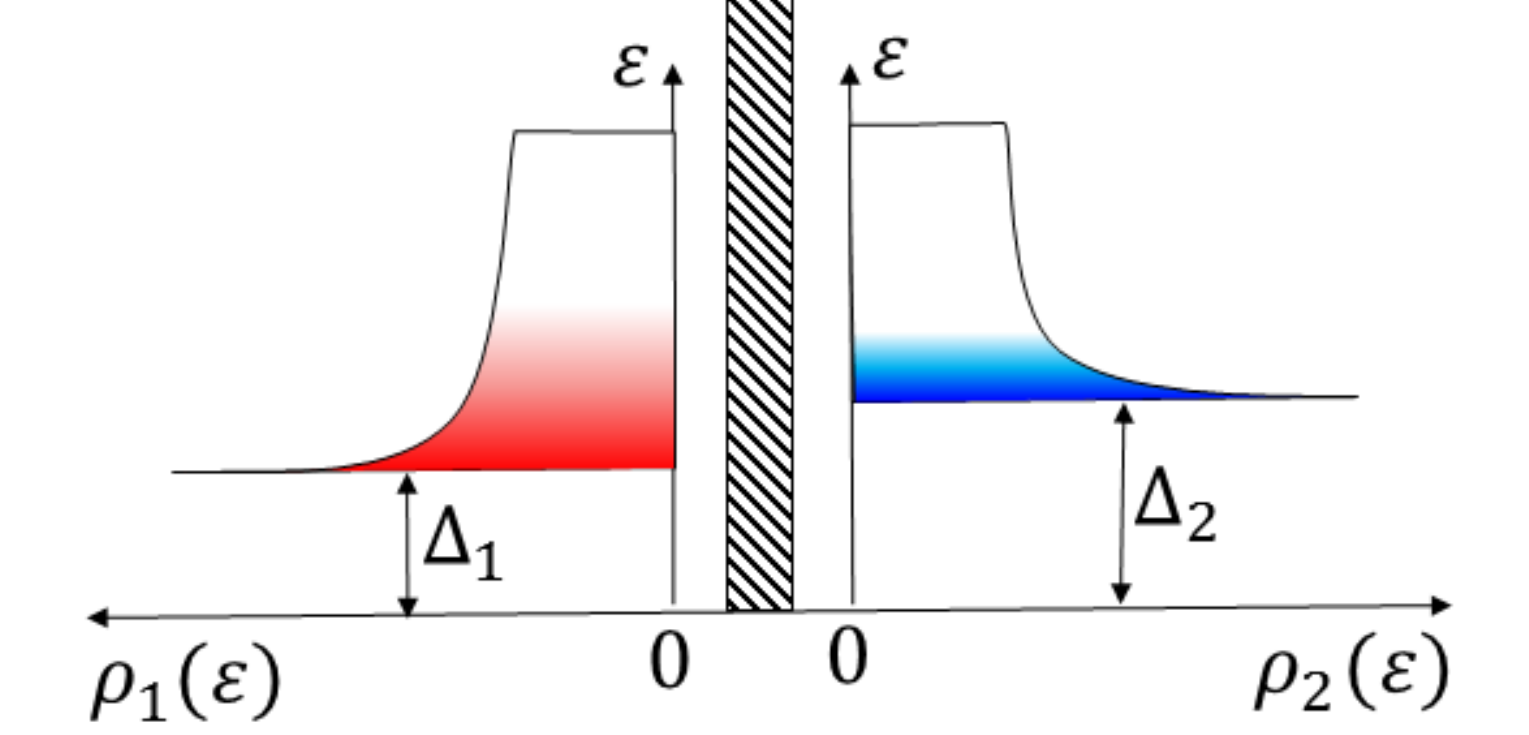} \caption{(a) Geometry of a superconducting point contact: two superconducting leads $S_1$ and $S_2$ are connected via a narrow constriction X. (b) The  energy dispersion of the quasiparticle excitations in the leads. At a given energy $\varepsilon$, both electron-like and hole-like quasiparticle branches are present. The quasiparticle states with the group velocity directed to (from) the scatterer X are shown in solid (dashed) lines. (c) Density of states $\rho_{1,2}(\varepsilon)$ of the quasiparticle excitations in the two superconductors in the junction. The superconductors have different temperatures $T_1$ and $T_2$; the corresponding difference in quasiparticle distributions drives the particle and entropy currents.} 
\label{fig:junction}
\end{figure}
%%%%%%%%%%%%%%%%%%%%%%%%%%%%%%%%%%%%%%%%%%%%%%%%%%%%%%%%%%%%%%%%%%%%%%%%%%%%%
%%%%%%%%%%%%%%%%%%%%%%%%%%%%%%%%%%%%%%%%%%%%%%%%%%%%%%%%%%%%%%%%%%%%%%%%%%%%%%%%%%%%%%
\section{Bogoliubov-de Gennes scattering states in 1D} \label{sec:2}
%%%%%%%%%%%%%%%%%%%%%%%%%%%%%%%%%%%%%%%%%%%%%%%%%%%%%%%%%%%%%%%%%%%%%%%%%%%%%%%%%%%%%%
Before evaluating the thermoelectric effects let us briefly review details of a scattering problem. In the spirit of the Landauer approach, we consider a one-dimensional one-channel problem illustrated schematically in Fig.~\ref{fig:junction}(a). We refer to the two superconducting leads as ``left'' and ``right'' and label with corresponding index $l\in\{1,2\}$. The superconductors may have different gaps, $\Delta_1 \neq \Delta_2$. We adopt a convention in which the quasiparticle energies $\varepsilon = \sqrt{\xi_l^2+\Delta_l^2}$ are positive,
variable $\xi_l$ denotes the kinematic part of the energy measured from the Fermi level.
%  where $\xi_s = \sqrt{\varepsilon^2-\Delta^2_s}$ 
The energy spectra of quasiparticles are illustrated on both sides of the junction in Fig.~\ref{fig:junction}(b). At a given energy, there are multiple states corresponding to the distinct particle-like and hole-like quasiparticle branches, which we label as $b\in \{e,h\}$. The scattering region consists of a scatterer X embedded in the normal regions $N_1$ and $N_2$. Even if the normal regions are not present in the physical device, we introduce them into the model for an appropriate formulation of a boundary condition for the scatterer X. We expect that the latter artificial construction is consistent in the leading order in $\varepsilon/E_F$, where $E_F$ is the Fermi energy \footnote{In other words, we expect that the scattering properties of the junction with (SNXNS) and without (SXS) the narrow regions N are equivalent in the leading order in $\varepsilon/E_F$.}. We address the effect of the terms $\propto\varepsilon/E_F$ in Appendix \ref{sec:bdg_appendix}.

In the Bogoliubov-de Gennes (BdG) formalism, a typical scattering wavefunction in the two leads may be written as
\begin{align*}
	&\Psi_{S_1} = \\ 
	& \,\,\left( \begin{array}{c}
		  u_1 \\
		  v_1
	\end{array}\right) e^{iq_ex} + r^{he}_{11} \left( \begin{array}{c}
		  v_1 \\
		  u_1
	  \end{array}\right) e^{iq_hx}  +   r^{ee}_{11} \left( \begin{array}{c}
		  u_1 \\
		  v_1
	  \end{array}\right) e^{-iq_ex}   ,  \\
	  &\Psi_{S_2} = \\ 
	  & \,\,{ t^{ee}_{21} \left( \begin{array}{c}
		  u_2 \\
		  v_2
	  \end{array}\right) e^{iq_ex}} +{ t^{he}_{21} \left( \begin{array}{c}
		  v_2 \\
		  u_2
	  \end{array}\right) e^{-iq_hx}}.
\end{align*}
Here, the coherence factors are defined as usual,
\begin{equation}
    u^2_{l} = 1-v^2_{l} {= \frac{1}{2}\left(1+\frac{\xi_{l}}{\varepsilon}\right)}.    
\end{equation}
The scattering amplitudes $r^{ee}_{11}$, $t^{ee}_{21}$, $r^{he}_{11}$, $t^{he}_{21}$ are the basic parameters in the Landauer transport theory. In our nomenclature, $r$ denotes an amplitude of reflection into the same lead, whereas $t$ is an inter-lead transmission amplitude. The upper indices (e.g. $ee,\,he$) denote the quasiparticle type, and the lower indices (e.g. $11,\,21$) label the lead. For example, the term $t^{he}_{21}$ denotes the scattering amplitude of the electron-like quasiparticle incident from the left lead into a hole-like quasiparticle in the right lead. 

The scatterer $X$ is modeled by the following energy-dependent 2-by-2 unitary scattering matrix
\begin{align}
	s^{\phantom\dagger}_\xi = e^{i\gamma^{\phantom\dagger}_\xi} \left( \begin{array}{cc}
		e^{i  {\eta_\xi}}\, r^{\phantom\dagger}_{\xi}\, & i\, e^{-i\varphi/2}\, t^{\phantom\dagger}_\xi  \\
		i \,  e^{i\varphi/2} \,t^{\phantom\dagger}_{\xi} &  e^{-i {\eta_\xi}}\, r^{\phantom\dagger}_\xi
	\end{array}\right). 
	\label{s01}
\end{align} 
Here, the parameters $r_\xi$ and $t_\xi$ are the magnitudes of the electron reflection and transmission amplitudes; the unitarity of the scattering matrix requires that $r_\xi^2 + t_\xi^2 = 1$. The phase $\gamma_{\xi}$ is the Friedel phase, which determines a modulation of the density of states in the vicinity of the scatterer. {The phase $\eta_\xi$ models absence of the inversion symmetry. } We work in a gauge, where the superconducting gaps $\Delta_{1,2}$ are real, and the Josephson phase difference $\varphi$ is absorbed in the scattering matrix. The scattering matrix $s_{\xi}$ acts in the particle sector of the wavefunction, and $s_{-\xi}^{\ast}$ acts in the hole sector. {The parameters $r_\xi$, $t_\xi$, $\eta_{\xi}$ and $\gamma_{\xi}$ may have an arbitrary dependence on $\xi$. For example, the particle-hole symmetry/asymmetry is encoded in the parity of the scattering matrix parameters with respect to the reversal of $\xi \rightarrow -\xi$, i.e. the system lacks a particle-hole symmetry if any of the conditions $t_{\xi} \neq t_{-\xi}$, $\eta_{\xi} \neq \eta_{-\xi}$ or $\gamma_{\xi} \neq \gamma_{-\xi}$ are satisfied.} 

Assuming the superconducting coherence length is much greater than the Fermi wavelength in the leads, we may express the scattering matrix for the Bogoliubov quasiparticles in terms of $s_\xi$. For that, 
we follow Ref.~[\onlinecite{Beenakker1991}] and use the boundary conditions induced by the scatterer to derive (see details in Appendix \ref{sec:amplitudes_ap})
\begin{widetext}
\begin{equation}
    \begin{aligned}
        r^{ee}_{11} &= \frac{\xi_1}{2 D_{\varepsilon}}\left[ (\varepsilon+\xi_2)\,r_\varepsilon e^{i(\gamma_{-\varepsilon}+ \eta_\varepsilon)} - (\varepsilon-\xi_2)\,r_{-\varepsilon}e^{i(\gamma_\varepsilon+ \eta_{-\varepsilon})}\right], \\
        r^{he}_{11} & = \frac{1}{2D_{\varepsilon}} \left[-\Delta_1(\varepsilon\, \cos\delta\gamma_{\varepsilon}-i\xi_2\sin \delta \gamma_\varepsilon) + \Delta_1 (\varepsilon\, { \cos \delta \eta_{\varepsilon}+ i \,\xi_2 \sin\delta \eta_\varepsilon})\, r_{\varepsilon} r_{-\varepsilon}+ \Delta_2(\varepsilon \cos \varphi + i\,\xi_1 \sin \varphi)\, t_{\varepsilon} t_{-\varepsilon}\right], \\
        t^{ee}_{21} & = \frac{i\xi_1}{2 D_{\varepsilon}}\left[\sqrt{(\varepsilon+\xi_1)(\varepsilon+\xi_2)}\,t_{\varepsilon} e^{i(\varphi/2+\gamma_{-\varepsilon})}- \sqrt{(\varepsilon-\xi_1)(\varepsilon-\xi_2)}\,t_{-\varepsilon} e^{i(-\varphi/2+\gamma_{\varepsilon})}\right], \\
        t^{he}_{21} &= \frac{i\xi_1}{2D_{\varepsilon}} \left[\sqrt{(\varepsilon +\xi_1)(\varepsilon-\xi_2)}\, t_{\varepsilon} r_{-\varepsilon}\,e^{i(\varphi/2+ \eta_{-\varepsilon})} -  \sqrt{(\varepsilon-\xi_1)(\varepsilon+\xi_2)}\, t_{-\varepsilon} r_{\varepsilon}\,e^{i(-\varphi/2+ \eta_\varepsilon)}\right],
    \end{aligned} \label{amplitudes}
\end{equation}
where we introduced the following notations:
\begin{equation}
    \begin{aligned}
        D_{\varepsilon} &= \frac{1}{2} \left[ (\varepsilon^2+\xi_1\xi_2)\cos\delta\gamma_\varepsilon - i\varepsilon (\xi_1 + \xi_2)\sin \delta \gamma_\varepsilon - (\varepsilon^2-\xi_1\xi_2) \, r_{\varepsilon}r_{-\varepsilon}\, {\cos \delta\eta_{\varepsilon}} -  {i \varepsilon( \xi_2 - \xi_1) r_{\varepsilon} r_{-\varepsilon} \sin \delta \eta_\varepsilon} - \Delta_1 \Delta_2 \,t_{\varepsilon} t_{-\varepsilon} \cos \varphi \right], \label{denom} \\
        \delta \gamma_{\varepsilon} &= \gamma_\varepsilon - \gamma_{-\varepsilon},\quad {\delta \eta_{\varepsilon} = \eta_\varepsilon - \eta_{-\varepsilon}}. 
    \end{aligned}
\end{equation}
\end{widetext}
The amplitudes in Eqs.~(\ref{amplitudes}) are written for a particle-like quasiparticle incident from the left superconducting lead. The rest of the amplitudes can be obtained from Eqs.~(\ref{amplitudes}) as follows: (i) To obtain the amplitudes for a hole-like quasiparticle, one replaces $s_{\varepsilon} \leftrightarrow s^\ast_{-\varepsilon}$ (i.e. replacing $\gamma_{\varepsilon} \leftrightarrow -\gamma_{-\varepsilon}$, {$\eta_{\varepsilon} \leftrightarrow -\eta_{-\varepsilon}$}, $\varphi \leftrightarrow -\varphi$, $t_{\varepsilon} \leftrightarrow -t_{-\varepsilon}$, $r_{\varepsilon} \leftrightarrow r_{-\varepsilon}$), (ii) The amplitudes for the quasiparticles incident from the right are be obtained 
%from Eqs.~(\ref{amplitudes}) 
by the reversal of phases $\varphi \leftrightarrow -\varphi$, {$\eta_{\varepsilon} \leftrightarrow -\eta_{\varepsilon}$}  and gaps $\Delta_1 \leftrightarrow \Delta_2$. If the gaps are equal $\Delta_1 = \Delta_2 = \Delta$, and if there is no particle-hole asymmetry, the amplitudes~(\ref{amplitudes}) simplify 
\begin{align}
	r^{ee}_{11} &= \frac{e^{i\gamma}r\, \xi^2}{\xi^2+t^2\Delta^2\sin^2\varphi/2}, \nonumber \\
	r^{he}_{11} &= \frac{i\Delta\, t^2 \left( \xi\cos\varphi/2+i\varepsilon\sin\varphi/2 \right)\sin\varphi/2}{\xi^2+t^2\Delta^2\sin^2\varphi/2}, \nonumber   \\
	t^{ee}_{21} & = \frac{e^{i\gamma}t\, i\xi  \left( \xi\cos\varphi/2+i\varepsilon\sin\varphi/2 \right)}{\xi^2+t^2\Delta^2\sin^2\varphi/2}, \label{noAsym}\\
	t^{he}_{21} &= \frac{-\xi \Delta\,rt\, \sin\varphi/2 }{\xi^2+t^2\Delta^2\sin^2\varphi/2}. \nonumber
\end{align}

%%%%%%%%%%%%%%%%%%%%%%%%%%%%%%%%%%%%%%%%%%%%%%%%%%%%%%%%%%%%%%%%%%%%%%%%%%%%%%%%%%%%%%
\section{General expressions for the heat and particle currents generated by the temperature difference applied to a junction.} \label{sec:3}
%%%%%%%%%%%%%%%%%%%%%%%%%%%%%%%%%%%%%%%%%%%%%%%%%%%%%%%%%%%%%%%%%%%%%%%%%%%%%%%%%%%%%%

A complementary view of the superconducting junction is given in Fig.~\ref{fig:junction}(c), where we show the density of states of the two superconductors. The distinct temperatures in the two leads $T_1 \neq T_2$ induce distinct quasiparticle occupations that drive the thermoelectric charge and heat currents. In addition, the temperatures implicitly control the gaps of the superconductors $\Delta_{1,2}$. Variation of the gaps $\delta \Delta_{1,2}$ with respect to shift of temperatures $\delta T$ may also induce adjustment of currents.

%%%%%%%%%%%%%%%%%%%%%%%%%%%%%%%%%%%%%%%%%%%%%%%%%%%%%%%%%%%%%%%%%%%%%%%%%%%%%%%%%%%%%%
\subsection{Heat current} \label{sec:heatcur}
%%%%%%%%%%%%%%%%%%%%%%%%%%%%%%%%%%%%%%%%%%%%%%%%%%%%%%%%%%%%%%%%%%%%%%%%%%%%%%%%%%%%%%

First, let us examine the heat current. As shown in the Appendix~\ref{sec:heatcur_ap}, the heat current may be written as a balance of currents flowing from left-to-right $J_{1}$ and right-to-left $J_{2}$,
\begin{align}
& J = J_{1} - J_{2}, \label{ful_h_cur} \\
& J_l = \frac{2}{h} \int_{\Delta_l}^\infty \frac{\varepsilon^2\, d\varepsilon}{\xi_l}\, \left[ j^{e}_l(\varepsilon)  + j^{h}_l(\varepsilon) \right] f\left(\varepsilon/T_l\right),    \label{h_cur_from_one_lead} \\
& {\rm where\,\,\,}j^b_l(\varepsilon) = \frac{\xi_l}{\varepsilon}(1-|r^{bb}_{ll}(\varepsilon)|^2-|r^{\bar bb}_{ll}(\varepsilon)|^2). \label{J_bdg}
\end{align}
The two currents $J_1$ and $J_2$ correspond to the quasiparticles originating from the left and right leads respectively (subscript index $l \in \{ 1,2\}$ labels leads as before). We assume that the quasiparticles are in thermal equilibrium with the lead from which they originate. Therefore  the Fermi occupation function of the quasiparticles $f(\varepsilon/T_l) = (e^{\varepsilon/T_l}+1)^{-1}$ is controlled by the corresponding temperatures $T_{l}$ (in our convention, temperature has units of energy, i.e. we set $k_B = 1$). Let us comment on other terms appearing in Eq.~(\ref{h_cur_from_one_lead}). The prefactor 2 corresponds to the spin degeneracy. A single factor of $\varepsilon$ arises because we evaluate the transport of energy across the junction. The factor $\varepsilon/\xi_l$ is due to the quasiparticle density of states in a superconductor. Notice that the expression in the brackets in Eq.~(\ref{h_cur_from_one_lead}) contains two terms $j_{l}^{b}(\varepsilon)$ corresponding to particle-like and hole-like quasiparticle branches labeled by the superscript $b \in \{e,h\}$.  The term $j_{l}^{b}(\varepsilon)$ has a physical meaning of a quasiparticle density current and is defined in Eq.~(\ref{J_bdg}); the factor $\xi_l/\varepsilon$ in Eq.~(\ref{J_bdg}) cancels with the inversely proportional term in Eq.~(\ref{h_cur_from_one_lead}). Equation~(\ref{J_bdg}) is written via the normal $r^{bb}_{ll}$ and Andreev $r^{\bar bb}_{ll}$ reflection amplitudes, but may be equivalently represented via the normal $t^{bb}_{\bar ll}$ and Andreev $t^{\bar bb}_{\bar ll}$ transmission amplitudes as discussed in Appendix~\ref{sec:heatcur_ap}. Here the ``bar'' above the indices denotes negation, e.g. $\bar e = h$ and $\bar 1 = 2$. 

Equation~(\ref{ful_h_cur}) is valid at arbitrary temperatures $T_{1,2}$ and gaps $\Delta_{1,2}$ of superconductors. Now let us consider the case where the temperature difference $\delta T=\delta T_{1}-\delta T_{2}$ is small, $T_{\rm 1,2} = T + \delta T_{1,2}$, and extract the heat current proportional to $\delta T$ from Eq.~(\ref{ful_h_cur}). In superconductors, the gaps may vary by some $\delta \Delta_1$, $\delta \Delta_2$ with temperature $\delta T_{1,2}$, and one may ask whether such a variation has an effect on  current~(\ref{ful_h_cur})-(\ref{J_bdg}). We argue that this effect vanishes to the linear order in $\delta T$. Indeed, a virtual variation of gaps $\delta \Delta_{1,2}$ at fixed $\delta T = 0$ does not lead to the heat current because it would violate the second law of thermodynamics. 
%Indeed, these terms describe the heat current in response to variation of gaps $\delta \Delta_{L,R}$ at fixed and equal temperatures $T$ of the two superconductors. If they did not vanish, we would get a heat current between the two systems at equal temperatures in violation of the second law of thermodynamics. 
The second law of thermodynamics also requires that the heat current vanishes if $\delta T = 0$ at arbitrary $T$, i.e. $J_1 = J_2$. Therefore the integrands in Eq.~(\ref{h_cur_from_one_lead}) corresponding to $l=1$ and $l=2$ must be equal to each other. At $\delta T\neq 0$, this symmetry allows one to rewrite Eqs.~(\ref{ful_h_cur}) and (\ref{h_cur_from_one_lead}) only via the parameters corresponding, {\sl e.g.}, to the left lead \begin{align}
    J =  \frac{\delta T\,2}{T^2\,h} \int_{\Delta_{\rm max}}^\infty \frac{\varepsilon^3\, d\varepsilon}{\xi_1}\, \left[ j^{e}_1(\varepsilon)  + j^{h}_1(\varepsilon) \right] [-f'(x)]_{x = \varepsilon/T}\,,  \label{ful_h_cur_expanded}
\end{align}
where $\Delta_{\rm max} = {\rm max}(\Delta_1,\Delta_2)$. We substitute the scattering amplitudes~(\ref{amplitudes}) in Eq.~(\ref{ful_h_cur_expanded}) and introduce the heat conductance by relation $J=G_T^{SS}\,\delta T$, to find (see Appendix~\ref{sec:heatcur_ap} for details)
\begin{align}
    &G_T^{SS} =\frac{2}{T^2\,h}  \int_{\Delta_{\rm max}}^\infty\,d\varepsilon\,\frac{\varepsilon^2\xi_1\xi_2}{|D_{\varepsilon}|^2}\left[ \varepsilon^2\left(1-r_\varepsilon^2r_{-\varepsilon}^2\right) +\xi_1\xi_2\, t_{\varepsilon}^2 t_{-\varepsilon}^2 \right. \nonumber \\
    & -\left.\Delta_1\Delta_2\, t_{\varepsilon} t_{-\varepsilon} \left( \cos\delta\gamma_\varepsilon+r_{\varepsilon}\,r_{-\varepsilon}\,  {\cos \delta \eta_\varepsilon} \right)\cos\varphi\right] \left[ -f'(x)\right]_{x = \varepsilon/T}. \label{heatcurrent1}
\end{align}
Here, the superscript $SS$ denotes the superconductor-superconductor contact, and the subscript $T$ is used to distinguish the heat conductance $G_T^{SS}$ and the electric conductance $G$.  Equation~(\ref{heatcurrent1}) is written at arbitrary phase $\varphi$, particle-hole asymmetry, as well as possibly non-equal gaps, $\Delta_1 \neq \Delta_2$, at equilibrium; the denominator $D_{\varepsilon}$ is defined in Eq.~(\ref{denom}). 
%\begin{align}
%	J = \frac{\delta T}{\pi T^2} \int_{\Delta_{max}}^\infty d\varepsilon\, \varepsilon^2  \left[ \left( |t_{ee}^{21}|^2+|t_{he}^{21}|^2 \right)  + \left( |t_{hh}^{21}|^2+|t_{eh}^{21}|^2 \right) \right] \left[ - f'(\varepsilon/T)\right]. \label{heatcurrent}
%\end{align}
%The expressions~(\ref{heatcurrent0}) and (\ref{heatcurrent}) are equivalent, although the latter is more concise. 

%%%%%%%%%%%%%%%%%%%%%%%%%%%%%%%%%%%%%%%%%%%%%%%%%%%%%%%%%%%%%%%%%%%%%%%%%%%%%%%%%%%%%%
\subsection{Particle current} \label{sec:thermocur}
%%%%%%%%%%%%%%%%%%%%%%%%%%%%%%%%%%%%%%%%%%%%%%%%%%%%%%%%%%%%%%%%%%%%%%%%%%%%%%%%%%%%%%
The presence of a non-dissipative Josephson component of the current [\onlinecite{Josephson1962}] complicates the discussion of the particle current caused by a temperature gradient applied to a superconductor. The total current in a superconducting junction may be written as a sum of a dissipative $\tilde I(\varphi)$  and non-dissipative $\tilde{\tilde I} (\varphi)$ parts \footnote{In our work, we focus on the conventional Josephson junctions, where Eq.~(\ref{even_odd}) is applicable. We leave the analysis of more exotic cases, e.g. $\phi_0$ - junctions \cite{BuzdinPRB2003}, for future works. In such junctions, the current depends on the additional phase $\phi_0$ that breaks time-reversal symmetry. Then equation Eq.~(\ref{even_odd}) may be generalized  $I(\varphi,\varphi_0) = \tilde I(\varphi,\varphi_0) + \tilde{\tilde I} (\varphi,\varphi_0)$, where the dissipative and non-dissipative components satisfy the following parity conditions $\tilde I(-\varphi,-\varphi_0) = \tilde I(\varphi,\varphi_0)$ and $\tilde{\tilde I}(-\varphi,-\varphi_0) = -\tilde{\tilde I}(\varphi,\varphi_0)$.},
\begin{equation}
	I(\varphi) = \tilde I(\varphi) + \tilde{\tilde I} (\varphi) \label{even_odd}\,.
\end{equation}
One may distinguish the two contributions by their parity with respect to the  phase $\varphi$ reversal. The dissipative part $\tilde I(-\varphi) = \tilde I(\varphi)$ is an even, while the non-dissipative one, $\tilde{\tilde I}(-\varphi) = -\tilde{\tilde I}(\varphi)$, is an odd function of $\varphi$. Before focusing on the dissipative component of the current, which is the main subject of this work, we briefly discuss the non-dissipative component of the thermoelectric current.

{\it Non-dissipative currents.} As discussed in Sec.~\ref{sec:heatcur}, the temperature has a two-fold effect in superconductors: first, it induces variation of the superconducting gap, and second, it controls the quasiparticle occupation factors. 

Let us first illustrate the former effect using the weak-tunneling regime as an example. In that case, the non-dissipative Josephson current  may be written as \begin{equation}
    \tilde {\tilde I} = I_c(\Delta_1,\Delta_2) \sin \varphi, \label{Josephson_cur} 
\end{equation}
where $\varphi$ is the Josephson phase, and $I_c(\Delta_1,\Delta_2)$ is the critical current depending on the gaps in the leads. In response to the temperature variation $\delta T$, the superconducting gaps in respective leads may vary by $\delta \Delta_1$ and $\delta \Delta_2$ and induce a variation of the Josephson current, $\delta I = \left(\frac{\partial I_c}{\partial \Delta_1}\delta \Delta_1 + \frac{\partial I_c}{\partial \Delta_2}\delta \Delta_2\right)\sin \varphi$. Such a thermoelectric effect exists even in the case of a perfect particle-hole symmetry. In contrast, the conventional thermoelectric effect in normal metals relies on the particle-hole asymmetry.  

To appreciate the effect of the quasiparticle occupation factors, we notice first that a short weak link at a finite phase bias supports localized Andreev states, in addition to the propagating ones, coming from the opposite leads. An Andreev state contributes to the non-dissipative current across the junction, $I_A=-(2e/\hbar)(1-2f_A)(d\varepsilon_A/d\varphi)$. Here $\varepsilon_A(\varphi)<\Delta_{\rm L,R}$ is the energy of Andreev level,
%\footnote{The energy of Andreev level must be below the smallest of $\Delta_1$ and $\Delta_2$, in which case
%$\varepsilon^2=4\Delta_1\Delta_2\tau^2\sin^2\varphi [1-\tau^2\sin^2\varphi]/[4\tau^2\sin^2\varphi +(\sqrt{\Delta_1/\Delta_2}-\sqrt{\Delta_2/\Delta_1})^2]$
%}
and $f_A$ is the occupation factor. In equilibrium, $f_A=[1+\exp(-\varepsilon_A/T)]^{-1}$. At finite $\delta T$, the occupation factor $f_A$ of the localized state depends on the relaxation mechanism establishing the steady-state distribution or, in the absence of relaxation, on the heating protocol. In either case, the corresponding contribution to the non-dissipative current is not universal and is beyond the scope of this work.

%The effect of temperature variation $\delta T$ on the quasiparticle occupation is more involved. The Josepshon current~(\ref{Josephson_cur}) is carried by the localized Andreev states as well as delocalized quasiparticle states [\#\#\#\# TODO: insert proper reference]. The occupation of the Andreev states in the presence of the temperature gradient $\delta T$ depends on the heating protocol and requires the knowledge of the relaxation mechanisms.  It is beyond the scope of this work.

{\it Dissipative currents.} In this work, we focus on the dissipative part of the current fully determined by the delocalized quasiparticle states. This current may be evaluated using the Landauer scattering theory. Similar to Eqs.~(\ref{ful_h_cur})-(\ref{J_bdg}), we write the total charge current  as
\begin{align}
&	 I =  I_1 -  I_2,  \label{ful_c_cur} \\
&  I_l = \frac{2e}{h} \int_{\Delta_l}^\infty \frac{\varepsilon\, d\varepsilon}{\xi_l}\, \left[ \tilde i^{e}_l(\varepsilon)  - \tilde i^{h}_l(\varepsilon) \right] f\left(\varepsilon/T_s\right), \label{c_cur_one_lead} \\
&	i^b_l(\varepsilon) =   1 - |r^{bb}_{ll}(\varepsilon)|^2 + |r^{\bar bb}_{ll}(\varepsilon)|^2 +  \frac{2\Delta_l}{\varepsilon} {\rm Re}\left[r^{\bar b b}_{ll}(\varepsilon)\right]. \label{I_bdg}
\end{align}
Note that Eqs.~(\ref{c_cur_one_lead})-(\ref{I_bdg}) are written to the lowest-order in $\varepsilon/E_F$ (we address the role of the dropped terms $\propto\varepsilon/E_F$ in Appendix \ref{sec:bdg_appendix}). As in Sec.~\ref{sec:heatcur}, the two terms $I_1$ and $I_2$ correspond to the quasiparticles originating in the left and right leads labeled by the subscript $l\in \{1,2\}$.   Equation~(\ref{I_bdg}) has a meaning of a dimensionless current induced by an excited quasiparticle of type $b \in \{e,h\}$  ($\bar b$ denotes a particle-hole inversion of a quasiparticle branch, so $\bar e = h$ and $\bar h = e$). The tilde $\sim$ above the terms in Eq.~(\ref{c_cur_one_lead}) stands for taking an even-in-$\varphi$ part of the functions to obtain the dissipative current [see discussion below Eq.~(\ref{even_odd})]. The first three terms in Eq.~(\ref{I_bdg}) agree with the well-known expressions for NS junctions~[\onlinecite{BTK1982}]. 
% The last term in Eq.~(\ref{I_bdg}) produces the odd in $\varphi$ contribution to the current and, in practice, drops from the equations for the  dissipative currents. 

We assume that the temperature difference between the two superconductors $\delta T$ is small, $T_{1,2} = T \pm \delta T/2$, and evaluate the current proportional to $\delta T$. Similar to Section~\ref{sec:heatcur}, in the linear order in $\delta T$, we may disregard the influence of the temperature variation on the gaps in the leads. Furthermore, the parts of integrand in Eq.~(\ref{c_cur_one_lead}) corresponding, respectively, to the left and right leads must be equal each other at $\delta T = 0$. This allows us to rewrite Eqs.~(\ref{ful_c_cur}) and (\ref{c_cur_one_lead}) via the parameters corresponding to a single lead and expand in $\delta T$ (see  the Appendix \ref{sec:chargecur_ap} for details),
\begin{align}
	& I =  \frac{\delta T\,2e}{T^2\,h} \int_{\Delta_{\rm max}}^\infty \frac{\varepsilon^2\, d\varepsilon}{\xi_{1}}\, \left[ \tilde i^{e}_{1}(\varepsilon)   - \tilde i^{h}_{1}(\varepsilon) \right] [-f'(x)]_{x = \varepsilon/T}, \label{c_cur_expanded}
\end{align}
where $\Delta_{\rm max} = \max (\Delta_1,\Delta_2)$. Recall that Eq.~(\ref{c_cur_expanded}) is only the dissipative part of the current, and the notation $\sim$ stands for taking the even-in-$\varphi$ part of the functions. Finally, we substitute the scattering amplitudes (\ref{amplitudes}) in Eq.~(\ref{c_cur_expanded}) and obtain a simple expression
\begin{align}
	 I = \frac{\delta T\,2e}{T^2\,h} \int_{\Delta_{\rm max}}^\infty\,d\varepsilon\,\frac{\varepsilon^3\xi_1 \xi_2}{|D_{\varepsilon}|^2} \left( t^2_{\varepsilon} - t^2_{-\varepsilon}\right) \left[ -f'(x)\right]_{x = \varepsilon/T}, \label{thermopower01}
\end{align}
written for the arbitrary phase $\varphi$, particle-hole asymmetry, as well as possibly non-equal gaps $\Delta_1,\Delta_2$. 

A conventional Seebeck effect is impossible in a superconductor because of the presence of the superfluid condensate [\onlinecite{Ginzburg1944},\onlinecite{GinzburgRMP2004}]: a small temperature bias applied to a junction between two superconductors does not lead to a build-up of the chemical potential difference. It causes, however, a dissipative particle current, if the system lacks particle-hole symmetry. We will characterize the thermoelectric linear response by a ``current Seebeck coefficient'' $S_I^{SS}$ defined by a relation $I=S_I^{SS}\delta T$ (the superscript $SS$ stands for the superconductor-superconductor contact; the subscript $I$ denotes the current). Therefore, using Eq.~(\ref{thermopower01}), we obtain
\begin{align}
	 S_I^{SS} = \frac{1}{T^2}\frac{2e}{h} \int_{\Delta_{\rm max}}^\infty\,d\varepsilon\,\frac{\varepsilon^3\xi_1 \xi_2}{|D_{\varepsilon}|^2} \left( t^2_{\varepsilon} - t^2_{-\varepsilon}\right) \left[ -f'(x)\right]_{x = \varepsilon/T}, \label{thermopower1}
\end{align}
In the normal state, the current Seebeck coefficient is $S_I^N=GS$, where $S$ is the conventionally-defined Seebeck coefficient.

\section{Symmetric junction}
\label{sec:sj}
%%%%%%%%%%%%%%%%%%%%%%%%%%%%%%%%%%%%%%%%%%%%%%%%%%%%%%%%%%%%%%%%%%%%%%%%%%%%%%%%%%%%%%

We set $\Delta_1 = \Delta_2 = \Delta$ for a symmetric $S-S$ junction. We also assume a weak particle-hole asymmetry, { which we specify below}. 
%so we retain it only in the numerator of Eq.~(\ref{thermopower1}) and neglect it elsewhere, {\sl i.e.} set $t_{\varepsilon} = t =\sqrt \tau = {\rm const}$, where $\tau$ is the transmission coefficient of a single-channel contact.

%%%%%%%%%%%%%%%%%%%%%%%%%%%%%%%%%%%%%%%%%%%%%%%%%%%%%%%%%%%%%%%%%%%%%%%%%%%%%%%%%%%%%%
\subsection{Thermal conductance}
%%%%%%%%%%%%%%%%%%%%%%%%%%%%%%%%%%%%%%%%%%%%%%%%%%%%%%%%%%%%%%%%%%%%%%%%%%%%%%%%%%%%%%
\label{sqpc_heat_conductance}
%%%%%%%%%%%%%%%%%%%%%%%%%%%%%%%%%%%%%%%%%%%%%%%%%%%%%%%%%%%%%%%%%%%%%%%%%%%%%
\begin{figure}
(a) \includegraphics[width=0.9\linewidth]{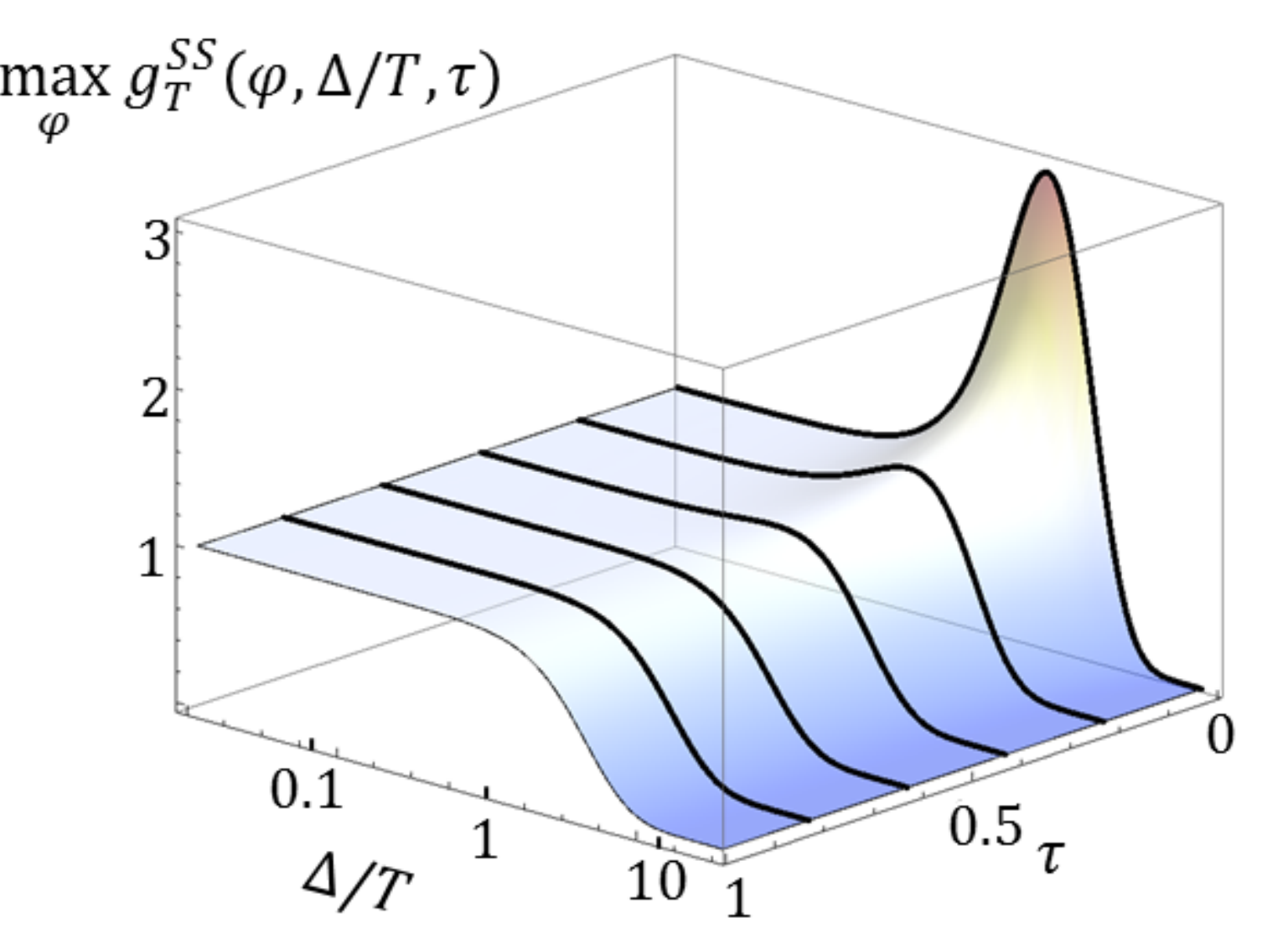} \\
\vspace{1cm}
(b) \includegraphics[width=0.9\linewidth]{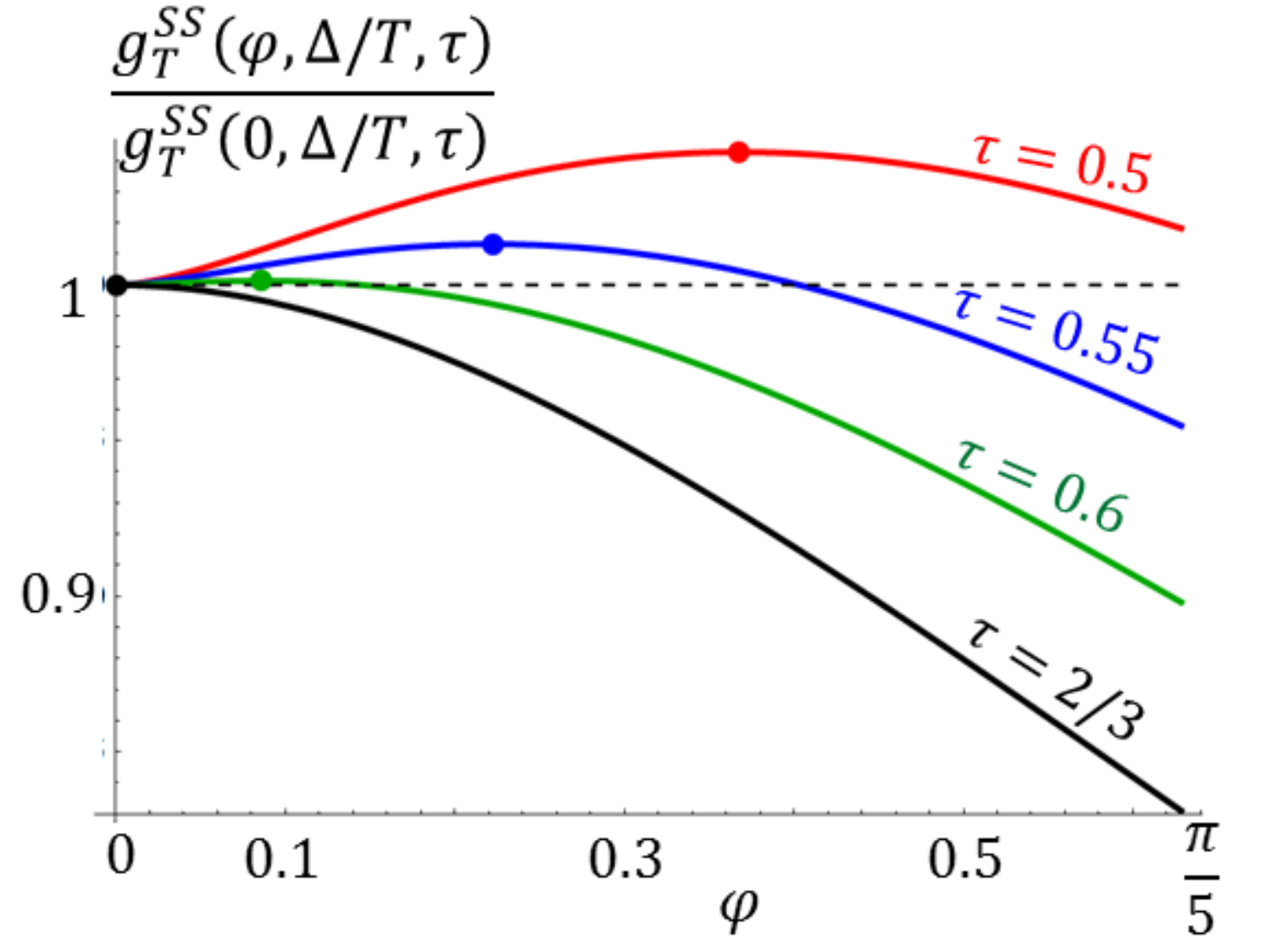}
\caption{Heat conductance of a superconducting point contact, see Eq.~(\ref{J_small_asymmetry}). (a) Normalized heat conductance ${\rm max}_{\varphi}g_T^{SS}(\varphi,\Delta/T,\tau)$ maximized over Josephson phase $\varphi$ {\sl vs.} transmission coefficient $\tau$ and ratio $\alpha = \Delta/T$. For $\tau<2/3$, the dependence is non-monotonic in $\alpha$. (b) Dependence of $g^{SS}_T(\varphi,\Delta/T,\tau)$ on $\varphi$ close to $\tau = 2/3$ at fixed $\alpha = \Delta/T = 10$.} 
\label{fig:heatcurrent_vs_delta}
\end{figure}
%%%%%%%%%%%%%%%%%%%%%%%%%%%%%%%%%%%%%%%%%%%%%%%%%%%%%%%%%%%%%%%%%%%%%%%%%%%%%
Superconductivity opens a gap in the excitations spectrum. Since the heat current is carried by quasiparticles, and the presence of superconducting gap reduces their density (at a given temperature),  one could expect that superconductivity also suppresses the heat conductance relative to its value in the absence of the gap. Contrary to this intuition, the heat current of a superconducting contact at $T\sim\Delta$ may even exceed that  of the contact in its normal state, as we discuss below. 

{The small particle-hole asymmetry is not essential for the heat conductance, so, in the leading order, we neglect it within the present subsection. In other words, we assume that the parameters entering the scattering matrix~(\ref{s01}) are energy-independent, i.e. we set $t_{\xi} = \sqrt \tau$=const, $\gamma_\xi =$const, $\eta_{\xi}=$const,  where $\tau$ is the transmission coefficient of a single-channel contact.} We normalize the heat conductance Eq.~(\ref{heatcurrent1}) by its normal-state value $G_T$:
\begin{align}
&G^{SS}_T=G_T\, g_T^{SS}(\varphi,\Delta/T,\tau), \nonumber\\
&g_T^{SS}(\varphi,\alpha,\tau)=-\frac{6}{\pi^2}\int_\alpha^\infty dx\,  x^2 \left[1+\frac{\alpha^2(2-3\tau)\sin^2\varphi/2}{x^2-\alpha^2+\alpha^2\tau\sin^2\varphi/2}\right.\nonumber\\
&\left.-\frac{2\alpha^4\tau(1-\tau)\sin^4\varphi/2}{(x^2-\alpha^2+\alpha^2\tau\sin^2\varphi/2)^2}\right]f'(x)\,,\,\,\,\alpha=\frac{\Delta}{T}.
\label{J_small_asymmetry} 
\end{align}
Here $G_T$ is the normal-state heat conductance, satisfying the Wiedemann-Franz law,
\begin{equation}
G_T=\frac{\pi^2}{3}\frac{T G}{e^2}\,,\,\,G =\frac{2 e^2}{h}\tau\,,\,\, L=\frac{\pi^2}{3e^2}\,,
\label{WFL} 
\end{equation}
and $L$ is the conventionally-defined Lorenz number of a normal-state conductor. 

The heat conductance $G^{SS}_T$ at a finite gap $\Delta$ differs from the normal-state value $G_T$ at the same temperature by the dimensionless factor $g_T^{SS}(\varphi,\Delta/T,\tau)$. Henceforth, the capitalized  variables,  e.g. $G_T$ and $G$, denote dimensionful quantities, whereas the variables in lower-case, e.g. $g_T$ and $g$, denote their dimensionless variants. After some algebra, function $g_T^{SS}(\varphi,\Delta/T,\tau)$ can be reduced to the respective expression obtained in Refs.~[\onlinecite{SaulsPRL2003},\onlinecite{SaulsPRB2004}]. The latter was derived within a semiclassical theory, formally applicable only to point contacts containing a large number of quantum channels. A similar correspondence between the results of semiclassical theory and the scattering-matrix quantum theory for single-channel contacts was established quite some time ago for the equilibrium Josephson current~[\onlinecite{Beenakker1991}]. For a detailed analysis of the function $g_T^{SS}(\varphi,\Delta/T,\tau)$, we refer the reader to Refs.~[\onlinecite{SaulsPRL2003},\onlinecite{SaulsPRB2004}] (see also Ref. [\onlinecite{Virtanen2015}], where the heat current noise was analyzed). Here we only mention several noteworthy observations evident from Eq.~(\ref{J_small_asymmetry}).~\footnote{Equation~(\ref{J_small_asymmetry}) generalizes the result previously derived using the  tunneling Hamiltonian approach \cite{Maki1966,GuttmanPRB1997b} to arbitrary transparency $\tau$. Note that an incorrect sign was obtained in front of the phase $\varphi$ dependent term in Ref.~[\onlinecite{GuttmanPRB1997b}] and was subsequently corrected by Refs.~[\onlinecite{SaulsPRL2003},\onlinecite{SaulsPRB2004}].}

The first two terms in the square brackets of the integrand of Eq.~(\ref{J_small_asymmetry}) give the leading terms in the  asymptotic behavior of $g_T^{SS}(\varphi,\Delta/T,\tau)$ at $\varphi\to 0$ or at $\Delta/T\to 0$. The leading asymptote at $\Delta/T\to 0$ and fixed $\varphi$ is
\begin{equation}
g_T^{SS}(\varphi,\Delta/T,\tau) =  1 + (2-3\,\tau)\,\frac{3}{\pi^2}\left(\frac{\Delta}{T}\right)^2\sin^2\frac{\varphi}{2}.
\label{F_asymp_small_alpha}
\end{equation}
At $\tau<2/3$ and $\varphi\neq 0$, the opening of the gap results in an increase of thermal conductance.
%; the effect is maximal at the phase difference $\varphi=\pi$. 
Upon further increase of the gap, quasiparticles freeze-out, so the overall temperature dependence of $g_T^{SS}(\varphi,\Delta/T, \tau)$ is non-monotonic at $\tau<2/3$. 
%At fixed $\tau$, $T$, and $\Delta$, the maximal value ${\rm max}_{\varphi}\left[g^{SS}_T(\varphi,\Delta,T,\tau)\right]$. 
%is reached at the phase difference $\varphi=\pi$. 
We plot this maximal value ${\rm max}_{\varphi}\left[g^{SS}_T(\varphi,\Delta/T,\tau)\right]$ as a function of $\tau$ and $\alpha = \Delta/T$ in Fig.~\ref{fig:heatcurrent_vs_delta}(a).

The leading term of the phase dependence of $g_T^{SS}(\varphi,\Delta/T, \tau)$ at $\varphi\to 0$ and fixed $\alpha = \Delta/T$ scales $\propto \frac{3}{2\pi^2}\alpha^3 f'(\alpha) (2-3\tau)\varphi^2\ln(\varphi)$ and is not analytical  in $\varphi$ at any finite $\Delta$. A more detailed analysis shows that the phase dependence becomes non-monotonic in the vicinity of $\varphi=0$ once $\tau$ becomes smaller than $2/3$ as shown in Fig.~\ref{fig:heatcurrent_vs_delta}(b). 

In the end of this section we note that, with two modifications, Eq. (\ref{J_small_asymmetry}) is applicable to a contact of two s-wave superconductors connected by a short junction formed by a single helical edge of a topological insulator. The first modification is that one has to set $\tau=1$ in that equation, assuming there are no magnetic barriers which would cause backscattering of the edge electrons [\onlinecite{FuPRB2008}]. 
The second modification is the need of an overall factor $1/2$ in Eq. (\ref{J_small_asymmetry}) reflecting the absence of spin degeneracy for a helical channel. These two modifications indeed reduce Eq. (\ref{J_small_asymmetry}) to the result of Ref. [\onlinecite{Sothman2016}] devoted to the thermal transport across a topological Josephson junction.

%{\bf maybe restore numerical coefficients in the asymptote and write in a separate line; maybe add a figure for the phase dependence at $\tau=2/3$ and several $\alpha$; or fixed $\alpha$ and several $\tau$ around $\tau=2/3$}

%%%%%%%%%%%%%%%%%%%%%%%%%%%%%%%%%%%%%%%%%%%%%%%%%%%%%%%%%%%%%%%%%%%%%%%%%%%%%%%%%%%%%%
\subsection{Particle current response to temperature bias} \label{sec:particle_current_SXS}
%%%%%%%%%%%%%%%%%%%%%%%%%%%%%%%%%%%%%%%%%%%%%%%%%%%%%%%%%%%%%%%%%%%%%%%%%%%%%%%%%%%%%%

{The presence of the particle-hole asymmetry of the scattering amplitude $t_{\xi}$ is essential for thermopower. This is reflected in the numerator of the integrand in Eq~(\ref{thermopower1}), which vanishes if $t_{\varepsilon} = t_{-\varepsilon}$. In this subsection we assume that the particle-hole asymmetry is weak, $\Delta||\partial s_{\xi}/\partial\xi||\ll 1$, i.e., the scattering matrix varies slowly on the energy scale of the superconducting gap $\Delta$. We seek to evaluate the thermopower coefficient $S_I^{SS}$ in the leading order in $\Delta||\partial s_{\xi}/\partial\xi||\ll 1$. That amounts to accounting for the particle-hole asymmetry in the numerator of the integrand of Eq.~(\ref{thermopower1}), where we write $t^2_{\varepsilon} - t^2_{-\varepsilon} = 2\varepsilon \frac{\partial \tau_{\varepsilon}}{\partial \varepsilon}$, but disregarding it in the denominator, where we set $t_{\varepsilon} = t_{-\varepsilon} = \sqrt{\tau}=$const, $\eta_{\varepsilon} =$const and $\gamma_{\varepsilon}=$const.} 

%In our case, the particle-hole symmetry is violated by the energy dependence of the normal-state scattering matrix Eq.~(\ref{s01}). We assume here a weak violation of the symmetry, $\Delta||\partial s_{\xi}/\partial\xi||\ll 1$, and evaluate the particle current in the leading order in $||\partial s_{\xi}/\partial\xi||$. That amounts to accounting for the particle-hole asymmetry in the numerator of the integrand of Eq.~(\ref{thermopower1}), but disregarding it in the denominator. Such consideration excludes a small range of phase differences around $\varphi=0$. 

The failure of the leading-order approximation near $\varphi=0$ is due to the appearance of a shallow Andreev level in the spectrum of excitations. Its energy $\varepsilon_A$ is a root of the denominator  $D_\varepsilon(\varphi)$ appearing in Eq.~(\ref{thermopower1}). If $s_\xi$ of Eq.~(\ref{s01}) is independent of energy, the level merges with the continuum at $\varphi=0$. This leads to $D_{\varepsilon}(0)\propto (\varepsilon-\Delta)$ near the spectral edge and to the divergence of the integral in Eq.~(\ref{thermopower1}). Accounting for a small $\partial s_{\xi}/\partial\xi$ makes the difference $\Delta-\varepsilon_A$ finite at any $\varphi$, and approximately independent of the phase difference in the small-$\varphi$ domain of the width $\varphi_0\sim{\rm max}\{\Delta |d\ln\tau/d\xi|, \Delta |d\gamma/d\xi|,\Delta |d\eta/d\xi|\}$. Below we concentrate on $|\varphi|$ outside the domain $\varphi_0$ where the results are independent of the minute details of the scattering matrix. Estimates of the particle current within that domain can be obtained by setting $|\varphi|\sim\varphi_0$ in Eqs.~(\ref{I_small_asymmetry}) and (\ref{G_asymp_low_tau}) of this Section.

%To retain the particle-hole asymmetry terms only in the numerator,  we expand $\tau_{\varepsilon}-\tau_{-\varepsilon} = 2\varepsilon \frac{\partial \tau}{\partial \mu}$. 
Then, with the assumption of equal gaps, we obtain
\begin{align}
& S_I^{SS} = G\,S\,s^{SS}(\varphi,\Delta/T,\tau)\,,\,\,\, S=\frac{\pi^2}{3}\frac{\partial\ln G}{\partial \mu}\frac{T}{e} \,,\label{I_small_asymmetry} \\ 
& s^{SS}(\varphi,\alpha,\tau) = - \frac{6}{\pi^2}\int_\alpha^\infty   dx\,\frac{x^4(x^2-\alpha^2)\,f'(x)}{(x^2-\alpha^2+\alpha^2\tau\sin^2\varphi/2)^2}. \nonumber 
% Do we want to present H^{SS} in the extended form?
\end{align}
Here $\mu$ is the chemical potential, $S$ is the normal-state Seebeck coefficient given by the Mott formula, and function $s^{SS}$ describes the modification introduced by superconductivity; $s^{SS}(\varphi,0,\tau)=1$. 
%%%%%%%%%%%%%%%%%%%%%%%%%%%%%%%%%%%%%%%%%%%%%%%%%%%%%%%%%%%%%%%%%%%%%%%%%%%%%
\begin{figure}
 \includegraphics[width=0.9\linewidth]{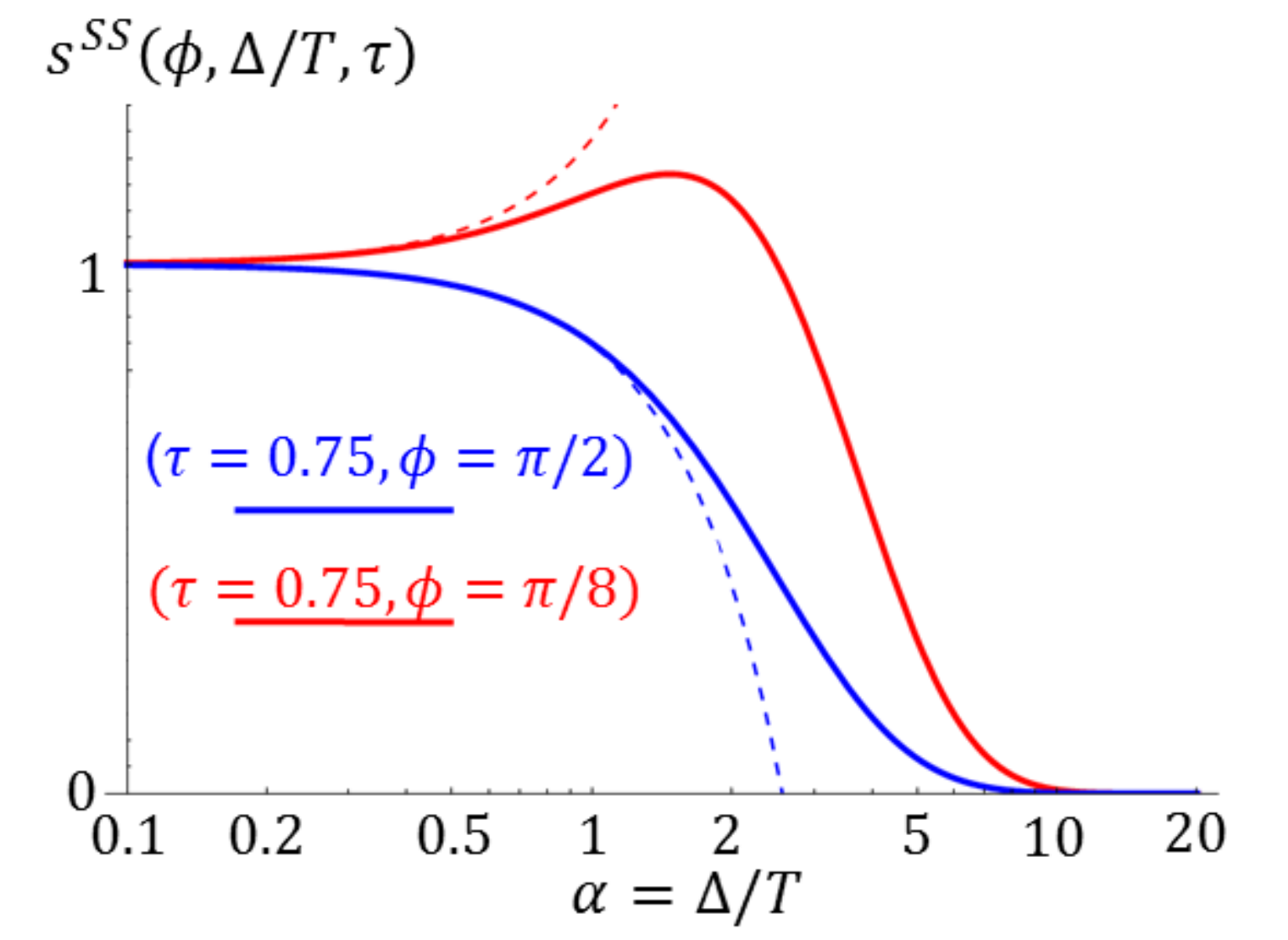} 
\caption{Normalized current Seebeck coefficient $s^{SS}(\varphi,\alpha,\tau)$ as a function of $\alpha = \Delta/T$, see Eq.~(\ref{I_small_asymmetry}). Asymptotes~(\ref{G_asymp_small_alpha}) are shown in faint dashed lines.} 
\label{fig:chargecurrent_vs_delta}
\end{figure}
%%%%%%%%%%%%%%%%%%%%%%%%%%%%%%%%%%%%%%%%%%%%%%%%%%%%%%%%%%%%%%%%%%%%%%%%%%%%%
A ``na\"ive'' tunneling limit corresponds to setting $\tau=0$ in the argument of $s^{SS}$,
\begin{align}
s^{SS}(\varphi,\alpha,0) = -\frac{6}{\pi^2}\int_\alpha^\infty  dx\,\frac{x^4}{x^2-\alpha^2}f'(x). \label{G_tau_dropped}
\end{align}
In this approximation, $s^{SS}$ is expectedly logarithmically divergent, in agreement with the result of Ref.~[\onlinecite{SmithPRB1980}]. Equation~(\ref{G_tau_dropped}) manifestly disagrees with [\onlinecite{GuttmanPRB1997a}], where a convergent factor, $s^{SS}(\varphi,\alpha,0) = - (6/\pi^2)\int_\alpha^\infty  dx\,x^2\,f'(x)$, was found within the tunneling Hamiltonian formalism. The root of this inconsistency lies in the disparate scattering amplitudes used in Refs. [\onlinecite{SmithPRB1980}] and  [\onlinecite{GuttmanPRB1997a}]. In Appendix \ref{sec:bdg_appendix}, we demonstrate that the scattering amplitudes imposed by the tunneling Hamiltonian approach used in Ref.~[\onlinecite{GuttmanPRB1997a}] correspond to the $k_Fd \rightarrow 0$ limit, where $d$ is the thickness of the tunneling barrier. This limit completely misses the appearance, even at zero phase bias, of shallow Andreev levels induced by the particle-hole asymmetry. We demonstrate that the  logarithmic terms are recovered already in the $\propto (k_Fd)$ correction to particle current. Given that $d$ is finite in any physical device, we favor the approach of Ref.~[\onlinecite{SmithPRB1980}].

In the proper   asymptotic evaluation of Eq.~(\ref{I_small_asymmetry}) at $\tau\rightarrow 0$,
\begin{align}
s^{SS}(\varphi,\Delta/T,\tau) = -\frac{3}{\pi^2}\!\left(\frac{\Delta}{T}\right)^3\!f'\left(\frac{\Delta}{T}\right){\rm ln} \left(\frac{2T}{\Delta\,\tau\sin^2\varphi/2}\right)\,; 
\label{G_asymp_low_tau}
\end{align}
the divergence at $\varphi=0$ is regularized by a finite difference $\Delta-\varepsilon_A$, as was mentioned in the beginning of this Section. 

Equation~(\ref{G_asymp_low_tau}) hints that the particle current response in the superconducting state may exceed that of the junction in its normal state. To show explicitly that $s^{SS}>1$ is possible, we present here the $\Delta/T\ll 1$ asymptote, valid at arbitrary $\tau$:
\begin{equation}
s^{SS}(\varphi,\Delta/T,\tau) =  1 + \left(1-2\tau \sin^2\frac{\varphi}{2}\right)\frac{3}{\pi^2}\left(\frac{\Delta}{T}\right)^2\,. 
\label{G_asymp_small_alpha} 
\end{equation}
Function $g^{SS}(\varphi,\Delta/T,\tau)$ for the full range of variation of $\Delta/T$ and two different sets of parameters $\tau$ and $\varphi$ is plotted in Fig.~\ref{fig:chargecurrent_vs_delta} and clearly shows the possibility of a non-monotonic variation.

%%%%%%%%%%%%%%%%%%%%%%%%%%%%%%%%%%%%%%%%%%%%%%%%%%%%%%%%%%%%%%%%%%%%%%%%%%%%%%%%%%%%%%
\section{NS junction, $\Delta_1=0$, $\Delta_2\neq 0$.} 
\label{sec:nsj}
%%%%%%%%%%%%%%%%%%%%%%%%%%%%%%%%%%%%%%%%%%%%%%%%%%%%%%%%%%%%%%%%%%%%%%%%%%%%%%%%%%%%%%

%%%%%%%%%%%%%%%%%%%%%%%%%%%%%%%%%%%%%%%%%%%%%%%%%%%%%%%%%%%%%%%%%%%%%%%%%%%%%
\begin{figure}
\includegraphics[width=0.9\linewidth]{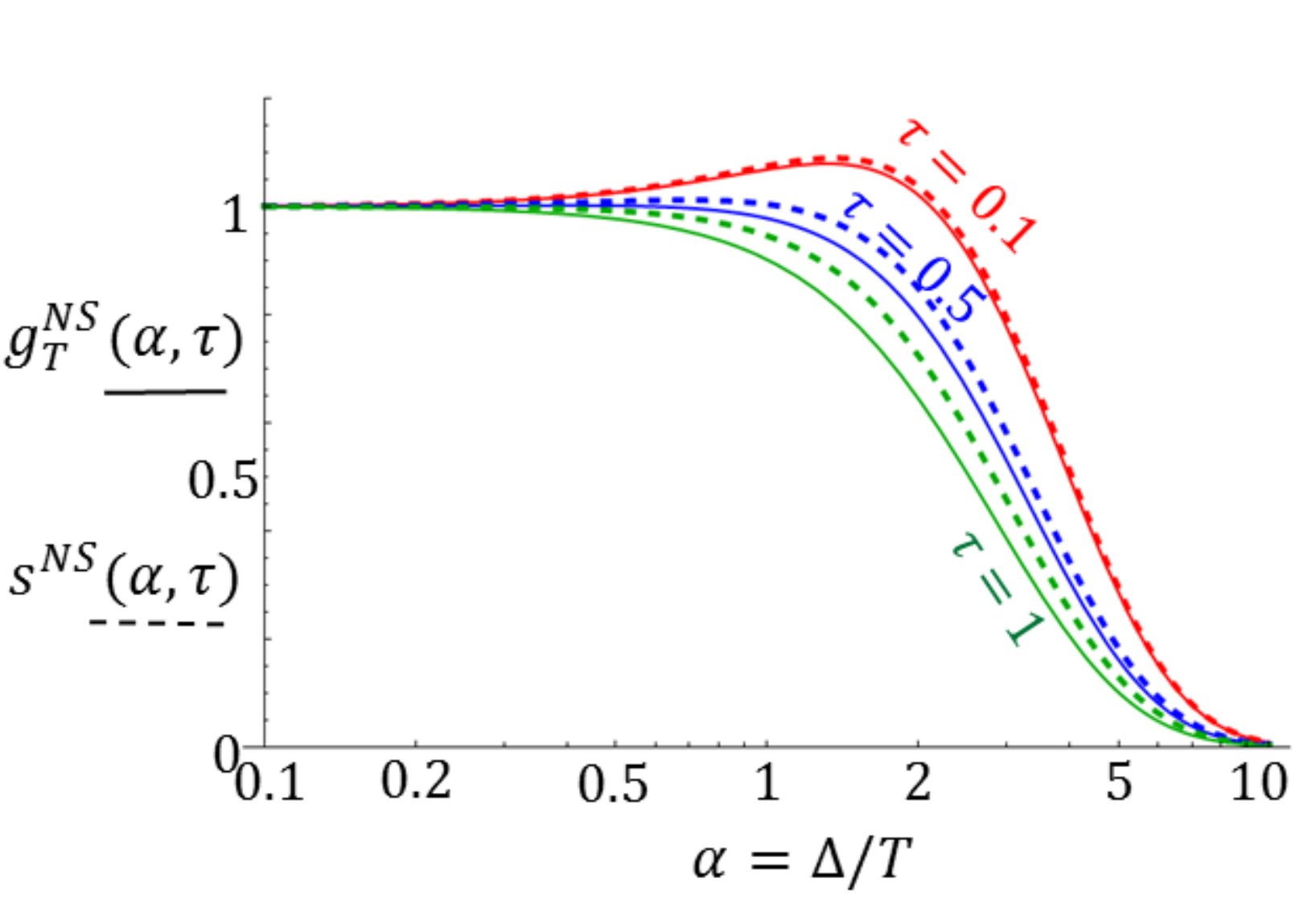}
\caption{Plots of the normalized heat conductance $g_T^{NS}(\Delta/T,\tau)$ (solid) and the thermoelectric coefficient $s^{NS}(\Delta/T,\tau)$ (dashed), see Eqs.~(\ref{J_small_asymmetry_NS}) and (\ref{I_small_asymmetry_NS}).} 
\label{fig:charge_and_heat_current_vs_delta_NS}
\end{figure}
%%%%%%%%%%%%%%%%%%%%%%%%%%%%%%%%%%%%%%%%%%%%%%%%%%%%%%%%%%%%%%%%%%%%%%%%%%%%%
To find the heat conductance and the current Seebeck coefficient of NS junction, we set $\Delta_1=0$ and $\xi_1=\varepsilon$ in Eqs.~(\ref{heatcurrent1}) and (\ref{thermopower1}). We also simplify notations by replacing $\Delta_2\to\Delta$. Furthermore, assuming weak particle-hole asymmetry, we keep the corresponding terms only in the numerator of Eq.~(\ref{thermopower1}).

%%%%%%%%%%%%%%%%%%%%%%%%%%%%%%%%%%%%%%%%%%%%%%%%%%%%%%%%%%%%%%%%%%%%%%%%%%%%%
\subsection{Heat conductance of NS junction}
%%%%%%%%%%%%%%%%%%%%%%%%%%%%%%%%%%%%%%%%%%%%%%%%%%%%%%%%%%%%%%%%%%%%%%%%%%%%%
After the said simplifications, we find
\begin{align}
    &G_T^{NS}(\Delta,T,\tau) = G_T \, g_T^{NS}(\Delta/T,\tau),   
    \nonumber \\ 
 &g_T^{NS}(\alpha,\tau) = -  \frac{12}{\pi^2}
   \label{J_small_asymmetry_NS}\\
%    &- \frac{6}{\pi^2} \int_{\alpha}^{\infty} dx  x^2 \left\{ 1 + \frac{2\,(2-3\tau)\sqrt{x^2-\alpha^2}(x-\sqrt{x^2-\alpha^2})}{[\tau x+(2-\tau)\sqrt{x^2-\alpha^2})]^2}\right. \nonumber \\
%    & \left. - \frac{\tau^2(x-\sqrt{x^2-\alpha^2})^2}{[\tau x+(2-\tau)\sqrt{x^2-\alpha^2})]^2}\right\}f'(x).
&\times \int_{\alpha}^{\infty} dx  \frac{x^2\sqrt{x^2-\alpha^2}\left[x(2-\tau)+\tau\sqrt{x^2-\alpha^2}\right]}{\left[\tau x+(2-\tau)\sqrt{x^2-\alpha^2}\right]^2} f'(x), 
 \nonumber
\end{align}
where the heat conductance in the normal-state $G_T$ is defined in Eq.~(\ref{WFL}). 
%the heat-conductance in the normal state, see the definition in Eq.~(\ref{J_small_asymmetry}). {\it \#\# For now, I keep $F^{NS}$ in two forms. The first form simplifies evaluation of asymptotics at small $\alpha$. The second form is more concise and is better suited to evaluate asymptotics at large $\alpha$.\#\#} 
Function $g_T^{NS}(\Delta/T,\tau)$ describes the deviation of the heat conductance from its value for the junction in the normal state at the same $T$. (For the case of the NS boundary, the phase $\varphi$ is absent because it can be gauged away from the problem.)

At small gap, $\alpha =\Delta/T\ll 1$, the leading asymptotic behavior of $g_T^{NS}$ is
\begin{align}
    g_T^{NS}(\Delta/T,\tau) &= 1 + (2-3\tau)\,\frac{3}{4\pi^2}\left(\frac{\Delta}{T}\right)^2 \,. 
    \label{F_NS_asymp_small_alpha}
%    \\
%    F_{NS}(\alpha,\tau) &= \frac{\pi^{1/2}\,2^{3/2}\,\alpha^{5/2}\,e^{-\alpha}}{2-\tau},\quad {\rm if}\,\, \frac{1}{\tau^2} \gg \alpha \gg 1, \\
%    F_{NS}(\alpha,\tau) &= \frac{\pi^{1/2}\,2^{3/2}\,\alpha^{3/2}\,(2-\tau)\,e^{-\alpha}}{\tau^2},\quad {\rm if}\,\,  \alpha \gg \frac{1}{\tau^2},
\end{align}
Therefore, we find that the ``high-temperature" heat conductance of the NS junction behaves in a similar way to the heat conductance of a superconducting quantum point contact given by Eq.~(\ref{F_asymp_small_alpha}). At $\tau<2/3$, the heat conductance grows when the gap opens. Combined with the fact that the heat conductance is exponentially suppressed at large $\Delta/T\gg 1$, we obtain a non-monotonic dependence as illustrated by Fig.~\ref{fig:charge_and_heat_current_vs_delta_NS}. In contrast, at a higher transmission coefficient, $\tau>2/3$, the heat conductance is a monotonic function of $\Delta/T$. To our surprise, we find the same as in Sec.~\ref{sqpc_heat_conductance} value $\tau = 2/3$ to separate the domains of a monotonic and non-monotonic behavior in $\Delta/T$.

The details of the low-temperature ($\Delta/T=\alpha\gg 1$) behavior of $g_T^{NS}$ depend on the relative smallness of the two parameters, $\tau$ and $1/\sqrt\alpha$; their ratio defines the quasiparticles energy interval most effective in the heat transfer. To capture the entire crossover behavior as a function of $\tau\sqrt\alpha$, we present the low-temperature asymptote of $g_T^{NS}$ in the form:
\begin{align}
  g_T^{NS}(\alpha,\tau) &=
    \frac{6\sqrt{2}\,\alpha^{5/2}\,e^{-\alpha}}{\pi^{3/2}(2-\tau)} \, h\left(\frac{\tau}{2-\tau}\sqrt{\frac{\alpha}{2}}\right);  \label{F_NS_asymp_large_alpha} \\
    h(\beta) &= \int_{0}^\infty \frac{dx}{\sqrt\pi}\,\frac{\sqrt x e^{-x}}{(\sqrt x +\beta)^2}, \,\,\beta =\frac{\tau}{2-\tau}\sqrt{\frac{\alpha}{2}}.  \nonumber
  %\sqrt{8\pi\alpha^3} e^{-\alpha} \left\{\begin{array}{cc}
  %        \frac{2-\tau}{\tau^2}, & {\rm if}\,\alpha \gg \tau^{-2}, \\
%		  \frac{\alpha}{2-\tau}, &\quad {\rm if}\,\tau^{-2}\gg \alpha \gg 1,
%	  \end{array}\right. 
	  %\\
 % F^{NS}(\alpha,\tau) &= \frac{\pi^{1/2}\,2^{3/2}\,\alpha^{5/2}\,e^{-\alpha}}{2-\tau},\quad {\rm if}\,\, \frac{1}{\tau^2} \gg \alpha \gg 1, \\
 %   F^{NS}(\alpha) &= \frac{\pi^{1/2}\,2^{3/2}\,\alpha^{3/2}\,(2-\tau)\,e^{-\alpha}}{\tau^2},\quad {\rm if}\,\,  \alpha \gg \frac{1}{\tau^2},
\end{align}
The crossover function $h(\beta)$ here varies from $h(0)=1$ to $h(\beta)=1/(2\beta^2)$ at $\beta\gg 1$.

%%%%%%%%%%%%%%%%%%%%%%%%%%%%%%%%%%%%%%%%%%%%%%%%%%%%%%%%%%%%%%%%%%%%%%%%%%%%%
\subsection{Particle current driven by temperature bias across NS junction}
%%%%%%%%%%%%%%%%%%%%%%%%%%%%%%%%%%%%%%%%%%%%%%%%%%%%%%%%%%%%%%%%%%%%%%%%%%%%%
Similar to $S_I^{SS}$ of Sec.~\ref{sec:particle_current_SXS}, we define the current Seebeck coefficient for $NS$ junction by relation $I=S_I^{NS}\delta T$ and then normalize it by the corresponding value in the normal state at the same temperature, $S_I^N=GS$. From Eq.~(\ref{thermopower1}), we obtain $S_{I}^{NS}$ for the NS junction 
\begin{align}
&S_I^{NS} =  GS\, s^{NS}(\Delta/T,\tau)\,,\,\,\, S=\frac{\pi^2}{3}\frac{\partial\ln G}{\partial \mu}\frac{T}{e} \,,\label{I_small_asymmetry_NS} \\
&s^{NS}(\alpha,\tau) 
%\nonumber \\ & - \frac{6}{\pi^2} \int_{\alpha}^{\infty} dx\, x^2\, \left\{1 + \frac{2\alpha^2 (1-\tau)}{\left[\tau x+(2-\tau)\sqrt{x^2-\alpha^2}\right]^2}\right. \nonumber \\
%    &\left.- \frac{(\tau^2-2\tau+2)(x-\sqrt{x^2-\alpha^2})^2}{\left[\tau x+(2-\tau)\sqrt{x^2-\alpha^2})\right]^2} \right\} f'(x)
%    \label{I_small_asymmetry_NS}
%  \\ &
  = -\frac{24}{\pi^2}\!\int_{\alpha}^{\infty}\!\!\! dx  \frac{x^3\sqrt{x^2-\alpha^2}}{\left[\tau x+(2-\tau)\sqrt{x^2-\alpha^2}\right]^2} f'(x). \nonumber
\end{align}
%where $G$ and $S$ are the conventional conductance and Seebeck coefficient in the normal state. {\it \#\# For now, I keep $H^{NS}$ in two forms. The first form simplifies evaluation of asymptotics at small $\alpha$. The second form is more concise and is better suited to evaluate asymptotics at large $\alpha$.\#\#}
Clearly, at $\alpha=0$, factor $s^{NS}=1$ regardless the value of $\tau$. Opening of a small gap results in a positive correction to the $s^{NS}=1$ value at any $\tau\neq 1$; the corresponding asymptote of $s^{NS}(\Delta/T,\tau)$ at $\Delta/T\ll 1$ is
\begin{align}
    s^{NS}(\Delta/T,\tau) &= 1 +  (1-\tau)\frac{3}{2\pi^2}\left(\frac{\Delta}{T}\right)^2\,. \label{H_NS_small_alpha}
%    H_{NS}(\alpha,\tau) &= \frac{\pi^{1/2}\,2^{5/2}\,\alpha^{5/2}\,e^{-\alpha}}{(2-\tau)^2},\quad {\rm if}\,\, \frac{1}{\tau^2} \gg \alpha \gg 1, \\
%    H_{NS}(\alpha,\tau) &= \frac{\pi^{1/2}\,2^{5/2}\,\alpha^{5/2}\,e^{-\alpha}}{\tau^2},\quad {\rm if}\,\,  \alpha \gg \frac{1}{\tau^2},
\end{align}
In the opposite limit of low temperatures, $\Delta/T\gg 1$, the small quasiparticle density results in an exponential suppression of $s^{NS}$,
\begin{align}
  s^{NS}(\alpha,\tau) &= \frac{12\sqrt{2}\,\alpha^{5/2}\,e^{-\alpha}}{\pi^{3/2}(2-\tau)^2} \, h\left(\frac{\tau}{2-\tau}\sqrt{\frac{\alpha}{2}}\right),
\label{H_NS_asymp_large_alpha}
\end{align}
with the same crossover function $h$ as in Eq.~(\ref{F_NS_asymp_large_alpha}).

By comparing the asymptotic behavior in Eqs.~(\ref{F_NS_asymp_large_alpha}) and (\ref{H_NS_asymp_large_alpha}), we find that $g_T^{NS}/s^{NS} = (2-\tau)/{2}$ for $\alpha \gg 1$. Moreover, the functions become identical, $g_T^{NS}(\alpha,\tau) = s^{NS}(\alpha,\tau)$ at any $\alpha$, if $\tau=0$. The comparison of numerically evaluated plots of $g_T^{NS}$ and $s^{NS}$ in Fig.~\ref{fig:charge_and_heat_current_vs_delta_NS} demonstrates that they behave similarly. (We mention in passing that at $\tau=1$ the leading correction shown in Eq.~(\ref{H_NS_small_alpha}) is replaced by $-(24/35\pi^2)(\Delta/T)^3$.)

Note that the Peltier effect in NS junctions, which is Onsager-reciprocal to the thermoelectric effect discussed in our work, was considered in Ref.~[\onlinecite{Averin1995}]. In that study, a non-monotonic dependence of the heat current on $\alpha = \Delta/T$ was also obtained.

%%%%%%%%%%%%%%%%%%%%%%%%%%%%%%%%%%%%%%%%%%%%%%%%%%%%%%%%%%%%%%%%%%%%%%%%%%%%%
\subsection{Lorenz number and Seebeck coefficient}
%%%%%%%%%%%%%%%%%%%%%%%%%%%%%%%%%%%%%%%%%%%%%%%%%%%%%%%%%%%%%%%%%%%%%%%%%%%%%
%%%%%%%%%%%%%%%%%%%%%%%%%%%%%%%%%%%%%%%%%%%%%%%%%%%%%%%%%%%%%%%%%%%%%%%%%%%%%
\begin{figure}
\includegraphics[width=0.9\linewidth]{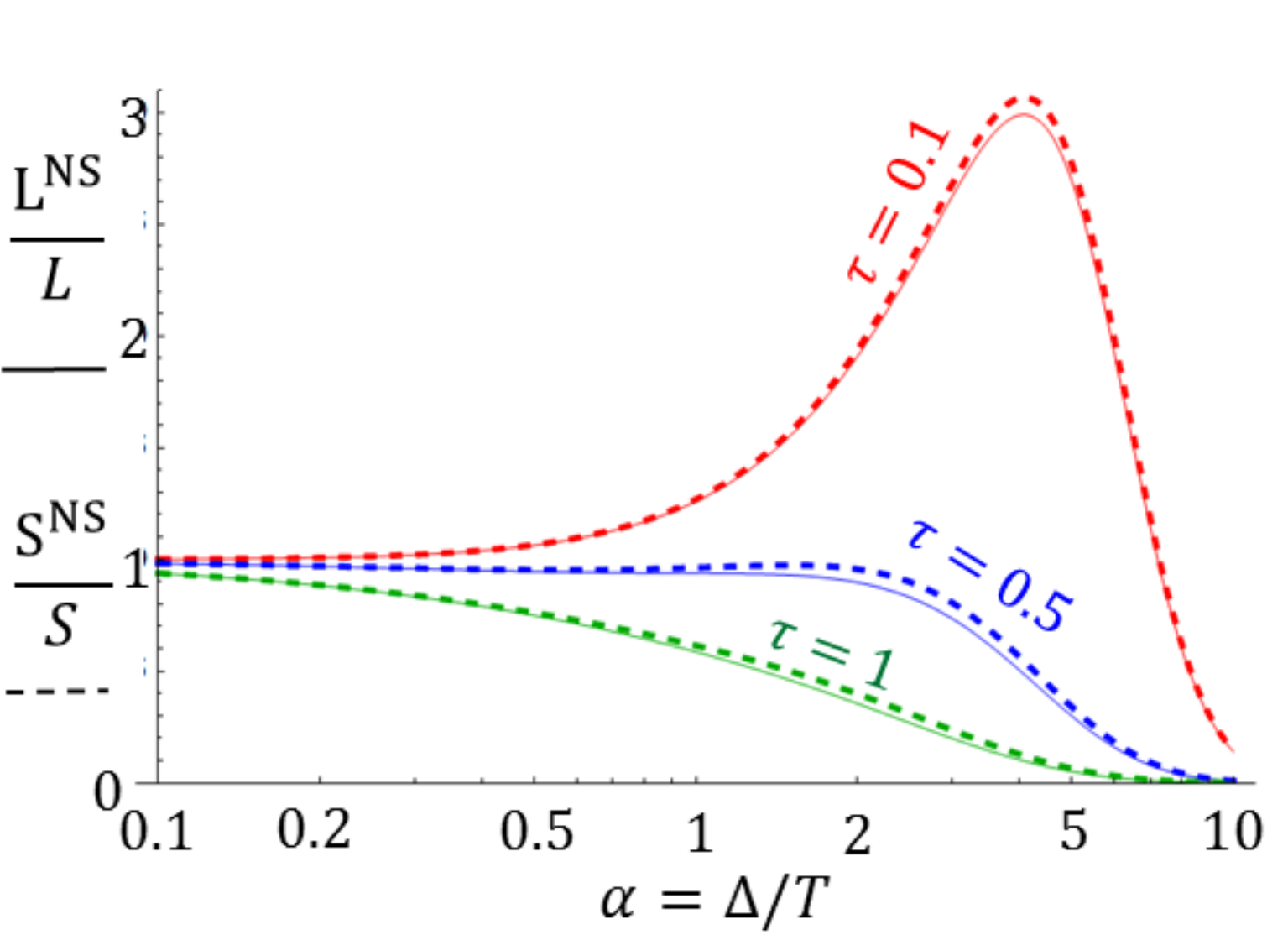}
\caption{Lorenz number (solid lines) and Seebeck coefficient (dashed lines) of NS contact normalized by the respective normal-state values, as a function of $\alpha = \Delta/T$.} 
\label{fig:lorenz_vs_seebeck}
\end{figure}
%%%%%%%%%%%%%%%%%%%%%%%%%%%%%%%%%%%%%%%%%%%%%%%%%%%%%%%%%%%%%%%%%%%%%%%%%%%%%

In order to define the Lorenz number and Seebeck coefficient, we need to introduce the conductance of the NS junction. Its relation to the scattering matrix is well-known from the seminal work~[\onlinecite{BTK1982}]. For the case of vanishing particle-hole asymmetry, it can be written as
\begin{equation}
     G^{NS} = \frac{4e^2}{h\,T} \int_{0}^\infty\!\! d\varepsilon \left(1-\left|r^{ee}_{11}\right|^2+\left|r^{he}_{11}\right|^2\right)[-f'(x)]_{x = \varepsilon/T}.
\end{equation}
Using here Eq.~(\ref{amplitudes}) we find, in agreement with Ref.~[\onlinecite{BTK1982}],
\begin{align}
     &G^{NS}(\Delta,T,\tau) = G\, g^{NS}(\Delta/T,\tau),  \label{conductance_NS}\\
     & \,\, g^{NS}(\alpha,\tau)= - 4\int_{0}^\infty dx \left[\frac{\alpha^2\tau\, \theta(\alpha-x)}{\alpha^2(2-\tau)^2-4x^2(1-\tau)}\right. \nonumber\\
     &\qquad\qquad\qquad\qquad\left.  +\frac{x\,\theta(x-\alpha)}{\tau x+(2-\tau)\sqrt{x^2-\alpha^2}}\right]f'(x). \nonumber
\end{align}
Here, in contrast with the expressions for heat conductance~(\ref{J_small_asymmetry_NS}) and thermoelectric~(\ref{I_small_asymmetry_NS}) coefficient, the sub-gap states contribute to the conductance due to the Andreev reflection, cf. the first term in the integrand of Eq.~(\ref{conductance_NS}). 

Opening of a small gap ($\Delta/T\ll 1$) leads to an increase of the conductance over its normal-state value,
\begin{align}
   &g^{NS} = 1 + k(\tau)\frac{\Delta}{T}, \label{K_NS_asymp_small_alpha}\\  & k(\tau) =  \frac{\tau}{4 (1-\tau)}\,\left[1-\frac{\tau ^2 }{2 (2-\tau ) \sqrt{1-\tau}}\ln \left(\frac{1+\sqrt{1-\tau }}{1-\sqrt{1-\tau }}\right)\right]. \nonumber
\end{align}
Note that quasiparticles with energies both below and above $\Delta$ contribute to $k(\tau)$. In the opposite limit of low temperatures, $T\ll\Delta$, the Andreev reflection contribution is the dominant one, resulting~[\onlinecite{BeenakkerPRB1992}] in $g^{NS}={2\tau}/{(2-\tau)^2}$, as long as $\tau\gg\exp(-\Delta/T)$, {\it i.e.}, is not exponentially small. 
% The statements in this paragraphs are correst (SP). 
%{\bf please check if the statements in this paragraph are right}

The Lorenz number for the NS junction reads
\begin{align}
    &L^{NS}(\Delta,T,\tau) = \frac{G^{NS}_T}{TG^{NS}} = L \frac{g_T^{NS}(\Delta/T,\tau)}{g^{NS}(\Delta/T,\tau)}, \label{NS_lorenz} 
\end{align}
where we used Eqs.~(\ref{J_small_asymmetry_NS}) and (\ref{conductance_NS}) for the heat and particle transport, respectively; $L$ is the Lorenz number for a normal-state conductor, see Eq.~(\ref{WFL}). The non-monotonic dependence of thermal conductance $G_T^{NS}$ on $\Delta/T$ at small transmission coefficients carries over to such dependence of $L^{NS}$. To see that, we use the leading terms of the small-$\alpha$ expansions, Eqs.~(\ref{F_NS_asymp_small_alpha}) and (\ref{K_NS_asymp_small_alpha}), and additionally restrict these expansions to the leading terms in the small-$\tau$ limit, \footnote{Here we also use the expansion of $g^{NS}$ at $\tau = 0$:  $g^{NS}(\alpha,0) = 1 - C\,\, \alpha^2; \,\,C = \frac{1}{8} \int_{0}^{+\infty} dx \frac{\tanh^2 x}{x^2} = \frac{7\zeta(3)}{4\pi^2}\approx  0.21$.
}
\begin{equation}
\frac{L^{NS}(\Delta,T,\tau)}{L}=1+\frac{6+7 \zeta(3)}{4\pi^2}\left(\frac{\Delta}{T}\right)^2-\frac{\tau}{4}\frac{\Delta}{T}\,.
\label{LNSsmallalpha}
\end{equation}
The ratio $L^{NS}(\Delta,T,\tau)/L>1$ at $\Delta/T>\pi^2\tau/[6+7\zeta(3)]$, safely within the domain of validity of the expansions (\ref{F_NS_asymp_small_alpha}) and (\ref{K_NS_asymp_small_alpha}), if $\tau\ll 1$. Upon further increase of $\Delta/T$, thermal current freezes out, while the particle current reaches a $T$-independent value supported by the Andreev reflection processes. As the result, $L^{NS}/L\propto\exp(-\Delta/T)$ at low temperatures. The pre-exponential factor depends on whether $\tau\sqrt\alpha$ is large or small. Considering for definiteness the latter case, we find
\begin{equation}
\frac{L^{NS}(\Delta,T,\tau)}{L}
=\frac{6\sqrt{2}}{\tau\,\pi^{3/2}}
\left(\frac{\Delta}{T}\right)^{\!5/2} \!\exp\left(\!-\frac{\Delta}{T}\right)\,.
\label{LNSlargealpha}
\end{equation}
The non-monotonic temperature dependence of $L^{NS}/L$ expected at small $\tau$ from the consideration of asymptotes is confirmed by the results of numerical evaluation, see Fig.~\ref{fig:lorenz_vs_seebeck}.

Next, taking similar steps, we evaluate the Seebeck coefficient. It is defined as the ratio of the thermoelectric coefficient (\ref{I_small_asymmetry_NS}) and conductance (\ref{conductance_NS}),
\begin{align}
    S^{NS} &= \frac{S^{NS}_I}{G^{NS}} = S \frac{s^{NS}(\Delta/T,\tau)}{g^{NS}(\Delta/T,\tau)}\,.
\end{align}
Here $S$ is the normal-state Seebeck coefficient which satisfies the Mott law, see Eqs.~(\ref{I_small_asymmetry}) and (\ref{I_small_asymmetry_NS}). At small transmission coefficient $\tau$, the ratio $S^{NS}/S$ is a non-monotonic function of $\Delta/T$. This can be seen by considering the opposite limits of that function. At $\Delta/T\ll 1$, the analysis uses Eqs.~(\ref{H_NS_small_alpha}) and (\ref{K_NS_asymp_small_alpha}), and leads to a result identical to Eq.~(\ref{LNSsmallalpha}) describing the behavior of $L^{NS}/L$. The same is true for the $\Delta/T\gg 1$ asymptote, which follows Eq.~(\ref{LNSlargealpha}) derived above. Numerical evaluation of $S^{NS}/S$ shows that it is quite close to the dimensionless ratio $L^{NS}/L$ in the entire domain of parameters $\tau$ and $\Delta/T$, see Fig.~\ref{fig:lorenz_vs_seebeck}; each may significantly exceed the normal-state value of $1$, if the transmission coefficient $\tau$ is small.

%%%%%%%%%%%%%%%%%%%%%%%%%%%%%%%%%%%%%%%%%%%%%%%%%%%%%%%%%%%%%%%%%%%%%%%%%%%%%
\section{NSN junction}
\label{sec:nsnj}
%%%%%%%%%%%%%%%%%%%%%%%%%%%%%%%%%%%%%%%%%%%%%%%%%%%%%%%%%%%%%%%%%%%%%%%%%%%%%
%%%%%%%%%%%%%%%%%%%%%%%%%%%%%%%%%%%%%%%%%%%%%%%%%%%%%%%%%%%%%%%%%%%%%%%%%%%%%
\begin{figure}
\includegraphics[width=0.9\linewidth]{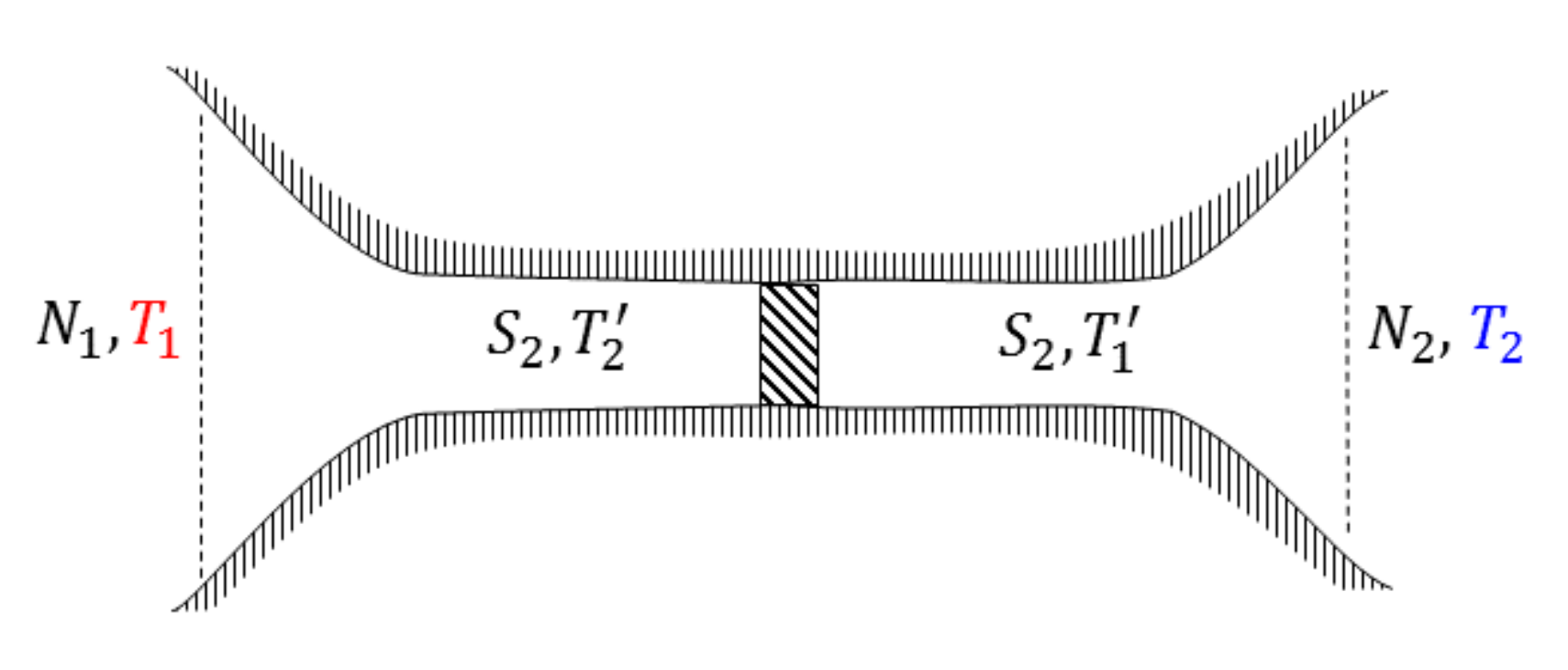}
\caption{Schematics of the NSN junction. Dashed lines indicate the boundaries between the normal and superconducting parts.} 
\label{fig:NSN_schematics}
\end{figure}
%%%%%%%%%%%%%%%%%%%%%%%%%%%%%%%%%%%%%%%%%%%%%%%%%%%%%%%%%%%%%%%%%%%%%%%%%%%%%

Lastly, we consider thermal conductance and thermopower of NSN junction sketched in Fig.~\ref{fig:NSN_schematics}, with NS boundaries situated in the wider parts of the channel. This is a typical geometry of mesoscopic transport experiments with cold atoms [\onlinecite{Stadler2012,BrantutScience2013,HusmannScience2015,Husmann-PNAS2018}]. Such geometry can be also implemented for the electron transport in mesoscopic solids. The two  questions we want to address here, is whether the Lorentz number $L^{NSN}$ and Seebeck coefficient $S^{NSN}$ of the NSN junction are sensitive to the properties of the quantum point contact constraining the transport through the superconducting (or superfluid in the case of atomic point contact) part of the device.

In the following, we assume a sufficiently fast equilibration within the Bogoliubov quasiparticles subsystem, so that we may use the notion of local temperature and view NSN junction as a sequence of NS, SS and NS junctions connected in series. It requires inelastic relaxation length be much shorter than the distance between the NS and SN interfaces. This is not a stringent condition for a cold atoms gas close to the unitary limit~[\onlinecite{Husmann-PNAS2018}], but may require further analysis in the case of electron transport [\onlinecite{Goffman2017,Mourik2012,DellaRocca2007,Bretheau2013},\onlinecite{Braggio2018}]. We also assume a sufficiently short BCS coherence length which simplifies~[\onlinecite{Kanasz-Nagy2016}] consideration of the conductance of NSN junction. For definiteness, we focus on a symmetric junction and build upon the elements discussed in the previous Sections. 

Treating the heat resistances as additive quantities, $\left(G_T^{NSN}\right)^{-1} = 2\left(G_T^{NS}\right)^{-1}+\left(G_T^{SS}\right)^{-1}$, we find
\begin{equation}
G_T^{NSN} = G^{SS}_T
\left[1+2\frac{G^{SS}_T}{G^{NS}_T}\right]^{-1}.
\label{GTNSN}
\end{equation}
It is clear that $G_T^{NSN}$ does represent the thermal conductance of the superconducting point contact as long as the cross-sectional area of the NS boundary is wide enough, so that $G_T^{NS}\gg G_T^{SS}$. 

To evaluate the Lorenz number of the NSN structure, we recall [\onlinecite{Kanasz-Nagy2016}] that only the NS boundaries contribute to the resistance, $G^{NSN}=G^{NS}/2$, which yields
\begin{equation}
\frac{L^{NSN}}{L} = \frac{G^{SS}_T}{G_T}\frac{2G}{G^{NS}} \left[1+2\frac{G^{SS}_T}{G^{NS}_T}\right]^{-1}.
\label{LNSN}
\end{equation}
The temperature dependence of $G^{NS}$ complicates the behavior of the Lorenz number $L^{NSN}$, compared with that of $G_T^{NSN}$.

To further specify $L^{NSN}$, we introduce the numbers of fully-transmitting channels in the point contact, $N_n$, and at the cross-section of the NS interface, $N_w$ (subscripts $n$ and $w$ stand for ``narrow'' and ``wide"), and assume no partial transmission is present in the system. Under these assumptions, 
\begin{eqnarray}
&&\frac{G_T^{NSN}}{G_T} = g_T^{SS}(0,\Delta/T,1)
\left[1+2\frac{N_n}{N_w}\frac{g_T^{SS}(0,\Delta/T,1)}{g_T^{NS}(\Delta/T,1)}\right]^{-1},
\nonumber\\
&&G_T=\frac{\pi^2 T}{3e^2}G\,,\,\,\, G=\frac{2e^2}{h} N_n\,,
\nonumber\\
&&\frac{L^{NSN}}{L} = \frac{L^{NS}}{L}
\left[1+\frac{1}{2}\frac{N_w}{N_n}\frac{g_T^{NS}(\Delta/T,1)}{g_T^{SS}(0,\Delta/T,1)}\right]^{-1}\,.
\label{LNSNL}
\end{eqnarray}
Here the definitions of functions $g_T^{SS}(0,\Delta/T,1)$, $g_T^{NS}(\Delta/T,1)$, and $L^{NSN}/L$ and their various limits are presented, respectively, in Eqs.~(\ref{J_small_asymmetry}) and (\ref{F_asymp_small_alpha}), Eqs.~(\ref{J_small_asymmetry_NS})-(\ref{F_NS_asymp_large_alpha}), and Eqs.~(\ref{NS_lorenz})-(\ref{LNSlargealpha}). The thermal conductance and Lorenz number for an NSN structure depend, in addition, on the ratio of the channel numbers $N_n/N_w$. In the limiting case $N_n/N_w\ll 1$, the ratio $G^{NSN}/G_T$ closely follows $g_T^{SS}(0,\Delta/T,1)$ presented in Fig.~\ref{fig:heatcurrent_vs_delta}a. At a finite but small value $\Delta/T\ll 1$, the thermal conductance in superconducting state is close to its normal-state value, $G_T^{NSN}\approx G_T$. The behavior of $L^{NSN}$ is different: it falls off drastically upon entering the superconducting state because of the shunting effect of the superfluid condensate represented by the last factor\footnote{Two comments are in place here. First, while considering $\Delta/T\ll 1$, we still assume the BCS coherence length shorter than the superfluid domain in the NSN structure. Second, the ratio $g_T^{NS}/g_T^{SS}$ appearing in the last factor of Eq.~(\ref{LNSNL}) becomes small and may compensate the large factor $N_w/(2N_n)$ at low temperatures, $T/\Delta\sim (N_w/2N_n)^2$; however, at such low temperatures is exponentially small, $L^{NSN}\sim\exp\{-(N_w/2N_n)^2\}$} Eq.~(\ref{LNSNL}). This suppression occurs on top of the Lorenz number reduction at $\Delta/T\sim 1$ brought by the ratio $L^{NS}/L$, see Fig.~\ref{fig:lorenz_vs_seebeck}. We note that the suppression of the Lorenz number upon the transition to a superfluid state of ${}^6$Li cold atoms confined to a quantum channel was indeed observed in Ref.~[\onlinecite{Husmann-PNAS2018}]. %\\{\bf we may want to plot the ratio $F^{SS}(0,\Delta/T,1)/F^{NS}(\Delta/T,1)$}

The applied temperature gradient induces a heat current $J = G_T^{NSN}(T_1-T_2)$,  where $T_1$ and $T_2$ are the temperatures of the left and right normal parts. In order to evaluate the temperatures $T'_1$ and $T'_2$ within the superconducting parts of the structure (see Fig.~\ref{fig:NSN_schematics}), we equate $J$ and the corresponding expression for the NS boundary, which gives $(T_1-T_1') = J/G_T^{NS} = (T_1-T_2)G^{SS}/(2G^{SS}+G^{NS})$. As usual, in order to evaluate the Seebeck coefficient, we assume that there is no net particle current flowing through the structure. Thus the particle current due to the induced voltage $V^{NS}$ on the NS boundary is compensated by the thermoelectric current $G^{NS} V^{NS} = G^{NS}S^{NS} (T_1-T_1')$. The latter equation gives the Seebeck coefficient of the entire structure,
\begin{equation}
    S^{NSN} = \frac{2V^{NS}}{T_1-T_2} = S^{NS}
    \left[1+\frac{1}{2}\frac{G^{NS}_T}{G^{SS}_T}\right]^{-1}
    \label{SNSN}
\end{equation}
The comparison of $S^{NSN}$ with the Seebeck coefficient $S$ of the system in the normal state strongly depends on the details of the potential confining the motion of fermions. One may produce a crude estimate in terms of the number of opened channels $N_n$, $N_w$, and their rate of their change with the change of the chemical potential $\mu$. Assuming $N_w\gg N_n$ and considering only temperatures high compared to $\Delta\neq 0$ and to the level spacing for the quantized transverse motion of fermions in any part of the device, we find 
\begin{equation}
    \frac{S^{NSN}}{S} \sim 2\frac{dN_w/d\mu}{dN_n/d\mu} \left(\frac{N_n}{N_w}\right)^2\,.
    \label{SNSNest}
\end{equation}
For a simplest harmonic confining potential, the estimate indicates $S^{NSN}<S$.

\section{Conclusions}

Applications of scattering theory are ubiquitous in mesoscopic physics. Surprisingly, the strengths of this method were not fully exploited in the study of thermal effects of superconducting devices. We fill this apparent void by relating the heat conductance and thermally-induced particle current to the normal-state scattering matrix, see Eqs. (\ref{heatcurrent1}) and (\ref{thermopower1}), and specify these general results to the practically-important cases of superconducting quantum point contacts (SQPC), NS boundaries, and NSN ballistic devices.
 
Considering the quasiparticle transport in SQPC within the scattering formalism, we elucidated the role of Andreev levels in thermally-induced currents, resolved the discrepancy between the two perturbative in tunneling calculations [\onlinecite{GuttmanPRB1997a,SmithPRB1980}], and obtained results valid at arbitrary transmission coefficients, see Section~\ref{sec:sj}.

Analyzing the SQPC alone, one is able to find particle and entropy currents. The conventional characteristics of thermally-induced linear transport, the Lorenz number and Seebeck coefficient, are not defined due to the shunting effect of the superfluid condensate. That prompted us to develop the theory for NS boundaries, see Section~\ref{sec:nsj}, and NSN devices (Section~\ref{sec:nsnj}) where these quantities are well-defined (the latter geometry is of special interest because of the experiments with cold ${}^6$Li atoms~[\onlinecite{Husmann-PNAS2018}]).
The practical conclusion of that study, is while the thermal conductance Eq.~(\ref{GTNSN}) in NSN geometry is proportional to the thermal conductance of SQPC, the thermopower Eq.~(\ref{SNSN}) is not. Instead, it is sensitive to the details of the confining potential away from the narrowest cross-section of the channel and to the thermal conductance of the SQPC. A crude estimate Eq.~(\ref{SNSNest}) indicates that NSN Seebeck coefficient $S^{NSN}$ is lower than the one in the normal state, $S$, at sufficiently high temperatures. However, lowering the temperature below the energy separation between the quantized levels of the transverse motion at the narrowest cross-section, may revert the relation between $S^{NSN}$ and $S$. A full analysis of the experiment~[\onlinecite{Husmann-PNAS2018}] is beyond this work.
\\
\\
{\sl Acknowledgements}\\
We thank A. Akhmerov and A. Braggio for useful comments, and L. Corman, D. Husmann, S. H\"ausler, and T. Esslinger  for numerous discussions of their transport experiments. This work is supported by the DOE contract DE-FG02-08ER46482 (LIG), and by the ARO grant W911NF-18-1-0212 (SSP).

%%%%%%%%%%%%%%%%%%%%%%%%%%%%%%%%%%%%%%%%%%%%%%%%%%%%%%%%%%%%%%%%%%%%%%%%%%%%%
\bibliography{biblio}

\begin{thebibliography}{45}%
\makeatletter
\providecommand \@ifxundefined [1]{%
 \@ifx{#1\undefined}
}%
\providecommand \@ifnum [1]{%
 \ifnum #1\expandafter \@firstoftwo
 \else \expandafter \@secondoftwo
 \fi
}%
\providecommand \@ifx [1]{%
 \ifx #1\expandafter \@firstoftwo
 \else \expandafter \@secondoftwo
 \fi
}%
\providecommand \natexlab [1]{#1}%
\providecommand \enquote  [1]{``#1''}%
\providecommand \bibnamefont  [1]{#1}%
\providecommand \bibfnamefont [1]{#1}%
\providecommand \citenamefont [1]{#1}%
\providecommand \href@noop [0]{\@secondoftwo}%
\providecommand \href [0]{\begingroup \@sanitize@url \@href}%
\providecommand \@href[1]{\@@startlink{#1}\@@href}%
\providecommand \@@href[1]{\endgroup#1\@@endlink}%
\providecommand \@sanitize@url [0]{\catcode `\\12\catcode `\$12\catcode
  `\&12\catcode `\#12\catcode `\^12\catcode `\_12\catcode `\%12\relax}%
\providecommand \@@startlink[1]{}%
\providecommand \@@endlink[0]{}%
\providecommand \url  [0]{\begingroup\@sanitize@url \@url }%
\providecommand \@url [1]{\endgroup\@href {#1}{\urlprefix }}%
\providecommand \urlprefix  [0]{URL }%
\providecommand \Eprint [0]{\href }%
\providecommand \doibase [0]{http://dx.doi.org/}%
\providecommand \selectlanguage [0]{\@gobble}%
\providecommand \bibinfo  [0]{\@secondoftwo}%
\providecommand \bibfield  [0]{\@secondoftwo}%
\providecommand \translation [1]{[#1]}%
\providecommand \BibitemOpen [0]{}%
\providecommand \bibitemStop [0]{}%
\providecommand \bibitemNoStop [0]{.\EOS\space}%
\providecommand \EOS [0]{\spacefactor3000\relax}%
\providecommand \BibitemShut  [1]{\csname bibitem#1\endcsname}%
\let\auto@bib@innerbib\@empty
%</preamble>
\bibitem [{\citenamefont {Mendelssohn}(1953)}]{Mendelssohn1953}%
  \BibitemOpen
  \bibfield  {author} {\bibinfo {author} {\bibfnamefont {K.}~\bibnamefont
  {Mendelssohn}},\ }\bibfield  {title} {\enquote {\bibinfo {title} {Thermal
  conductivity of superconductors},}\ }\href@noop {} {\bibfield  {journal}
  {\bibinfo  {journal} {Physica}\ }\textbf {\bibinfo {volume} {19}},\ \bibinfo
  {pages} {775} (\bibinfo {year} {1953})}\BibitemShut {NoStop}%
\bibitem [{\citenamefont {Bardeen}\ \emph {et~al.}(1959)\citenamefont
  {Bardeen}, \citenamefont {Rickayzen},\ and\ \citenamefont
  {Tewordt}}]{Bardeen1959}%
  \BibitemOpen
  \bibfield  {author} {\bibinfo {author} {\bibfnamefont {J.}~\bibnamefont
  {Bardeen}}, \bibinfo {author} {\bibfnamefont {G.}~\bibnamefont {Rickayzen}},
  \ and\ \bibinfo {author} {\bibfnamefont {L.}~\bibnamefont {Tewordt}},\
  }\bibfield  {title} {\enquote {\bibinfo {title} {Theory of the thermal
  conductivity of superconductors},}\ }\href@noop {} {\bibfield  {journal}
  {\bibinfo  {journal} {Phys. Rev.}\ }\textbf {\bibinfo {volume} {113}},\
  \bibinfo {pages} {982} (\bibinfo {year} {1959})}\BibitemShut {NoStop}%
\bibitem [{\citenamefont {Ginzburg}(1944)}]{Ginzburg1944}%
  \BibitemOpen
  \bibfield  {author} {\bibinfo {author} {\bibfnamefont {V.~L.}\ \bibnamefont
  {Ginzburg}},\ }\bibfield  {title} {\enquote {\bibinfo {title} {The
  thermoelectric phenomena in superconductors},}\ }\href@noop {} {\bibfield
  {journal} {\bibinfo  {journal} {J. Phys. (USSR)}\ }\textbf {\bibinfo {volume}
  {8}},\ \bibinfo {pages} {148} (\bibinfo {year} {1944})}\BibitemShut {NoStop}%
\bibitem [{\citenamefont {Ginzburg}(2004)}]{GinzburgRMP2004}%
  \BibitemOpen
  \bibfield  {author} {\bibinfo {author} {\bibfnamefont {V.~L.}\ \bibnamefont
  {Ginzburg}},\ }\href {\doibase 10.1103/RevModPhys.76.981} {\bibfield
  {journal} {\bibinfo  {journal} {Rev. Mod. Phys.}\ }\textbf {\bibinfo {volume}
  {76}},\ \bibinfo {pages} {981} (\bibinfo {year} {2004})}\BibitemShut
  {NoStop}%
\bibitem [{\citenamefont {Van~Harlingen}\ \emph {et~al.}(1980)\citenamefont
  {Van~Harlingen}, \citenamefont {Heidel},\ and\ \citenamefont
  {Garland}}]{VanHarlingenPRB1980}%
  \BibitemOpen
  \bibfield  {author} {\bibinfo {author} {\bibfnamefont {D.~J.}\ \bibnamefont
  {Van~Harlingen}}, \bibinfo {author} {\bibfnamefont {D.~F.}\ \bibnamefont
  {Heidel}}, \ and\ \bibinfo {author} {\bibfnamefont {J.~C.}\ \bibnamefont
  {Garland}},\ }\bibfield  {title} {\enquote {\bibinfo {title} {Experimental
  study of thermoelectricity in superconducting indium},}\ }\href {\doibase
  10.1103/PhysRevB.21.1842} {\bibfield  {journal} {\bibinfo  {journal} {Phys.
  Rev. B}\ }\textbf {\bibinfo {volume} {21}},\ \bibinfo {pages} {1842--1857}
  (\bibinfo {year} {1980})}\BibitemShut {NoStop}%
\bibitem [{\citenamefont {Shelly}\ \emph {et~al.}(2016)\citenamefont {Shelly},
  \citenamefont {Matrozova},\ and\ \citenamefont
  {Petrashov}}]{Shellye-Matrozova-Petrashov2016}%
  \BibitemOpen
  \bibfield  {author} {\bibinfo {author} {\bibfnamefont {C.~D.}\ \bibnamefont
  {Shelly}}, \bibinfo {author} {\bibfnamefont {E.~A.}\ \bibnamefont
  {Matrozova}}, \ and\ \bibinfo {author} {\bibfnamefont {V.~T.}\ \bibnamefont
  {Petrashov}},\ }\bibfield  {title} {\enquote {\bibinfo {title} {{Resolving
  thermoelectric {\textquotedblleft}paradox{\textquotedblright} in
  superconductors}},}\ }\href
  {http://advances.sciencemag.org/content/2/2/e1501250} {\bibfield  {journal}
  {\bibinfo  {journal} {Sci. Adv.}\ }\textbf {\bibinfo {volume} {2}},\ \bibinfo
  {pages} {e1501250} (\bibinfo {year} {2016})}\BibitemShut {NoStop}%
\bibitem [{\citenamefont {Mamin}\ \emph {et~al.}(1984)\citenamefont {Mamin},
  \citenamefont {Clarke},\ and\ \citenamefont
  {Van~Harlingen}}]{Mamin-Clarke-VanHarlingen-PRB1984}%
  \BibitemOpen
  \bibfield  {author} {\bibinfo {author} {\bibfnamefont {H.~J.}\ \bibnamefont
  {Mamin}}, \bibinfo {author} {\bibfnamefont {J.}~\bibnamefont {Clarke}}, \
  and\ \bibinfo {author} {\bibfnamefont {D.~J.}\ \bibnamefont
  {Van~Harlingen}},\ }\bibfield  {title} {\enquote {\bibinfo {title} {Charge
  imbalance induced by a temperature gradient in superconducting aluminum},}\
  }\href {\doibase 10.1103/PhysRevB.29.3881} {\bibfield  {journal} {\bibinfo
  {journal} {Phys. Rev. B}\ }\textbf {\bibinfo {volume} {29}},\ \bibinfo
  {pages} {3881} (\bibinfo {year} {1984})}\BibitemShut {NoStop}%
\bibitem [{\citenamefont {Maki}\ and\ \citenamefont
  {Griffin}(1965)}]{Maki1966}%
  \BibitemOpen
  \bibfield  {author} {\bibinfo {author} {\bibfnamefont {K.}~\bibnamefont
  {Maki}}\ and\ \bibinfo {author} {\bibfnamefont {A.}~\bibnamefont {Griffin}},\
  }\bibfield  {title} {\enquote {\bibinfo {title} {{Entropy Transport Between
  Two Superconductors by Electron Tunneling}},}\ }\href
  {https://link.aps.org/doi/10.1103/PhysRevLett.15.921} {\bibfield  {journal}
  {\bibinfo  {journal} {Phys. Rev. Lett.}\ }\textbf {\bibinfo {volume} {15}},\
  \bibinfo {pages} {921} (\bibinfo {year} {1965})}\BibitemShut {NoStop}%
\bibitem [{\citenamefont {Guttman}\ \emph
  {et~al.}(1997{\natexlab{a}})\citenamefont {Guttman}, \citenamefont
  {Nathanson}, \citenamefont {Ben-Jacob},\ and\ \citenamefont
  {Bergman}}]{GuttmanPRB1997a}%
  \BibitemOpen
  \bibfield  {author} {\bibinfo {author} {\bibfnamefont {G.~D.}\ \bibnamefont
  {Guttman}}, \bibinfo {author} {\bibfnamefont {B.}~\bibnamefont {Nathanson}},
  \bibinfo {author} {\bibfnamefont {E.}~\bibnamefont {Ben-Jacob}}, \ and\
  \bibinfo {author} {\bibfnamefont {D.~J.}\ \bibnamefont {Bergman}},\
  }\bibfield  {title} {\enquote {\bibinfo {title} {{Thermoelectric and
  thermophase effects in Josephson junctions}},}\ }\href {\doibase
  10.1103/PhysRevB.55.12691} {\bibfield  {journal} {\bibinfo  {journal} {Phys.
  Rev. B}\ }\textbf {\bibinfo {volume} {55}},\ \bibinfo {pages} {12691}
  (\bibinfo {year} {1997}{\natexlab{a}})}\BibitemShut {NoStop}%
\bibitem [{\citenamefont {Guttman}\ \emph
  {et~al.}(1997{\natexlab{b}})\citenamefont {Guttman}, \citenamefont
  {Nathanson}, \citenamefont {Ben-Jacob},\ and\ \citenamefont
  {Bergman}}]{GuttmanPRB1997b}%
  \BibitemOpen
  \bibfield  {author} {\bibinfo {author} {\bibfnamefont {G.~D.}\ \bibnamefont
  {Guttman}}, \bibinfo {author} {\bibfnamefont {B.}~\bibnamefont {Nathanson}},
  \bibinfo {author} {\bibfnamefont {E.}~\bibnamefont {Ben-Jacob}}, \ and\
  \bibinfo {author} {\bibfnamefont {D.~J.}\ \bibnamefont {Bergman}},\
  }\bibfield  {title} {\enquote {\bibinfo {title} {{Phase-dependent thermal
  transport in Josephson junctions}},}\ }\href {\doibase
  10.1103/PhysRevB.55.3849} {\bibfield  {journal} {\bibinfo  {journal} {Phys.
  Rev. B}\ }\textbf {\bibinfo {volume} {55}},\ \bibinfo {pages} {3849}
  (\bibinfo {year} {1997}{\natexlab{b}})}\BibitemShut {NoStop}%
\bibitem [{\citenamefont {Giazotto}\ and\ \citenamefont
  {Martínez-Pérez}(2012)}]{GiazottoAPL2012}%
  \BibitemOpen
  \bibfield  {author} {\bibinfo {author} {\bibfnamefont {F.}~\bibnamefont
  {Giazotto}}\ and\ \bibinfo {author} {\bibfnamefont {M.~J.}\ \bibnamefont
  {Martínez-Pérez}},\ }\bibfield  {title} {\enquote {\bibinfo {title}
  {Phase-controlled superconducting heat-flux quantum modulator},}\ }\href
  {\doibase 10.1063/1.4750068} {\bibfield  {journal} {\bibinfo  {journal} {App.
  Phys. Lett.}\ }\textbf {\bibinfo {volume} {101}},\ \bibinfo {pages} {102601}
  (\bibinfo {year} {2012})}\BibitemShut {NoStop}%
\bibitem [{\citenamefont {Zhao}\ \emph {et~al.}(2003)\citenamefont {Zhao},
  \citenamefont {L\"ofwander},\ and\ \citenamefont {Sauls}}]{SaulsPRL2003}%
  \BibitemOpen
  \bibfield  {author} {\bibinfo {author} {\bibfnamefont {E.}~\bibnamefont
  {Zhao}}, \bibinfo {author} {\bibfnamefont {T.}~\bibnamefont {L\"ofwander}}, \
  and\ \bibinfo {author} {\bibfnamefont {J.~A.}\ \bibnamefont {Sauls}},\
  }\bibfield  {title} {\enquote {\bibinfo {title} {{Phase Modulated Thermal
  Conductance of Josephson Weak Links}},}\ }\href {\doibase
  10.1103/PhysRevLett.91.077003} {\bibfield  {journal} {\bibinfo  {journal}
  {Phys. Rev. Lett.}\ }\textbf {\bibinfo {volume} {91}},\ \bibinfo {pages}
  {077003} (\bibinfo {year} {2003})}\BibitemShut {NoStop}%
\bibitem [{\citenamefont {Zhao}\ \emph {et~al.}(2004)\citenamefont {Zhao},
  \citenamefont {L\"ofwander},\ and\ \citenamefont {Sauls}}]{SaulsPRB2004}%
  \BibitemOpen
  \bibfield  {author} {\bibinfo {author} {\bibfnamefont {E.}~\bibnamefont
  {Zhao}}, \bibinfo {author} {\bibfnamefont {T.}~\bibnamefont {L\"ofwander}}, \
  and\ \bibinfo {author} {\bibfnamefont {J.~A.}\ \bibnamefont {Sauls}},\
  }\bibfield  {title} {\enquote {\bibinfo {title} {{Heat transport through
  Josephson point contacts}},}\ }\href {\doibase 10.1103/PhysRevB.69.134503}
  {\bibfield  {journal} {\bibinfo  {journal} {Phys. Rev. B}\ }\textbf {\bibinfo
  {volume} {69}},\ \bibinfo {pages} {134503} (\bibinfo {year}
  {2004})}\BibitemShut {NoStop}%
\bibitem [{\citenamefont {Giazotto}\ and\ \citenamefont
  {Mart{\'i}nez-P{\'e}rez}(2012)}]{Giazotto-Nature2012}%
  \BibitemOpen
  \bibfield  {author} {\bibinfo {author} {\bibfnamefont {F.}~\bibnamefont
  {Giazotto}}\ and\ \bibinfo {author} {\bibfnamefont {M.~J.}\ \bibnamefont
  {Mart{\'i}nez-P{\'e}rez}},\ }\bibfield  {title} {\enquote {\bibinfo {title}
  {{The Josephson heat interferometer}},}\ }\href
  {http://dx.doi.org/10.1038/nature11702} {\bibfield  {journal} {\bibinfo
  {journal} {Nature}\ }\textbf {\bibinfo {volume} {492}},\ \bibinfo {pages}
  {401} (\bibinfo {year} {2012})}\BibitemShut {NoStop}%
\bibitem [{\citenamefont {Gurevich}\ \emph {et~al.}(2006)\citenamefont
  {Gurevich}, \citenamefont {Kozub},\ and\ \citenamefont
  {Shelankov}}]{GurevichEPJB2006}%
  \BibitemOpen
  \bibfield  {author} {\bibinfo {author} {\bibfnamefont {V.~L.}\ \bibnamefont
  {Gurevich}}, \bibinfo {author} {\bibfnamefont {V.~I.}\ \bibnamefont {Kozub}},
  \ and\ \bibinfo {author} {\bibfnamefont {A.~L.}\ \bibnamefont {Shelankov}},\
  }\bibfield  {title} {\enquote {\bibinfo {title} {Thermoelectric effects in
  superconducting nanostructures},}\ }\href {\doibase
  10.1140/epjb/e2006-00218-6} {\bibfield  {journal} {\bibinfo  {journal} {Eur.
  Phys. J. B}\ }\textbf {\bibinfo {volume} {51}},\ \bibinfo {pages} {285}
  (\bibinfo {year} {2006})}\BibitemShut {NoStop}%
\bibitem [{\citenamefont {Josephson}(1962)}]{Josephson1962}%
  \BibitemOpen
  \bibfield  {author} {\bibinfo {author} {\bibfnamefont {B.D.}\ \bibnamefont
  {Josephson}},\ }\bibfield  {title} {\enquote {\bibinfo {title} {Possible new
  effects in superconductive tunnelling},}\ }\href {\doibase
  10.1016/0031-9163(62)91369-0} {\bibfield  {journal} {\bibinfo  {journal}
  {Phys. Lett.}\ }\textbf {\bibinfo {volume} {1}},\ \bibinfo {pages} {251}
  (\bibinfo {year} {1962})}\BibitemShut {NoStop}%
\bibitem [{\citenamefont {Smith}\ \emph {et~al.}(1980)\citenamefont {Smith},
  \citenamefont {Tinkham},\ and\ \citenamefont {Skocpol}}]{SmithPRB1980}%
  \BibitemOpen
  \bibfield  {author} {\bibinfo {author} {\bibfnamefont {A.~D.}\ \bibnamefont
  {Smith}}, \bibinfo {author} {\bibfnamefont {M.}~\bibnamefont {Tinkham}}, \
  and\ \bibinfo {author} {\bibfnamefont {W.~J.}\ \bibnamefont {Skocpol}},\
  }\bibfield  {title} {\enquote {\bibinfo {title} {New thermoelectric effect in
  tunnel junctions},}\ }\href {\doibase 10.1103/PhysRevB.22.4346} {\bibfield
  {journal} {\bibinfo  {journal} {Phys. Rev. B}\ }\textbf {\bibinfo {volume}
  {22}},\ \bibinfo {pages} {4346} (\bibinfo {year} {1980})}\BibitemShut
  {NoStop}%
\bibitem [{\citenamefont {Hwang}\ \emph {et~al.}(2016)\citenamefont {Hwang},
  \citenamefont {L\'opez},\ and\ \citenamefont {S\'anchez}}]{HwangPRB2016}%
  \BibitemOpen
  \bibfield  {author} {\bibinfo {author} {\bibfnamefont {S.-Y.}\ \bibnamefont
  {Hwang}}, \bibinfo {author} {\bibfnamefont {R.}~\bibnamefont {L\'opez}}, \
  and\ \bibinfo {author} {\bibfnamefont {D.}~\bibnamefont {S\'anchez}},\
  }\bibfield  {title} {\enquote {\bibinfo {title} {Large thermoelectric power
  and figure of merit in a ferromagnetic--quantum dot--superconducting
  device},}\ }\href {\doibase 10.1103/PhysRevB.94.054506} {\bibfield  {journal}
  {\bibinfo  {journal} {Phys. Rev. B}\ }\textbf {\bibinfo {volume} {94}},\
  \bibinfo {pages} {054506} (\bibinfo {year} {2016})}\BibitemShut {NoStop}%
\bibitem [{\citenamefont {Trocha}\ and\ \citenamefont
  {Barna\ifmmode~\acute{s}\else \'{s}\fi{}}(2017)}]{TrochaPRB2017}%
  \BibitemOpen
  \bibfield  {author} {\bibinfo {author} {\bibfnamefont {P.}~\bibnamefont
  {Trocha}}\ and\ \bibinfo {author} {\bibfnamefont {J.}~\bibnamefont
  {Barna\ifmmode~\acute{s}\else \'{s}\fi{}}},\ }\bibfield  {title} {\enquote
  {\bibinfo {title} {Spin-dependent thermoelectric phenomena in a quantum dot
  attached to ferromagnetic and superconducting electrodes},}\ }\href {\doibase
  10.1103/PhysRevB.95.165439} {\bibfield  {journal} {\bibinfo  {journal} {Phys.
  Rev. B}\ }\textbf {\bibinfo {volume} {95}},\ \bibinfo {pages} {165439}
  (\bibinfo {year} {2017})}\BibitemShut {NoStop}%
\bibitem [{\citenamefont {Bezuglyi}\ and\ \citenamefont
  {Vinokur}(2003)}]{Bezuglyi2003}%
  \BibitemOpen
  \bibfield  {author} {\bibinfo {author} {\bibfnamefont {E.~V.}\ \bibnamefont
  {Bezuglyi}}\ and\ \bibinfo {author} {\bibfnamefont {V.}~\bibnamefont
  {Vinokur}},\ }\bibfield  {title} {\enquote {\bibinfo {title} {Heat transport
  in proximity structures},}\ }\href {\doibase 10.1103/PhysRevLett.91.137002}
  {\bibfield  {journal} {\bibinfo  {journal} {Phys. Rev. Lett.}\ }\textbf
  {\bibinfo {volume} {91}},\ \bibinfo {pages} {137002} (\bibinfo {year}
  {2003})}\BibitemShut {NoStop}%
\bibitem [{\citenamefont {Yokoyama}\ \emph {et~al.}(2005)\citenamefont
  {Yokoyama}, \citenamefont {Tanaka}, \citenamefont {Golubov},\ and\
  \citenamefont {Asano}}]{Golubov2005}%
  \BibitemOpen
  \bibfield  {author} {\bibinfo {author} {\bibfnamefont {T.}~\bibnamefont
  {Yokoyama}}, \bibinfo {author} {\bibfnamefont {Y.}~\bibnamefont {Tanaka}},
  \bibinfo {author} {\bibfnamefont {A.~A.}\ \bibnamefont {Golubov}}, \ and\
  \bibinfo {author} {\bibfnamefont {Y.}~\bibnamefont {Asano}},\ }\bibfield
  {title} {\enquote {\bibinfo {title} {Theory of thermal and charge transport
  in diffusive normal metal/superconductor junctions},}\ }\href {\doibase
  10.1103/PhysRevB.72.214513} {\bibfield  {journal} {\bibinfo  {journal} {Phys.
  Rev. B}\ }\textbf {\bibinfo {volume} {72}},\ \bibinfo {pages} {214513}
  (\bibinfo {year} {2005})}\BibitemShut {NoStop}%
\bibitem [{\citenamefont {Goffman}\ \emph {et~al.}(2017)\citenamefont
  {Goffman}, \citenamefont {Urbina}, \citenamefont {Pothier}, \citenamefont
  {Nyg\r{a}rd}, \citenamefont {Marcus},\ and\ \citenamefont
  {Krogstrup}}]{Goffman2017}%
  \BibitemOpen
  \bibfield  {author} {\bibinfo {author} {\bibfnamefont {M.~F.}\ \bibnamefont
  {Goffman}}, \bibinfo {author} {\bibfnamefont {C.}~\bibnamefont {Urbina}},
  \bibinfo {author} {\bibfnamefont {H.}~\bibnamefont {Pothier}}, \bibinfo
  {author} {\bibfnamefont {J.}~\bibnamefont {Nyg\r{a}rd}}, \bibinfo {author}
  {\bibfnamefont {C.~M.}\ \bibnamefont {Marcus}}, \ and\ \bibinfo {author}
  {\bibfnamefont {P.}~\bibnamefont {Krogstrup}},\ }\bibfield  {title} {\enquote
  {\bibinfo {title} {{Conduction channels of an InAs-Al nanowire Josephson weak
  link}},}\ }\href {http://stacks.iop.org/1367-2630/19/i=9/a=092002} {\bibfield
   {journal} {\bibinfo  {journal} {New J. Phys.}\ }\textbf {\bibinfo {volume}
  {19}},\ \bibinfo {pages} {092002} (\bibinfo {year} {2017})}\BibitemShut
  {NoStop}%
\bibitem [{\citenamefont {Mourik}\ \emph {et~al.}(2012)\citenamefont {Mourik},
  \citenamefont {Zuo}, \citenamefont {Frolov}, \citenamefont {Plissard},
  \citenamefont {Bakkers},\ and\ \citenamefont {Kouwenhoven}}]{Mourik2012}%
  \BibitemOpen
  \bibfield  {author} {\bibinfo {author} {\bibfnamefont {V.}~\bibnamefont
  {Mourik}}, \bibinfo {author} {\bibfnamefont {K.}~\bibnamefont {Zuo}},
  \bibinfo {author} {\bibfnamefont {S.~M.}\ \bibnamefont {Frolov}}, \bibinfo
  {author} {\bibfnamefont {S.~R.}\ \bibnamefont {Plissard}}, \bibinfo {author}
  {\bibfnamefont {E.~P. A.~M.}\ \bibnamefont {Bakkers}}, \ and\ \bibinfo
  {author} {\bibfnamefont {L.~P.}\ \bibnamefont {Kouwenhoven}},\ }\bibfield
  {title} {\enquote {\bibinfo {title} {{Signatures of Majorana Fermions in
  Hybrid Superconductor-Semiconductor Nanowire Devices}},}\ }\href {\doibase
  10.1126/science.1222360} {\bibfield  {journal} {\bibinfo  {journal}
  {Science}\ }\textbf {\bibinfo {volume} {336}},\ \bibinfo {pages} {1003}
  (\bibinfo {year} {2012})}\BibitemShut {NoStop}%
\bibitem [{\citenamefont {Della~Rocca}\ \emph {et~al.}(2007)\citenamefont
  {Della~Rocca}, \citenamefont {Chauvin}, \citenamefont {Huard}, \citenamefont
  {Pothier}, \citenamefont {Esteve},\ and\ \citenamefont
  {Urbina}}]{DellaRocca2007}%
  \BibitemOpen
  \bibfield  {author} {\bibinfo {author} {\bibfnamefont {M.~L.}\ \bibnamefont
  {Della~Rocca}}, \bibinfo {author} {\bibfnamefont {M.}~\bibnamefont
  {Chauvin}}, \bibinfo {author} {\bibfnamefont {B.}~\bibnamefont {Huard}},
  \bibinfo {author} {\bibfnamefont {H.}~\bibnamefont {Pothier}}, \bibinfo
  {author} {\bibfnamefont {D.}~\bibnamefont {Esteve}}, \ and\ \bibinfo {author}
  {\bibfnamefont {C.}~\bibnamefont {Urbina}},\ }\bibfield  {title} {\enquote
  {\bibinfo {title} {{Measurement of the Current-Phase Relation of
  Superconducting Atomic Contacts}},}\ }\href {\doibase
  10.1103/PhysRevLett.99.127005} {\bibfield  {journal} {\bibinfo  {journal}
  {Phys. Rev. Lett.}\ }\textbf {\bibinfo {volume} {99}},\ \bibinfo {pages}
  {127005} (\bibinfo {year} {2007})}\BibitemShut {NoStop}%
\bibitem [{\citenamefont {Bretheau}\ \emph {et~al.}(2013)\citenamefont
  {Bretheau}, \citenamefont {Girit}, \citenamefont {Urbina}, \citenamefont
  {Esteve},\ and\ \citenamefont {Pothier}}]{Bretheau2013}%
  \BibitemOpen
  \bibfield  {author} {\bibinfo {author} {\bibfnamefont {L.}~\bibnamefont
  {Bretheau}}, \bibinfo {author} {\bibfnamefont {\ifmmode \mbox{\c{C}}\else
  \c{C}\fi{}.~\"O.}\ \bibnamefont {Girit}}, \bibinfo {author} {\bibfnamefont
  {C.}~\bibnamefont {Urbina}}, \bibinfo {author} {\bibfnamefont
  {D.}~\bibnamefont {Esteve}}, \ and\ \bibinfo {author} {\bibfnamefont
  {H.}~\bibnamefont {Pothier}},\ }\bibfield  {title} {\enquote {\bibinfo
  {title} {{Supercurrent Spectroscopy of Andreev States}},}\ }\href {\doibase
  10.1103/PhysRevX.3.041034} {\bibfield  {journal} {\bibinfo  {journal} {Phys.
  Rev. X}\ }\textbf {\bibinfo {volume} {3}},\ \bibinfo {pages} {041034}
  (\bibinfo {year} {2013})}\BibitemShut {NoStop}%
\bibitem [{\citenamefont {Stadler}\ \emph {et~al.}(2012)\citenamefont
  {Stadler}, \citenamefont {Krinner}, \citenamefont {Meineke}, \citenamefont
  {Brantut},\ and\ \citenamefont {Esslinger}}]{Stadler2012}%
  \BibitemOpen
  \bibfield  {author} {\bibinfo {author} {\bibfnamefont {D.}~\bibnamefont
  {Stadler}}, \bibinfo {author} {\bibfnamefont {S.}~\bibnamefont {Krinner}},
  \bibinfo {author} {\bibfnamefont {J.}~\bibnamefont {Meineke}}, \bibinfo
  {author} {\bibfnamefont {J.-P.}\ \bibnamefont {Brantut}}, \ and\ \bibinfo
  {author} {\bibfnamefont {T.}~\bibnamefont {Esslinger}},\ }\bibfield  {title}
  {\enquote {\bibinfo {title} {{Observing the drop of resistance in the flow of
  a superfluid Fermi gas}},}\ }\href {https://doi.org/10.1038/nature11613}
  {\bibfield  {journal} {\bibinfo  {journal} {Nature}\ }\textbf {\bibinfo
  {volume} {491}},\ \bibinfo {pages} {736} (\bibinfo {year}
  {2012})}\BibitemShut {NoStop}%
\bibitem [{\citenamefont {Brantut}\ \emph {et~al.}(2013)\citenamefont
  {Brantut}, \citenamefont {Grenier}, \citenamefont {Meineke}, \citenamefont
  {Stadler}, \citenamefont {Krinner}, \citenamefont {Kollath}, \citenamefont
  {Esslinger},\ and\ \citenamefont {Georges}}]{BrantutScience2013}%
  \BibitemOpen
  \bibfield  {author} {\bibinfo {author} {\bibfnamefont {J.-P.}\ \bibnamefont
  {Brantut}}, \bibinfo {author} {\bibfnamefont {C.}~\bibnamefont {Grenier}},
  \bibinfo {author} {\bibfnamefont {J.}~\bibnamefont {Meineke}}, \bibinfo
  {author} {\bibfnamefont {D.}~\bibnamefont {Stadler}}, \bibinfo {author}
  {\bibfnamefont {S.}~\bibnamefont {Krinner}}, \bibinfo {author} {\bibfnamefont
  {C.}~\bibnamefont {Kollath}}, \bibinfo {author} {\bibfnamefont
  {T.}~\bibnamefont {Esslinger}}, \ and\ \bibinfo {author} {\bibfnamefont
  {A.}~\bibnamefont {Georges}},\ }\bibfield  {title} {\enquote {\bibinfo
  {title} {{A Thermoelectric Heat Engine with Ultracold Atoms}},}\ }\href
  {\doibase 10.1126/science.1242308} {\bibfield  {journal} {\bibinfo  {journal}
  {Science}\ }\textbf {\bibinfo {volume} {342}},\ \bibinfo {pages} {713}
  (\bibinfo {year} {2013})}\BibitemShut {NoStop}%
\bibitem [{\citenamefont {Husmann}\ \emph {et~al.}(2015)\citenamefont
  {Husmann}, \citenamefont {Uchino}, \citenamefont {Krinner}, \citenamefont
  {Lebrat}, \citenamefont {Giamarchi}, \citenamefont {Esslinger},\ and\
  \citenamefont {Brantut}}]{HusmannScience2015}%
  \BibitemOpen
  \bibfield  {author} {\bibinfo {author} {\bibfnamefont {D.}~\bibnamefont
  {Husmann}}, \bibinfo {author} {\bibfnamefont {S.}~\bibnamefont {Uchino}},
  \bibinfo {author} {\bibfnamefont {S.}~\bibnamefont {Krinner}}, \bibinfo
  {author} {\bibfnamefont {M.}~\bibnamefont {Lebrat}}, \bibinfo {author}
  {\bibfnamefont {T.}~\bibnamefont {Giamarchi}}, \bibinfo {author}
  {\bibfnamefont {T.}~\bibnamefont {Esslinger}}, \ and\ \bibinfo {author}
  {\bibfnamefont {J.-P.}\ \bibnamefont {Brantut}},\ }\bibfield  {title}
  {\enquote {\bibinfo {title} {Connecting strongly correlated superfluids by a
  quantum point contact},}\ }\href {\doibase 10.1126/science.aac9584}
  {\bibfield  {journal} {\bibinfo  {journal} {Science}\ }\textbf {\bibinfo
  {volume} {350}},\ \bibinfo {pages} {1498} (\bibinfo {year}
  {2015})}\BibitemShut {NoStop}%
\bibitem [{\citenamefont {Husmann}\ \emph {et~al.}(2018)\citenamefont
  {Husmann}, \citenamefont {Lebrat}, \citenamefont {H{\"a}usler}, \citenamefont
  {Brantut}, \citenamefont {Corman},\ and\ \citenamefont
  {Esslinger}}]{Husmann-PNAS2018}%
  \BibitemOpen
  \bibfield  {author} {\bibinfo {author} {\bibfnamefont {D.}~\bibnamefont
  {Husmann}}, \bibinfo {author} {\bibfnamefont {M.}~\bibnamefont {Lebrat}},
  \bibinfo {author} {\bibfnamefont {S.}~\bibnamefont {H{\"a}usler}}, \bibinfo
  {author} {\bibfnamefont {J.-P.}\ \bibnamefont {Brantut}}, \bibinfo {author}
  {\bibfnamefont {L.}~\bibnamefont {Corman}}, \ and\ \bibinfo {author}
  {\bibfnamefont {T.}~\bibnamefont {Esslinger}},\ }\bibfield  {title} {\enquote
  {\bibinfo {title} {{Breakdown of the Wiedemann{\textendash}Franz law in a
  unitary Fermi gas}},}\ }\href {\doibase 10.1073/pnas.1803336115} {\bibfield
  {journal} {\bibinfo  {journal} {PNAS}\ }\textbf {\bibinfo {volume} {115}},\
  \bibinfo {pages} {8563} (\bibinfo {year} {2018})}\BibitemShut {NoStop}%
\bibitem [{\citenamefont {Lesovik}\ and\ \citenamefont
  {Sadovskyy}(2011)}]{LesovikUFN2011}%
  \BibitemOpen
  \bibfield  {author} {\bibinfo {author} {\bibfnamefont {G.~B.}\ \bibnamefont
  {Lesovik}}\ and\ \bibinfo {author} {\bibfnamefont {I.~A.}\ \bibnamefont
  {Sadovskyy}},\ }\bibfield  {title} {\enquote {\bibinfo {title} {Scattering
  matrix approach to the description of quantum electron transport},}\ }\href
  {\doibase 10.3367/UFNe.0181.201110b.1041} {\bibfield  {journal} {\bibinfo
  {journal} {Physics-Uspekhi}\ }\textbf {\bibinfo {volume} {54}},\ \bibinfo
  {pages} {1007} (\bibinfo {year} {2011})}\BibitemShut {NoStop}%
\bibitem [{Note1()}]{Note1}%
  \BibitemOpen
  \bibinfo {note} {In other words, we expect that the scattering properties of
  the junction with (SNXNS) and without (SXS) the narrow regions N are
  equivalent in the leading order in $\varepsilon /E_F$.}\BibitemShut {Stop}%
\bibitem [{\citenamefont {Beenakker}(1991)}]{Beenakker1991}%
  \BibitemOpen
  \bibfield  {author} {\bibinfo {author} {\bibfnamefont {C.~W.~J.}\
  \bibnamefont {Beenakker}},\ }\bibfield  {title} {\enquote {\bibinfo {title}
  {{Universal limit of critical-current fluctuations in mesoscopic Josephson
  junctions}},}\ }\href {\doibase 10.1103/PhysRevLett.67.3836} {\bibfield
  {journal} {\bibinfo  {journal} {Phys. Rev. Lett.}\ }\textbf {\bibinfo
  {volume} {67}},\ \bibinfo {pages} {3836} (\bibinfo {year}
  {1991})}\BibitemShut {NoStop}%
\bibitem [{Note2()}]{Note2}%
  \BibitemOpen
  \bibinfo {note} {In our work, we focus on the conventional Josephson
  junctions, where Eq.~(\ref {even_odd}) is applicable. We leave the analysis
  of more exotic cases, e.g. $\phi _0$ - junctions \cite {BuzdinPRB2003}, for
  future works. In such junctions, the current depends on the additional phase
  $\phi _0$ that breaks time-reversal symmetry. Then equation Eq.~(\ref
  {even_odd}) may be generalized $I(\varphi ,\varphi _0) = \protect
  \mathaccentV {tilde}07EI(\varphi ,\varphi _0) + \protect \mathaccentV
  {tilde}07E{\protect \mathaccentV {tilde}07EI} (\varphi ,\varphi _0)$, where
  the dissipative and non-dissipative components satisfy the following parity
  conditions $\protect \mathaccentV {tilde}07EI(-\varphi ,-\varphi _0) =
  \protect \mathaccentV {tilde}07EI(\varphi ,\varphi _0)$ and $\protect
  \mathaccentV {tilde}07E{\protect \mathaccentV {tilde}07EI}(-\varphi ,-\varphi
  _0) = -\protect \mathaccentV {tilde}07E{\protect \mathaccentV
  {tilde}07EI}(\varphi ,\varphi _0)$.}\BibitemShut {Stop}%
\bibitem [{\citenamefont {Blonder}\ \emph {et~al.}(1982)\citenamefont
  {Blonder}, \citenamefont {Tinkham},\ and\ \citenamefont
  {Klapwijk}}]{BTK1982}%
  \BibitemOpen
  \bibfield  {author} {\bibinfo {author} {\bibfnamefont {G.~E.}\ \bibnamefont
  {Blonder}}, \bibinfo {author} {\bibfnamefont {M.}~\bibnamefont {Tinkham}}, \
  and\ \bibinfo {author} {\bibfnamefont {T.~M.}\ \bibnamefont {Klapwijk}},\
  }\bibfield  {title} {\enquote {\bibinfo {title} {{Transition from metallic to
  tunneling regimes in superconducting microconstrictions: Excess current,
  charge imbalance, and supercurrent conversion}},}\ }\href {\doibase
  10.1103/PhysRevB.25.4515} {\bibfield  {journal} {\bibinfo  {journal} {Phys.
  Rev. B}\ }\textbf {\bibinfo {volume} {25}},\ \bibinfo {pages} {4515}
  (\bibinfo {year} {1982})}\BibitemShut {NoStop}%
\bibitem [{\citenamefont {Virtanen}\ and\ \citenamefont
  {Giazotto}(2015)}]{Virtanen2015}%
  \BibitemOpen
  \bibfield  {author} {\bibinfo {author} {\bibfnamefont {P.}~\bibnamefont
  {Virtanen}}\ and\ \bibinfo {author} {\bibfnamefont {F.}~\bibnamefont
  {Giazotto}},\ }\bibfield  {title} {\enquote {\bibinfo {title} {{Fluctuation
  of heat current in Josephson junctions}},}\ }\href {\doibase
  10.1063/1.4914077} {\bibfield  {journal} {\bibinfo  {journal} {AIP Adv.}\
  }\textbf {\bibinfo {volume} {5}},\ \bibinfo {pages} {027140} (\bibinfo {year}
  {2015})}\BibitemShut {NoStop}%
\bibitem [{Note3()}]{Note3}%
  \BibitemOpen
  \bibinfo {note} {Equation~(\ref {J_small_asymmetry}) generalizes the result
  previously derived using the tunneling Hamiltonian approach \cite
  {Maki1966,GuttmanPRB1997b} to arbitrary transparency $\tau $. Note that an
  incorrect sign was obtained in front of the phase $\varphi $ dependent term
  in Ref.~[\protect \rev@citealpnum {GuttmanPRB1997b}] and was subsequently
  corrected by Refs.~[\protect \rev@citealpnum {SaulsPRL2003},\protect
  \rev@citealpnum {SaulsPRB2004}].}\BibitemShut {Stop}%
\bibitem [{\citenamefont {Fu}\ and\ \citenamefont {Kane}(2009)}]{FuPRB2008}%
  \BibitemOpen
  \bibfield  {author} {\bibinfo {author} {\bibfnamefont {L.}~\bibnamefont
  {Fu}}\ and\ \bibinfo {author} {\bibfnamefont {C.~L.}\ \bibnamefont {Kane}},\
  }\bibfield  {title} {\enquote {\bibinfo {title} {Josephson current and noise
  at a superconductor/quantum-spin-hall-insulator/superconductor junction},}\
  }\href {\doibase 10.1103/PhysRevB.79.161408} {\bibfield  {journal} {\bibinfo
  {journal} {Phys. Rev. B}\ }\textbf {\bibinfo {volume} {79}},\ \bibinfo
  {pages} {161408(R)} (\bibinfo {year} {2009})}\BibitemShut {NoStop}%
\bibitem [{\citenamefont {Sothmann}\ and\ \citenamefont
  {Hankiewicz}(2016)}]{Sothman2016}%
  \BibitemOpen
  \bibfield  {author} {\bibinfo {author} {\bibfnamefont {B.}~\bibnamefont
  {Sothmann}}\ and\ \bibinfo {author} {\bibfnamefont {E.~M.}\ \bibnamefont
  {Hankiewicz}},\ }\bibfield  {title} {\enquote {\bibinfo {title} {{Fingerprint
  of topological Andreev bound states in phase-dependent heat transport}},}\
  }\href {\doibase 10.1103/PhysRevB.94.081407} {\bibfield  {journal} {\bibinfo
  {journal} {Phys. Rev. B}\ }\textbf {\bibinfo {volume} {94}},\ \bibinfo
  {pages} {081407(R)} (\bibinfo {year} {2016})}\BibitemShut {NoStop}%
\bibitem [{\citenamefont {Bardas}\ and\ \citenamefont
  {Averin}(1995)}]{Averin1995}%
  \BibitemOpen
  \bibfield  {author} {\bibinfo {author} {\bibfnamefont {A.}~\bibnamefont
  {Bardas}}\ and\ \bibinfo {author} {\bibfnamefont {D.}~\bibnamefont
  {Averin}},\ }\bibfield  {title} {\enquote {\bibinfo {title} {Peltier effect
  in normal-metal--superconductor microcontacts},}\ }\href {\doibase
  10.1103/PhysRevB.52.12873} {\bibfield  {journal} {\bibinfo  {journal} {Phys.
  Rev. B}\ }\textbf {\bibinfo {volume} {52}},\ \bibinfo {pages} {12873}
  (\bibinfo {year} {1995})}\BibitemShut {NoStop}%
\bibitem [{\citenamefont {Beenakker}(1992)}]{BeenakkerPRB1992}%
  \BibitemOpen
  \bibfield  {author} {\bibinfo {author} {\bibfnamefont {C.~W.~J.}\
  \bibnamefont {Beenakker}},\ }\bibfield  {title} {\enquote {\bibinfo {title}
  {Quantum transport in semiconductor-superconductor microjunctions},}\ }\href
  {\doibase 10.1103/PhysRevB.46.12841} {\bibfield  {journal} {\bibinfo
  {journal} {Phys. Rev. B}\ }\textbf {\bibinfo {volume} {46}},\ \bibinfo
  {pages} {12841} (\bibinfo {year} {1992})}\BibitemShut {NoStop}%
\bibitem [{Note4()}]{Note4}%
  \BibitemOpen
  \bibinfo {note} {Here we also use the expansion of $g^{NS}$ at $\tau = 0$:
  $g^{NS}(\alpha ,0) = 1 - C\protect \tmspace +\thinmuskip {.1667em}\protect
  \tmspace +\thinmuskip {.1667em} \alpha ^2; \protect \tmspace +\thinmuskip
  {.1667em}\protect \tmspace +\thinmuskip {.1667em}C = \protect \frac {1}{8}
  \DOTSI \intop \ilimits@ _{0}^{+\infty } dx \protect \frac {\protect \qopname
  \relax o{tanh}^2 x}{x^2} = \protect \frac {7\zeta (3)}{4\pi ^2}\approx
  0.21$.}\BibitemShut {Stop}%
\bibitem [{\citenamefont {Hussein}\ \emph {et~al.}(2019)\citenamefont
  {Hussein}, \citenamefont {Governale}, \citenamefont {Kohler}, \citenamefont
  {Belzig}, \citenamefont {Giazotto},\ and\ \citenamefont
  {Braggio}}]{Braggio2018}%
  \BibitemOpen
  \bibfield  {author} {\bibinfo {author} {\bibfnamefont {R.}~\bibnamefont
  {Hussein}}, \bibinfo {author} {\bibfnamefont {M.}~\bibnamefont {Governale}},
  \bibinfo {author} {\bibfnamefont {S.}~\bibnamefont {Kohler}}, \bibinfo
  {author} {\bibfnamefont {W.}~\bibnamefont {Belzig}}, \bibinfo {author}
  {\bibfnamefont {F.}~\bibnamefont {Giazotto}}, \ and\ \bibinfo {author}
  {\bibfnamefont {A.}~\bibnamefont {Braggio}},\ }\bibfield  {title} {\enquote
  {\bibinfo {title} {{Nonlocal thermoelectricity in a Cooper-pair splitter}},}\
  }\href {\doibase 10.1103/PhysRevB.99.075429} {\bibfield  {journal} {\bibinfo
  {journal} {Phys. Rev. B}\ }\textbf {\bibinfo {volume} {99}},\ \bibinfo
  {pages} {075429} (\bibinfo {year} {2019})}\BibitemShut {NoStop}%
\bibitem [{\citenamefont {Kan\'asz-Nagy}\ \emph {et~al.}(2016)\citenamefont
  {Kan\'asz-Nagy}, \citenamefont {Glazman}, \citenamefont {Esslinger},\ and\
  \citenamefont {Demler}}]{Kanasz-Nagy2016}%
  \BibitemOpen
  \bibfield  {author} {\bibinfo {author} {\bibfnamefont {M.}~\bibnamefont
  {Kan\'asz-Nagy}}, \bibinfo {author} {\bibfnamefont {L.}~\bibnamefont
  {Glazman}}, \bibinfo {author} {\bibfnamefont {T.}~\bibnamefont {Esslinger}},
  \ and\ \bibinfo {author} {\bibfnamefont {E.~A.}\ \bibnamefont {Demler}},\
  }\bibfield  {title} {\enquote {\bibinfo {title} {{Anomalous Conductances in
  an Ultracold Quantum Wire}},}\ }\href {\doibase
  10.1103/PhysRevLett.117.255302} {\bibfield  {journal} {\bibinfo  {journal}
  {Phys. Rev. Lett.}\ }\textbf {\bibinfo {volume} {117}},\ \bibinfo {pages}
  {255302} (\bibinfo {year} {2016})}\BibitemShut {NoStop}%
\bibitem [{Note5()}]{Note5}%
  \BibitemOpen
  \bibinfo {note} {Two comments are in place here. First, while considering
  $\Delta /T\ll 1$, we still assume the BCS coherence length shorter than the
  superfluid domain in the NSN structure. Second, the ratio $g_T^{NS}/g_T^{SS}$
  appearing in the last factor of Eq.~(\ref {LNSNL}) becomes small and may
  compensate the large factor $N_w/(2N_n)$ at low temperatures, $T/\Delta \sim
  (N_w/2N_n)^2$; however, at such low temperatures is exponentially small,
  $L^{NSN}\sim \protect \qopname \relax o{exp}\protect \{-(N_w/2N_n)^2\protect
  \}$}\BibitemShut {NoStop}%
\bibitem [{\citenamefont {Buzdin}\ and\ \citenamefont
  {Koshelev}(2003)}]{BuzdinPRB2003}%
  \BibitemOpen
  \bibfield  {author} {\bibinfo {author} {\bibfnamefont {A.}~\bibnamefont
  {Buzdin}}\ and\ \bibinfo {author} {\bibfnamefont {A.~E.}\ \bibnamefont
  {Koshelev}},\ }\bibfield  {title} {\enquote {\bibinfo {title} {{Periodic
  alternating 0- and $\ensuremath{\pi}$-junction structures as realization of
  $\ensuremath{\varphi}$-Josephson junctions}},}\ }\href {\doibase
  10.1103/PhysRevB.67.220504} {\bibfield  {journal} {\bibinfo  {journal} {Phys.
  Rev. B}\ }\textbf {\bibinfo {volume} {67}},\ \bibinfo {pages} {220504(R)}
  (\bibinfo {year} {2003})}\BibitemShut {NoStop}%
\end{thebibliography}%
%%%%%%%%%%%%%%%%%%%%%%%%%%%%%%%%%%%%%%%%%%%%%%%%%%%%%%%%%%%%%%%%%%%%%%%%%%%%%

\appendix

%%%%%%%%%%%%%%%%%%%%%%%%%%%%%%%%%%%%%%%%%%%%%%%%%%%%%%%%%%%%%%%%%%%%%%%%%%%%%%%%%%%%%
\section{Derivation of scattering amplitudes} \label{sec:amplitudes_ap}
%%%%%%%%%%%%%%%%%%%%%%%%%%%%%%%%%%%%%%%%%%%%%%%%%%%%%%%%%%%%%%%%%%%%%%%%%%%%%%%%%%%%%
%%%%%%%%%%%%%%%%%%%%%%%%%%%%%%%%%%%%%%%%%%%%%%%%%%%%%%%%%%%%%%%%%%%%%%%%%%%%%
\begin{figure*}
 \centering\includegraphics[width=0.7\linewidth]{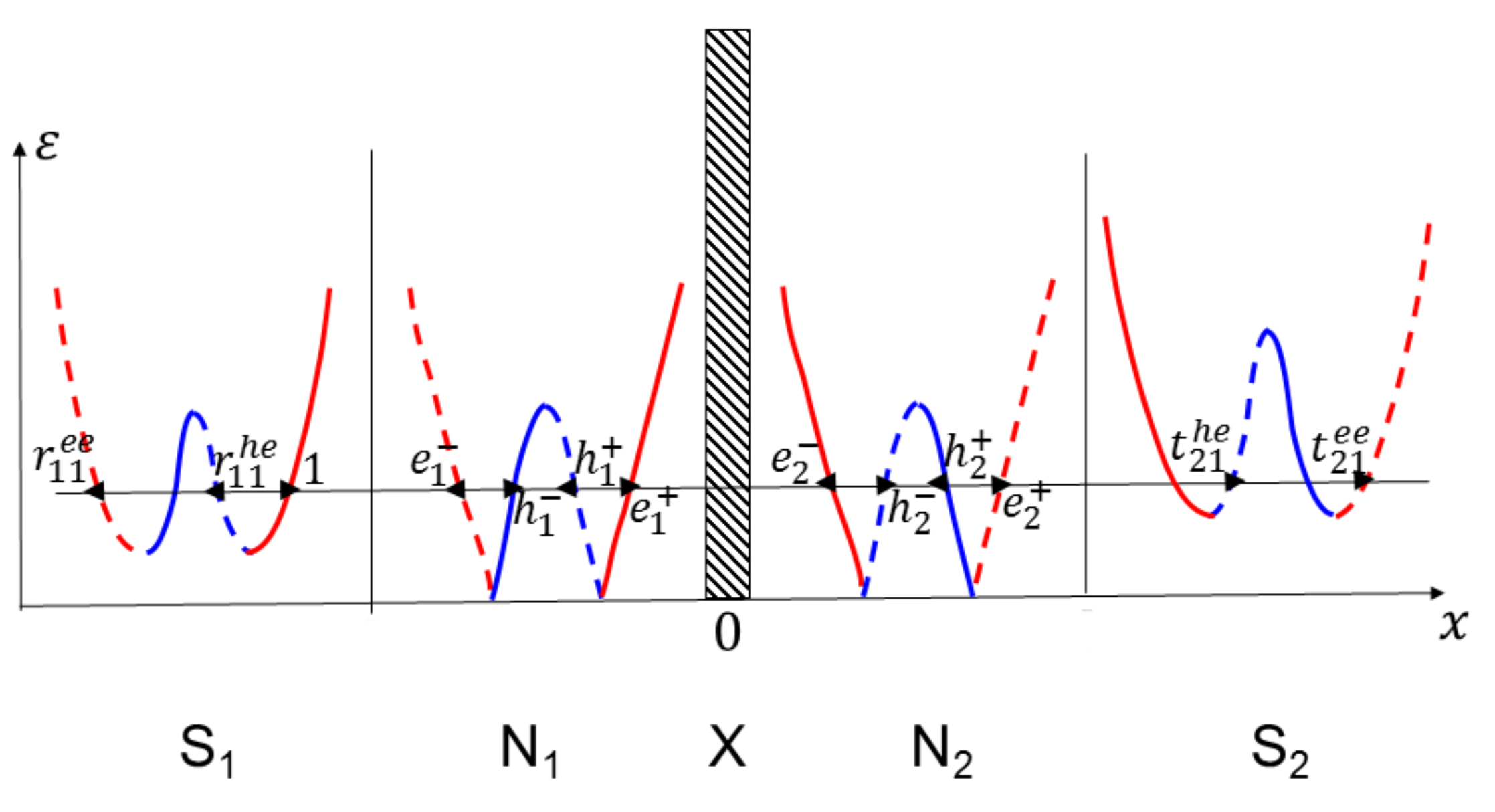} 
 \caption{Schematic representation of the scattering problem. The scatterer $X$ is surrounded by normal regions $N_1$ and $N_2$ which are adjacent to the superconducting regions $S_1$ and $S_2$. The real space is superimposed with the momentum space: the energy spectrum of excitations in each domain is shown. The incoming (outgoing) states, i.e. with group velocity directed to (from) the scatterer $X$, are shown in solid (dashed) lines. The electron-like (hole-like) states are shown in red (blue).} 
\label{fig_scat}
\end{figure*}
%%%%%%%%%%%%%%%%%%%%%%%%%%%%%%%%%%%%%%%%%%%%%%%%%%%%%%%%%%%%%%%%%%%%%%%%%%%%%

Consider a scattering wavefunction shown in Fig.~\ref{fig_scat} with four regions $S_1,\,N_1,\,N_2,\,S_2$. We follow the approach used in Ref.~[\onlinecite{Beenakker1991}]. The scattering amplitudes $r^{ee}_{11}$, $r^{he}_{11}$, $t^{ee}_{21}$, $t^{he}_{21}$, may be obtained in a few steps:

(i) We write the wavefunctions in superconducting regions $S_1$ and $S_2$
\begin{equation}
\begin{aligned}
	&\Psi_{S_1} = \\ 
	& \,\,\left( \begin{array}{c}
		  u_1 \\
		  v_1
	\end{array}\right) e^{iq_ex} + r^{he}_{11} \left( \begin{array}{c}
		  v_1 \\
		  u_1
	  \end{array}\right) e^{iq_hx}  +   r^{ee}_{11} \left( \begin{array}{c}
		  u_1 \\
		  v_1
	  \end{array}\right) e^{-iq_ex}   ,  \\
	  &\Psi_{S_2} = \\ 
	  & \,\,{ t^{ee}_{21} \left( \begin{array}{c}
		  u_2 \\
		  v_2
	  \end{array}\right) e^{iq_ex}} +{ t^{he}_{21} \left( \begin{array}{c}
		  v_2 \\
		  u_2
	  \end{array}\right) e^{-iq_hx}}.
\end{aligned}
\label{scat_wf}
\end{equation}
Here, the coherence factors 
\begin{equation}
    u^2_{1(2)} = 1-v^2_{1(2)} {= \frac{1}{2}\left(1+\frac{\xi_{1(2)}}{\varepsilon}\right)}    
\end{equation}
are energy-dependent, and the notation $\xi_{1(2)} = \sqrt{\varepsilon^2 - \Delta^2_{1(2)}}$ accounts for a possibility of non-equal gaps $\Delta_{1} \neq \Delta_2$. 

(ii) The wavefunctions in the normal regions are linear combinations of electron and hole wavefunctions
\begin{align}
    \Psi_{N_{1(2)}} =&\, e^{+}_{1(2)} \left( \begin{array}{c}
		  1 \\
		  0
	  \end{array}\right) e^{ik_ex} + 
	  h^{+}_{1(2)} \left( \begin{array}{c}
		  0 \\
		  1
	  \end{array}\right) e^{ik_hx} \label{wf_normal_region} \\
	  & + e^{-}_{1(2)} \left( \begin{array}{c}
		  1 \\
		  0
	  \end{array}\right) e^{-ik_ex} +   h^{-}_{1(2)} \left( \begin{array}{c}
		  0 \\
		  1
	  \end{array}\right) e^{-ik_hx} \nonumber
\end{align}
We work in the Andreev approximation, which is valid for energy $\varepsilon$ smaller than the Fermi energy, i.e. $\varepsilon \ll E_F$. In this approximation, the NS boundaries do not scatter the momentum across the Fermi sea, and one may equate the wavefunctions corresponding to the positive $k_F$ and negative $-k_F$ momentum separately. Such a boundary condition produces the following relation between the amplitudes in the normal and superconducting regions
\begin{equation}
  \begin{aligned}
    e^+_1 & = u_1+v_1\, r^{he}_{11}, \\
    h^+_1 & = v_1+u_1 \,r^{he}_{11}, \\
    e^-_1 & = u_1\, r^{ee}_{11},     \\
    h^-_1 & = v_1 \,r^{ee}_{11}, \\
    e^+_2 & = u_2\, t^{ee}_{21}, \\
    h^+_2 & = v_2\, t^{ee}_{21}, \\
    e^-_2 & = v_2\, t^{he}_{21}, \\
    h^-_2 & = u_2\, t^{he}_{21}. 
  \end{aligned} \label{norm_coef}
\end{equation}
     
(iii) Inside the normal region, the scattering amplitudes in $N_1$ and $N_2$ are related via the scattering matrix describing the scatterer
\begin{equation}
 s_0(\varepsilon) = e^{i\gamma_\varepsilon}\left( \begin{array}{cc}
		   e^{i\eta_{\varepsilon}}\,r_{\varepsilon} & i\, e^{-i\varphi/2}t_{\varepsilon} \\
		  i\, e^{i\varphi/2}t_{\varepsilon} & e^{-i\eta_{\varepsilon}}\,r_{\varepsilon}
		  \end{array}\right).
\end{equation}
The terms $r_{\varepsilon}$ and $t_{\varepsilon}$ are the real-valued relfection and transmission amplitudes; the phase $\varphi$ describes the time-reversal symmetry breaking; the phase $\eta_{\varepsilon}$ is related to the absence of the inversion symmetry; the overall-phase $\gamma_\varepsilon$ in the prefactor is the energy-dependent Friedel phase related to the modulation of the density of states in the presence of the scatterer. With the notations $ \bm s_e =\bm s_0(\varepsilon)$ and $\bm s_h = \bm s_0^\ast(-\varepsilon)$, the conditions for electron and hole quasiparticles in the normal region split 
\begin{equation}
    \begin{aligned}
       \left( \begin{array}{c}
		  e^-_1 \\
		  e^+_2
		  \end{array}\right) 
	   & = \bm s_e\,\left( \begin{array}{c}
		  e^+_1 \\
		  e^-_2
		  \end{array}\right) \\
	   \left( \begin{array}{c}
		  h^+_1 \\
		  h^-_2
		  \end{array}\right) 
	   & = \bm s_h \left( \begin{array}{c}
		  h^-_1 \\
		  h^+_2
		  \end{array}\right)	  
	\end{aligned} \label{scat_matr}
\end{equation}

(iv) We substitute Eqs.~(\ref{norm_coef}) in Eqs.~(\ref{scat_matr}) and obtain equations for the unknown scattering amplitudes $r^{ee}_{11}$, $r^{he}_{11}$, $t^{ee}_{21}$, $t^{he}_{21}$, which we write in a matrix form 
 \begin{equation}
     \begin{aligned}
       & \bm u\, \psi_n = \bm s_e\,(\bm v\, \psi_a + \bm u\, \psi_0), \\
       & \bm v \, \psi_0 + \bm u\, \psi_a = \bm s_h\,\bm v\,\psi_n,
     \end{aligned} \label{matr_eq}
 \end{equation}
where we absorbed the normal and Andreev scattering amplitudes in the 2-by-1 vectors $\psi_n = (r^{ee}_{11}\,t^{ee}_{21})^{\rm T}$ and $\psi_a = (r^{he}_{11}\,t^{he}_{21})^{\rm T}$; $\psi_0 = (1\, 0)^{\rm T}$ is a 2-by-1 vector; we also defined the 2-by-2 matrices 
 \begin{equation}
     \begin{aligned}
       & \bm u = \left( \begin{array}{cc}
		  u_1 & 0 \\
		  0 & u_2
		  \end{array}\right), \\
       & \bm v = \left( \begin{array}{cc}
		  v_1 & 0 \\
		  0 & v_2
		  \end{array}\right).
     \end{aligned} \label{uv}
 \end{equation}
 
(v) Solving the linear Eq.~(\ref{matr_eq}) in favor of $\psi_n$ and $\psi_a$, we obtain
 \begin{equation}
     \begin{aligned}
       \psi_n &= \bm u^{-1} (1-\bm s_e\bm a\bm s_h\bm a)^{-1} \bm s_e (1-\bm a^2)\bm u \, \psi_0, \\
       \psi_a &= \bm u^{-1} (1-\bm s_h\bm a\bm s_e\bm a)^{-1} (\bm s_h\bm a \bm s_e\bm u - \bm v) \, \psi_0. \\
     \end{aligned} \label{solution_scat_ampl}
\end{equation}
with $\bm a =\bm v \bm u^{-1}$. We perform the matrix multiplication in Eq.~(\ref{solution_scat_ampl}) and obtain the scattering amplitudes
\begin{widetext}
\begin{equation}
    \begin{aligned}
        r^{ee}_{11} &= \frac{\xi_1}{2 D_{\varepsilon}}\left[ (\varepsilon+\xi_2)\,r_\varepsilon e^{i(\gamma_{-\varepsilon}+ \eta_\varepsilon)} - (\varepsilon-\xi_2)\,r_{-\varepsilon}e^{i(\gamma_\varepsilon+ \eta_{-\varepsilon})}\right], \\
        r^{he}_{11} & = \frac{1}{2D_{\varepsilon}} \left[-\Delta_1(\varepsilon \cos\delta\gamma_{\varepsilon}-i\xi_2\sin \delta \gamma_\varepsilon) + \Delta_1 (\varepsilon { \cos \delta \eta_{\varepsilon}+ i \xi_2 \sin\delta \eta_\varepsilon})\, r_{\varepsilon} r_{-\varepsilon}+ \Delta_2(\varepsilon \cos \varphi + i\xi_1 \sin \varphi)\, t_{\varepsilon} t_{-\varepsilon}\right], \\
        t^{ee}_{21} & = \frac{i\xi_1}{2 D_{\varepsilon}}\left[\sqrt{(\varepsilon+\xi_1)(\varepsilon+\xi_2)}\,t_{\varepsilon} e^{i(\varphi/2+\gamma_{-\varepsilon})}- \sqrt{(\varepsilon-\xi_1)(\varepsilon-\xi_2)}\,t_{-\varepsilon} e^{i(-\varphi/2+\gamma_{\varepsilon})}\right], \\
        t^{he}_{21} &= \frac{i\xi_1}{2D_{\varepsilon}} \left[\sqrt{(\varepsilon +\xi_1)(\varepsilon-\xi_2)}\, t_{\varepsilon} r_{-\varepsilon}\,e^{i(\varphi/2+ \eta_{-\varepsilon})} -  \sqrt{(\varepsilon-\xi_1)(\varepsilon+\xi_2)}\, t_{-\varepsilon} r_{\varepsilon}\,e^{i(-\varphi/2+ \eta_\varepsilon)}\right],
    \end{aligned} \label{amplitudes_ap}
\end{equation}
where we introduced the following notations:
\begin{equation}
    \begin{aligned}
        D_{\varepsilon} &= \frac{1}{2} \left[ (\varepsilon^2+\xi_1\xi_2)\cos\delta\gamma_\varepsilon - i\varepsilon (\xi_1 + \xi_2)\sin \delta \gamma_\varepsilon - (\varepsilon^2-\xi_1\xi_2) \, r_{\varepsilon}r_{-\varepsilon}\, {\cos \delta\eta_{\varepsilon}} -  {i \varepsilon( \xi_2 - \xi_1) r_{\varepsilon} r_{-\varepsilon} \sin  \delta\eta_\varepsilon} - \Delta_1 \Delta_2 \,t_{\varepsilon} t_{-\varepsilon} \cos \varphi \right], \label{denom_ap} \\
        \delta \gamma_{\varepsilon} &= \gamma_\varepsilon - \gamma_{-\varepsilon},\quad {\delta \eta_{\varepsilon} = \eta_\varepsilon - \eta_{-\varepsilon}}. 
    \end{aligned}
\end{equation}
\end{widetext}

(vi) One may repeat the derivation of amplitudes for a {\it hole} like quasiparticle incident from the left lead. The equation for the amplitudes reads
 \begin{equation}
     \begin{aligned}
       & \bm u\, \psi_n = \bm s_h\,(\bm v\, \psi_a + \bm u\, \psi_0), \\
       & \bm v \, \psi_0 + \bm u\, \psi_a = \bm s_e\,\bm v\,\psi_n,
     \end{aligned} \label{matr_eq_h}
 \end{equation}
where $\psi_n = (r^{hh}_{11}\,t^{hh}_{21})^{\rm T}$ and $\psi_a = (r^{eh}_{11}\,t^{eh}_{21})^{\rm T}$ are the 2-by-1 vectors that encapsulate the normal and Andreev scattering amplitudes; and $\psi_0 = (1\, 0)^{\rm T}$. The matrices $\bm u$ and $\bm v$ are defined in Eq.~(\ref{uv}). By  inspection, equations for hole-like (\ref{matr_eq_h}) and electron-like Eq.~(\ref{matr_eq}) quasiparticles are related via the transformation $\bm s_e \leftrightarrow \bm s_h$. Thus, one may find the amplitudes $(r^{hh}_{11}\,t^{hh}_{21}\,r^{eh}_{11}\,t^{eh}_{21})$ by replacing $\varphi \leftrightarrow -\varphi$, $\gamma_{\varepsilon} \leftrightarrow -\gamma_{-\varepsilon}$, $\eta_{\varepsilon} \leftrightarrow -\eta_{-\varepsilon}$ (the latter transformations keep $\delta\gamma_{\varepsilon}$ and $\delta\eta_{\varepsilon}$ invariant), $r_{\varepsilon} \leftrightarrow r_{-\varepsilon}$, $t_{\varepsilon} \leftrightarrow -t_{-\varepsilon}$ in Eq.~(\ref{amplitudes_ap}).

(vii) The amplitudes corresponding to quasiparticles incoming from the {\it right} lead may be obtained by replacing the gaps $\Delta_1 \leftrightarrow \Delta_2$ as well as the phases $\varphi \leftrightarrow -\varphi$, $\eta_{\varepsilon} \leftrightarrow -\eta_{\varepsilon}$ in Eq.~(\ref{amplitudes_ap}).

(viii) For future reference, let us give the amplitudes in the tunneling limit $t_{\varepsilon} \ll 1$. For simplicity, we set $\gamma = \eta = 0$, $\Delta_1 = \Delta_2 = \Delta$ and obtain from Eq.~(\ref{amplitudes_ap})
\begin{equation}
\begin{aligned}
& t^{ee}_{21} = \frac{i}{2\xi}\left[(\varepsilon+\xi)t_{\varepsilon}e^{i\varphi/2} - (\varepsilon-\xi)t_{-
\varepsilon}e^{-i\varphi/2}\right],\\
& t^{he}_{21} = \frac{i\Delta}{2\xi}\left[t_{\varepsilon}e^{i\varphi/2} -t_{-\varepsilon}e^{-i\varphi/2}\right].\\
    \end{aligned} \label{tunnel_lim_Beenakker}
\end{equation}

%%%%%%%%%%%%%%%%%%%%%%%%%%%%%%%%%%%%%%%%%%%%%%%%%%%%%%%%%%%%%%%%%%%%%%%%%%%%%%%%%%%%%
\section{Details of derivation of Eqs.~(\ref{ful_h_cur})-(\ref{heatcurrent1}) for the heat current} \label{sec:heatcur_ap}
%%%%%%%%%%%%%%%%%%%%%%%%%%%%%%%%%%%%%%%%%%%%%%%%%%%%%%%%%%%%%%%%%%%%%%%%%%%%%%%%%%%%%

(i) We consider a scattering region shown in Fig.~\ref{fig_scat} with four regions $S_1,\,N_1,\,N_2,\,S_2$. The two superconducting leads  $S_1$ and $S_2$ are at temperatures $T_1$ and $T_2$ respectively. We make the standard assumption of the Landauer transport theory that the  quasiparticles emerging from each lead are in thermodynamic equilibrium with the corresponing lead. Then the total heat current $J$ through the contact may be expressed as a sum of independent contributions corresponding to the quasiparticles emerging from the distinct leads $J = J_1 - J_2$ as in Eq.~(\ref{ful_h_cur}), where 
\begin{align}
    J_l &= \frac{2}{h} \int_{0}^\infty d\xi \,\varepsilon_l \left[ j^{e}_l(\varepsilon_l)  + j^{h}_l(\varepsilon_l) \right] f\left(\varepsilon_l/T_l\right) \label{h_cur_from_one_lead_ap}
    \\ &= \frac{2}{h} \int_{\Delta_l}^\infty \frac{d\varepsilon\,\varepsilon^2}{\xi_l} \left[ j^{e}_l(\varepsilon)  + j^{h}_l(\varepsilon) \right] f\left(\varepsilon/T_l\right).    \label{h_cur_from_one_lead_ap1}
\end{align}
Here the subscript index $l \in\{1,2\}$ labels the leads; factor $2$ corresponds to spin degeneracy. The factor of energy $\varepsilon$ in the integrand~of Eq.~(\ref{h_cur_from_one_lead_ap}) indicates that we evaluate the energy current. Equations (\ref{h_cur_from_one_lead_ap}) and (\ref{h_cur_from_one_lead_ap1}) are related by a change of integration variable $\varepsilon = \sqrt{\xi^2 +\Delta_l^2}$. In the integral~(\ref{h_cur_from_one_lead_ap}), the integration variable is $\xi$ and $\varepsilon_l(\xi) = \sqrt{\xi^2+\Delta_l^2}$. In the integral~(\ref{h_cur_from_one_lead_ap1}), the integration variable is $\varepsilon$ and $\xi_l(\varepsilon) = \sqrt{\varepsilon^2-\Delta_l^2}$. We use Eq.~(\ref{h_cur_from_one_lead_ap1}) throughout the paper.

(ii) The terms $j^{e}_l$ and $j^{h}_l$ correspond to electron- and hole-like quasiparticles, labeled by the superscript $b \in\{e,h\}$. Each term $j^{b}_l(\varepsilon)$ may be evaluated using the corresponding BdG wavefunction $\Psi$, 
\begin{equation}
    j = \frac{1}{m v_F}\,{\rm Im}\left( \Psi^\dagger \sigma_z \nabla \Psi \right), \label{density_current}
\end{equation}
where $\sigma_z = {\rm diag}(1,-1)$ is the Pauli matrix acting in the Nambu space. Equation~(\ref{density_current}) has a physical meaning of the quasiparticle density current normalized by the Fermi velocity $v_F$, which renders it dimensionless. The density current~(\ref{density_current}) is conserved through the system, i.e. the quasiparticles do not dissappear. Thus, the current~(\ref{density_current}) evaluated at any spatial coordinate $x$ must yield the same result. As an example, let us evaluate Eq.~(\ref{density_current}) for the BdG wavefunction~(\ref{scat_wf}) corresponding to electron-like quasiparticle incident from the left lead. We evaluate it both to the left (i.e. for $\Psi_{S_1}$) and to the right (i.e. for $\Psi_{S_2}$) from the scatterer X
\begin{align}
  j^{e}_{1}(\varepsilon) & =\frac{\xi_1}{\varepsilon}\left[1- \left| r^{ee}_{11}\right|^2-\left| r^{he}_{11}\right|^2\right] \label{beforeX} \\
  & = \frac{\xi_2}{\varepsilon}\left[\left| t^{ee}_{21}\right|^2+\left| t^{he}_{21}\right|^2\right], \label{afterX}
\end{align}
where the equation $u_l^2-v_l^2 = \xi_s/\varepsilon$ was used. Equations~(\ref{beforeX}) and (\ref{afterX}) are equal as guaranteed by the unitarity. 

(iii) Next, we follow the steps as discussed in Sec.~\ref{sec:heatcur}. We assume that temperatures of the leads $T_{1,2} = T \pm \delta T/2$ differ by a small difference $\delta T$. At any $T$ and $\delta T = 0$, the heat currents flowing in opposite direction must cancel $J_1 = J_2$ to render the total heat current $J = J_1-J_2 = 0$. Thus, the integrands in Eq.~(\ref{h_cur_from_one_lead_ap}) corresponding to $l = 1$ and $l = 2$ are equal at $\delta T = 0$. At $\delta T \neq 0$, this condition allows to rewrite the total $J = J_1 - J_2$ only via the parameters corresponding to one lead (e.g. the left one) and expand in small~$\delta T$
\begin{align}
    J &= \frac{2}{h} \int_{\Delta_{max}}^\infty \frac{d\varepsilon\, \varepsilon^2}{\xi_1} \left[ j^{e}_1(\varepsilon)  + j^{h}_1(\varepsilon) \right] \left[f\left(\varepsilon/T_1\right)-f\left(\varepsilon/T_2\right)\right] \nonumber\\
    &= \frac{\delta T}{T^2}\frac{2}{h} \int_{\Delta_{max}}^\infty \frac{d\varepsilon\, \varepsilon^3}{\xi_1} \left[ j^{e}_1(\varepsilon)  + j^{h}_1(\varepsilon) \right] \left[-f'\left(x\right)\right]_{x = \varepsilon/T}, \label{J_total_ap}
\end{align}
where the lower integration limit is $\Delta_{max} = {\rm max}(\Delta_1,\Delta_2)$. The quasiparticles residing within the energy window  $\Delta_{max}>\varepsilon>\Delta_{min} = {\rm min}(\Delta_1,\Delta_2)$ do not contribute because they bounce back to the lead of their origin with probability 1 and, so, do not transfer energy between the leads. 

(iv) In order to evaluate the sum $j^{e}_1(\varepsilon)  + j^{h}_1(\varepsilon)$ appearing in the equation above, we use Eq.~(\ref{afterX}) (because it is more concise) and the amplitudes~(\ref{amplitudes_ap}) evaluated before. We obtain 
\begin{equation}
\begin{aligned}
 \left|t^{ee}_{21}\right|^2 = \frac{\xi_1^2}{4|D|^2}&\left[(\varepsilon+\xi_1)(\varepsilon+\xi_2) t_{\varepsilon}^2  \right. \\
 & \,\, + (\varepsilon-\xi_1)(\varepsilon-\xi_2) t_{-\varepsilon}^2 \\
 & \quad - \left. 2\Delta_1\Delta_2t_{\varepsilon}t_{-\varepsilon} \cos(\varphi-\delta \gamma_{\varepsilon})\right], \\
 \left|t^{he}_{21}\right|^2 = \frac{\xi_1^2}{4|D|^2}&\left[(\varepsilon+\xi_1)(\varepsilon-\xi_2) t_{\varepsilon}^2r_{-\varepsilon}^2  \right. \\
 & \,\, + (\varepsilon-\xi_1)(\varepsilon+\xi_2) t_{-\varepsilon}^2 r_{\varepsilon}^2 \\
 & \quad - \left. 2\Delta_1\Delta_2t_{\varepsilon}t_{-\varepsilon}r_{\varepsilon}r_{-\varepsilon} \cos(\varphi - \delta \eta_{\varepsilon})\right],
\end{aligned}
 \label{t2e}
\end{equation}
where we used the identitity $\sqrt{\varepsilon^2-\xi_s^2} = \Delta_s$. The analogous expressions for hole-like quasiparticles are obtained by replacing $r_{\varepsilon} \leftrightarrow r_{-\varepsilon}$, $t_{\varepsilon} \leftrightarrow -t_{-\varepsilon}$ and $\varphi \leftrightarrow -\varphi$ in the equations above,
\begin{equation}
\begin{aligned}
 \left|t^{hh}_{21}\right|^2 = \frac{\xi_1^2}{4|D|^2}&\left[(\varepsilon+\xi_1)(\varepsilon+\xi_2) t_{-\varepsilon}^2  \right. \\
 & \,\, + (\varepsilon-\xi_1)(\varepsilon-\xi_2) t_{\varepsilon}^2 \\
 & \quad - \left. 2\Delta_1\Delta_2t_{\varepsilon}t_{-\varepsilon} \cos(\varphi+\delta \gamma_{\varepsilon})\right], \\
 \left|t^{eh}_{21}\right|^2 = \frac{\xi_1^2}{4|D|^2}&\left[(\varepsilon+\xi_1)(\varepsilon-\xi_2) t_{-\varepsilon}^2r_{\varepsilon}^2  \right. \\
 & \,\, + (\varepsilon-\xi_1)(\varepsilon+\xi_2) t_{\varepsilon}^2 r_{-\varepsilon}^2 \\
 & \quad - \left. 2\Delta_1\Delta_2t_{\varepsilon}t_{-\varepsilon}r_{\varepsilon}r_{-\varepsilon} \cos(\varphi+\delta\eta_{\varepsilon})\right].
\end{aligned}
 \label{t2h}
\end{equation}
So, the sum $j^{e}_1(\varepsilon)  + j^{h}_1(\varepsilon)$ may be evaluated using Eq.~(\ref{afterX}),
\begin{align*}
    j^{e}_1(\varepsilon) & + j^{h}_1(\varepsilon) = \frac{\xi_2}{\varepsilon}\left[\left|t^{ee}_{21}\right|^2+\left|t^{he}_{21}\right|^2+\left|t^{hh}_{21}\right|^2+\left|t^{eh}_{21}\right|^2\right] \\
    & =\frac{\xi_2\xi_1^2}{2|D|^2\varepsilon} \left[\varepsilon^2(t_{\varepsilon}^2+t_{-\varepsilon}^2+t_{\varepsilon}^2r_{-\varepsilon}^2+t_{-\varepsilon}^2r_{\varepsilon}^2) \right. \\
    &\qquad\qquad + \xi_1 \xi_2(t_{\varepsilon}^2+t_{-\varepsilon}^2 - t_{\varepsilon}^2 r_{-\varepsilon}^2 - t_{-\varepsilon}^2 r_{\varepsilon}^2) \\
    & \qquad \qquad \left. -2\Delta_1\Delta_2 t_{\varepsilon}t_{-\varepsilon}(\cos \delta\gamma_{\varepsilon}+r_{\varepsilon}r_{-\varepsilon}\cos{\delta \eta_{\varepsilon}}) \cos \varphi \right] \\
    & =\frac{\xi_2\xi_1^2}{|D|^2\varepsilon} \left[\varepsilon^2(1-r_{\varepsilon}^2r_{-\varepsilon}^2 ) \right. \\
    &\qquad\qquad + \xi_1 \xi_2 t_{\varepsilon}^2t_{-\varepsilon}^2  \\
    & \qquad \qquad\left. -\Delta_1\Delta_2 t_{\varepsilon}t_{-\varepsilon}(\cos \delta\gamma_{\varepsilon}+r_{\varepsilon}r_{-\varepsilon} \cos{\delta \eta_{\varepsilon}}) \cos \varphi \right].
\end{align*}
We substitute the equation above in Eq.~(\ref{J_total_ap}) and obtain Eq.~(\ref{heatcurrent1}) in the main text.

%%%%%%%%%%%%%%%%%%%%%%%%%%%%%%%%%%%%%%%%%%%%%%%%%%%%%%%%%%%%%%%%%%%%%%%%%%%%%%%%%%%%%
\section{Details of derivation of Eqs.~(\ref{ful_c_cur})-(\ref{thermopower1}) for the particle current} \label{sec:chargecur_ap}
%%%%%%%%%%%%%%%%%%%%%%%%%%%%%%%%%%%%%%%%%%%%%%%%%%%%%%%%%%%%%%%%%%%%%%%%%%%%%%%%%%%%%

(i) Let us evaluate the charge current induced by a temperature difference $\delta T$ applied to the point contact. In the spirit of the Landauer transport theory, the charge current can be written as a balance of currents flowing from the opposite leads $I = I_1 - I_2$
\begin{align}
    I_l &= \frac{2e}{h} \int_{0}^\infty d\xi\, \left[  i^{e}_l(\varepsilon_l)  -  i^{h}_l(\varepsilon_l) \right] f\left(\varepsilon_l/T_l\right),  \label{c_cur_one_lead_ap0} \\
     &= \frac{2e}{h} \int_{\Delta_l}^\infty \frac{d\varepsilon\,\varepsilon}{\xi_l}\, \left[  i^{e}_l(\varepsilon)  -  i^{h}_l(\varepsilon) \right] f\left(\varepsilon/T_l\right), \label{c_cur_one_lead_ap1}
\end{align}
where $l \in \{1,2\}$ labels the leads. In Eq.~(\ref{c_cur_one_lead_ap0}), the integration variable is $\xi$, and $\varepsilon_l(\xi) = \sqrt{\xi^2+\Delta^2_l}$. In contrast, the integration variable is $\varepsilon$, and $\xi_l(\varepsilon) = \sqrt{\varepsilon^2-\Delta_l^2}$ in Eq.~(\ref{c_cur_one_lead_ap1}).

(ii) The two terms $i_{l}^{e}$ and $i_{l}^{h}$ correspond to the electron-like and hole-like quasiparticle currents, which contribute with the opposite signs. They have a physical meaning of a charge current induced by a quasiparticle, and may be evaluated with the knowledge of the two-component BdG wavefunction $\Psi$,
\begin{equation}
    i = \frac{1}{mv_F}{\rm Im}\left(\Psi^\dagger \nabla \Psi\right), \label{charge_current}
\end{equation}
where we normalized the expression by the Fermi velocity $v_F$ to render it dimensionless. In the BdG formalism, this current is not conserved because it does not take into account the contribution of the condensate. Nevertheless, one may evaluate this current in the normal regions $N_1$ and $N_2$. As an example, we evaluate Eq.~(\ref{charge_current}) for the BdG wavefunction~(\ref{scat_wf}) corresponding to the electron-like quasiparticle incident from the left superconductor. In Sec.\ref{sec:amplitudes_ap}, we used the Andreev boundary condition and obtained the wavefunction~(\ref{wf_normal_region}) in the normal region, with the amplitudes given in Eq.~(\ref{norm_coef}). Let us explicitly write it out:
\begin{align*}
& \Psi_{N_1} =   \\
    & (u_1+r_{11}^{he}v_1) \left( \begin{array}{c}
		  1 \\
		  0
	\end{array}\right) e^{ik_ex} + (v_1+r_{11}^{he}u_1) \left( \begin{array}{c}
		  0 \\
		  1
	  \end{array}\right) e^{ik_hx} \\
	  & \quad +   r^{ee}_{11} u_1 \left( \begin{array}{c}
		  1 \\
		  0
	  \end{array}\right) e^{-ik_ex}  +  r^{ee}_{11} v_1 \left( \begin{array}{c}
		  0 \\
		  1
	  \end{array}\right) e^{-ik_hx}  ,  \\
	& \Psi_{N_2} = \\ 
	  & t^{ee}_{21}u_2  \left( \begin{array}{c}
		  1 \\
		  0
	  \end{array}\right) e^{ik_ex} + 
          t^{ee}_{21} v_2  \left( \begin{array}{c}
		  0 \\
		  1
	  \end{array}\right) e^{ik_hx} \\
	  & \quad + t^{he}_{21}v_2 \left( \begin{array}{c}
		  1 \\
		  0
	  \end{array}\right) e^{-ik_e x}
	  +t^{he}_{21}u_2 \left( \begin{array}{c}
		  0 \\
		  1
	  \end{array}\right) e^{-ik_h x}.  
\end{align*}
The charge current~(\ref{charge_current}) is continuous in the normal regions, where there is no condensate. So, Eq.~(\ref{charge_current}) evaluated for both wavefunctions $\Psi_{N_1}$ and $\Psi_{N_2}$ must be equal 
\begin{align}
    i_1^e & = 1 - \left|r^{ee}_{11}\right|^2 + \left|r^{he}_{11}\right|^2 + 2 \frac{\Delta_1}{\varepsilon} {\rm Re} \left(r^{he}_{11}\right) \label{c_beforeX} \\
    & = \left|t^{ee}_{21}\right|^2 - \left|t^{he}_{21}\right|^2, \label{c_afterX}
\end{align}
where we used the identity $2\,u_l v_l = \frac{\Delta_l}{\varepsilon}$ and neglected the terms $\propto \varepsilon/E_F$. Although it is not immediately obvious that Eqs.~(\ref{c_beforeX}) and (\ref{c_afterX}) are equal, one may check that the equality holds for the derived amplitudes~(\ref{amplitudes_ap}).

(iii) As discussed in Sec.~\ref{sec:thermocur}, the current in superconductors may have a dissipative and non-dissipative Josephson components. The two components may be distinguished using their parity with respect to phase $\varphi$ reversal. The dissipative component is even and non-dissipative is odd under the phase $\varphi \rightarrow -\varphi$ reversal. In this work, we are interested in the dissipative component. To emphasize this, we write instead of Eq.~(\ref{c_cur_one_lead_ap1})
\begin{align}
    I_l &=  \frac{2e}{h} \int_{\Delta_l}^\infty \frac{d\varepsilon\,\varepsilon}{\xi_l}\, \left[  \tilde i^{e}_l(\varepsilon)  -  \tilde i^{h}_l(\varepsilon) \right] f\left(\varepsilon/T_l\right), \label{c_cur_one_lead_ap2}
\end{align}
where the notation $\sim$ above the terms in the brackets means taking an even in $\varphi$ part of the current. Hereinafter, all variables associated with particle current ($I,I_s$ etc.) denote the dissipative part of the current.

(iv) We set $T_{1,2} = T \pm \delta T/2$ and seek to evaluate the current proportional to $\delta T$. If $\delta T = 0$ (i.e. at thermodynamic equilibrium) the total dissipative current $I = I_1 - I_2$ must vanish because there are no ``kinematic forces'' that would drive the current. This condition allows to relate the integrands for $I_1$ and $I_2$ at $\delta T \neq 0$, and rewrite the total dissipative current $I$ only via the parameters corresponding, e.g., to the left lead
\begin{align}
    I &=  \frac{2e}{h} \int_{\Delta_{max}}^\infty \frac{d\varepsilon\,\varepsilon}{\xi_1}\, \left[  \tilde i^{e}_1(\varepsilon)  -  \tilde i^{h}_1(\varepsilon) \right] \left[f\left(\varepsilon/T_1\right)-f\left(\varepsilon/T_2\right)\right] \nonumber \\
    & = \frac{\delta T}{T^2}\frac{2e}{h} \int_{\Delta_{max}}^\infty \frac{d\varepsilon\,\varepsilon^2}{\xi_1}\, \left[  \tilde i^{e}_1(\varepsilon)  -  \tilde i^{h}_1(\varepsilon) \right] \left[-f'\left(x\right)\right]_{x =\varepsilon/T}, 
    \label{total_c_cur_ap}
\end{align}
where $\Delta_{max} = {\rm max}(\Delta_1,\Delta_2)$. Note that the subgap quasiparticles residing in the intermediate energy window~$\Delta_{max}>\varepsilon>\Delta_{min}$ do not contribute to the dissipative current (if the energy distribution of electron and hole-like quasiparticles are equal, the currents are equal in magnitude and opposite in sign). 

(v) Finally, we use the expressions~(\ref{t2e}) and (\ref{t2h}) for the amplitudes and Eq.~(\ref{c_afterX}) for the current and obtain a concise result
\begin{align}
     \tilde i^{e}_1(\varepsilon)  -  \tilde i^{h}_1(\varepsilon)  & =  \widetilde{\left|t^{ee}_{21}\right|^2}-\widetilde{\left|t^{he}_{21}\right|^2}-\widetilde{\left|t^{hh}_{21}\right|^2}+\widetilde{\left|t^{eh}_{21}\right|^2}\nonumber \\
    & =  \frac{\xi_1^2\xi_2\,\varepsilon}{|D|^2} \left(t_{\varepsilon}^2 - t_{-\varepsilon}^2\right) \label{c_cur_final_ap}.
\end{align}
Plugging it in Eq.~(\ref{total_c_cur_ap}), we recover Eq.~(\ref{thermopower1}) from the main part of the text. To double-check, we obtained the same result by using Eq.~(\ref{c_beforeX}) instead of Eq.~(\ref{c_afterX}).

%%%%%%%%%%%%%%%%%%%%%%%%%%%%%%%%%%%%%%%%%%%%%%%%%%%%%%%%%%%%%%%%%%%%%%%%%%%%%%%%%%%%%
\section{Comparison with the tunneling Hamiltonian approach.} \label{sec:tun_Ham_appendix}
%%%%%%%%%%%%%%%%%%%%%%%%%%%%%%%%%%%%%%%%%%%%%%%%%%%%%%%%%%%%%%%%%%%%%%%%%%%%%%%%%%%%%

Let us compare the obtained scattering amplitudes~(\ref{tunnel_lim_Beenakker}) with the tunneling Hamiltonian approach. The tunneling Hamiltonian model may be written as
\begin{align*}
	& H = H_1 + H_2 + H_T, \nonumber \\
	& H_1 = \sum_{k\sigma} \left(\frac{k^2}{2m}-E_F\right) c_{k\sigma}^\dagger c_{k\sigma} + \Delta \sum_{k} \left(c_{-k\downarrow}c_{k\uparrow} + {\rm h.c.}\right), \\ 
	& H_2 = \sum_{p\sigma} \left(\frac{p^2}{2m}-E_F\right) c_{p\sigma}^\dagger c_{p\sigma} + \Delta \sum_{p} \left(c_{-p\downarrow}c_{p\uparrow} + {\rm h.c.}\right), \\ 
	& H_T = \sum_{pk\sigma} w_{pk}\left( e^{i\varphi/2} c^\dagger_{p\sigma}c_{k\sigma} + e^{-i\varphi/2} c^\dagger_{k\sigma}c_{p\sigma} \right), 
	%\label{th} 
\end{align*}
where the momenta $k$ and $p$ label the states in the left and right leads; $w_{kp}$ is the tunneling matrix element. We perform the Bogoliubov transformation $c_{k\sigma}  = u_k \gamma_{k\sigma} - \sigma v_k \gamma_{\bar k\bar \sigma}^\dagger$ (similarly for $p$) and rewrite the tunneling part
\begin{align*}
	 H_T = \sum_{kp\sigma} w_{pk} \left( e^{i\varphi/2} u_p u_k - e^{-i\varphi/2} v_p v_k \right)  \gamma^\dagger_{p\sigma}\gamma_{\kappa\sigma}  \,\,+\,\,{\rm h.c.}  & \\
	 +\{\,{\rm terms}\,\,\propto \, \gamma \gamma, \gamma^\dagger \gamma^\dagger \}.  \nonumber
\end{align*}
Now, we find the scattering amplitudes in the lowest order in $w_{pk}$. First, we rewrite the momentum variables into the energy variables as follows $k_{e,h} = k_F \pm \xi/v_F$ (similarly for $p_{e,h}$). Further, we multiply by the factor accounting for the superconducting density of states $2\pi\rho_0\,\varepsilon/i\xi$ and obtain the scattering amplitudes
\begin{align}
    &\begin{aligned}
    & t_{21}^{ee} = -2\pi \rho_0\, w_{p_ek_e}\,\frac{i}{2\xi}\left[ (\varepsilon+\xi)e^{i\varphi/2}-(\varepsilon-\xi)e^{-i\varphi/2}\right],\\
    & t_{21}^{he} = -2\pi \rho_0\, w_{p_hk_e}\,\frac{i\Delta}{2\xi}\left[e^{i\varphi/2}-e^{-i\varphi/2}\right],
\label{amplitudes_tunHam}
\end{aligned}  \\
&{\rm where} \nonumber \\
& \begin{aligned}
    k_{e,h} = k_F \pm \xi/v_F, \\
    p_{e,h} = k_F \pm \xi/v_F.
\end{aligned} \nonumber
\end{align}

Let us compare the scattering amplitudes~(\ref{tunnel_lim_Beenakker}) obtained in the scattering approach [\onlinecite{Beenakker1991}] with the amplitudes~(\ref{amplitudes_tunHam}) obtained using the tunneling Hamiltonian method. Superficially, the amplitudes look similar, and they agree if the particle-hole asymmetry is dropped. However, they are distinct in the presence of the particle-hole asymmetry. In particular, in the limit $\varphi = 0$ and $\xi \rightarrow 0$, the scattering amplitudes~(\ref{tunnel_lim_Beenakker}) diverge, whereas the amplitudes in Eq.~(\ref{amplitudes_tunHam}) remain finite. Physically, this distinct behavior corresponds to the formation of the particle-hole-asymmetry-induced Andreev levels in the former case. This dichotomy also manifests itself in a  distinct behavior of the thermoelectric coefficient. In the former case of Eq. (\ref{tunnel_lim_Beenakker}), the thermoelectric coefficient is logarithmically large (see Eq.~(\ref{G_tau_dropped})). In the latter case of Eq.~(\ref{amplitudes_tunHam}), the thermoelectric coefficient remains finite $s^{SS} = -\frac{6}{\pi^2} \int_\alpha^\infty dx\,x^2\,f'(x)$.

We further explore the connection between the scattering amplitudes (\ref{tunnel_lim_Beenakker}) and (\ref{amplitudes_tunHam}) in the following section. We solve a "square barrier" BdG model and demonstrate how Eqs. (\ref{tunnel_lim_Beenakker}) and (\ref{amplitudes_tunHam}) emerge in two different limits. 

%%%%%%%%%%%%%%%%%%%%%%%%%%%%%%%%%%%%%%%%%%%%%%%%%%%%%%%%%%%%%%%%%%%%%%%%%%%%%%%%%%%%%
\section{Exact solution of a scattering problem in a BdG formalism.} \label{sec:bdg_appendix}
%%%%%%%%%%%%%%%%%%%%%%%%%%%%%%%%%%%%%%%%%%%%%%%%%%%%%%%%%%%%%%%%%%%%%%%%%%%%%%%%%%%%%

%%%%%%%%%%%%%%%%%%%%%%%%%%%%%%%%%%%%%%%%%%%%%%%%%%%%%%%%%%%%%%%%%%%%%%%%%%%%%
\begin{figure*}
 \centering\includegraphics[width=0.7\linewidth]{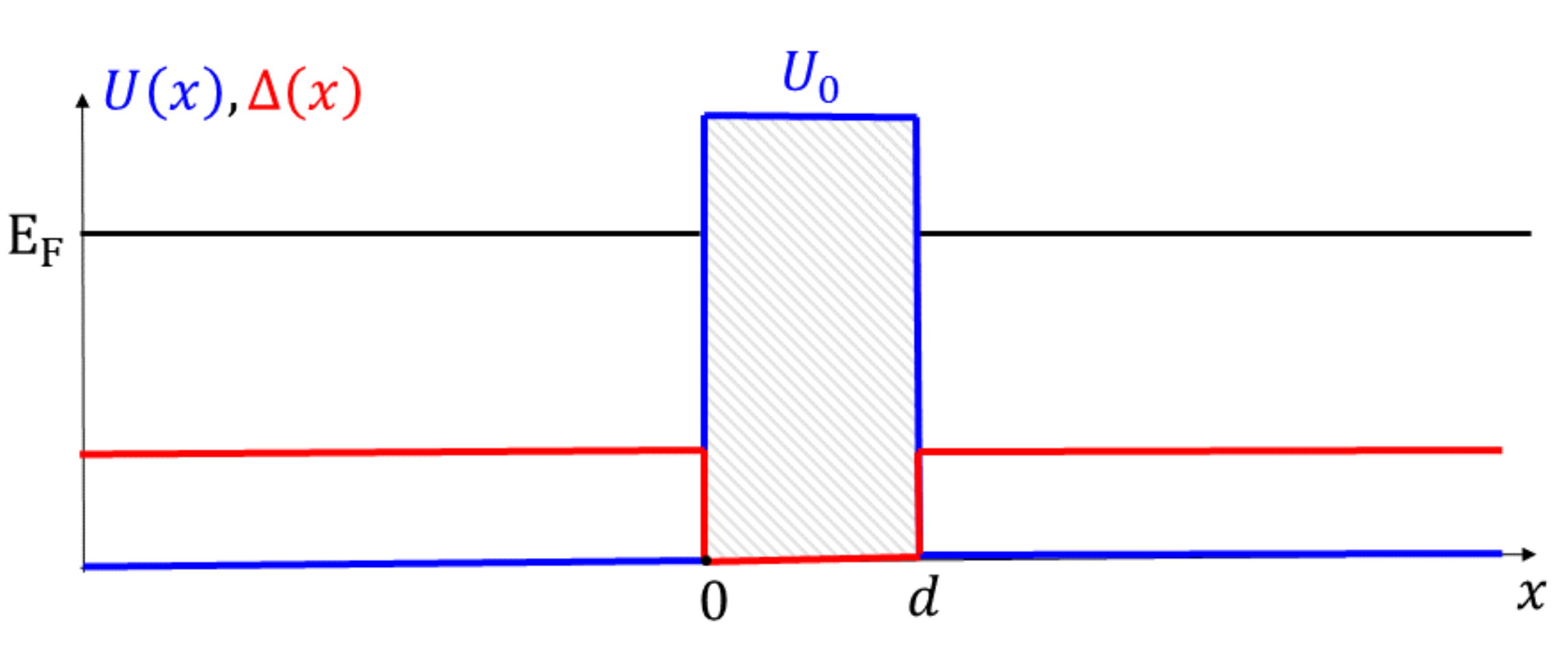} 
 \caption{One dimensional BdG scattering problem.} 
\label{fig:1D_bdg}
\end{figure*}
%%%%%%%%%%%%%%%%%%%%%%%%%%%%%%%%%%%%%%%%%%%%%%%%%%%%%%%%%%%%%%%%%%%%%%%%%%%%%

%First let us comment on the relation between the two approximations (a) Andreev  (semiclassical) approximation \cite{KopninBook} and (b) approximation  $\varepsilon\ll E_F$ \cite{Beenakker1991,BeenakkerPRB1992}. The former approximation (a) states that the parameters of the BdG equation such as superconducting gap $\Delta(x)$ and potential $U(x)$ vary smoothly on a scale of inverse Fermi length $x\sim k_F^{-1}$. This allows to treat the momentum $k_F$ as the largest variable and work in the leading order $(k-k_F)/k_F\ll 1$ (or equivalently $\varepsilon/E_F$). We would like to emphasize that this approximation (a) is essentially the same as the latter approximation (b), i.e. that $\varepsilon \ll E_F$. Even in the case where the gap $\Delta(x)$ or potential $U(x)$ in a specific model are not smooth enough, expanding in $\varepsilon/E_F$ (e.g. of the exact scattering amplitudes) and retaining only the leading order produces the same result as that of approximation (a). In other words, if we work close to $E_F$, we do not resolve fine details of the scattering potentials and may treat them as being effectively fuzzy on the scale of $k_F^{-1}$. 
Here, we solve a 1D BdG equation with a square barrier potential. We recover the scattering amplitudes~(\ref{tunnel_lim_Beenakker}) and (\ref{amplitudes_tunHam}) in two different limits. In addition, we demonstrate that the presence of the normal $N$ parts (introduced for convenience in Sec.~\ref{sec:amplitudes_ap}) is not essential. 

The BdG equation is $H\Psi(x) = \varepsilon \Psi(x)$ with
\begin{align}
     H = \left( \begin{array}{cc}
		  \frac{(-i\partial_x)^2}{2m}-E_F + U(x) & \Delta^\ast(x)\\
		  \Delta(x) &  -\frac{(-i\partial_x)^2}{2m}+E_F - U(x)
	  \end{array}\right), 
	  \label{bdgEq}
\end{align}
where the profiles of the gap $\Delta(x) = \Delta e^{i\varphi}\,\theta(x-d)+\Delta\,\theta(-x)$ and potential $U(x) = U[1-\theta(-x)-\theta(x-d)]$ are illustrated in Fig.~\ref{fig:1D_bdg}. In the left and right domains, we use the following ansatz for the scattering wavefunction
\begin{align} 
  \Psi_1 &= \left( \begin{array}{c}
		  u\\
		  v
	  \end{array}\right)e^{iq_ex} + r^{ee}_{11}\left( \begin{array}{c}
		  u\\
		  v
	  \end{array}\right)e^{-iq_ex}+ r^{he}_{11}\left( \begin{array}{c}
		  v\\
		  u
	  \end{array}\right)e^{-iq_hx},\nonumber\\
  \Psi_2 & =t^{ee}_{21} \left( \begin{array}{c}
		  u\, e^{-i\varphi/2}\\
		  v\, e^{i\varphi/2}
	  \end{array}\right)e^{iq_e(x-d)} \nonumber
	  \\& \qquad\qquad\qquad
	  + t^{he}_{21}\left( \begin{array}{c}
		  u\, e^{-i\varphi/2}\\
		  v\, e^{i\varphi/2}
	  \end{array}\right)e^{-iq_h(x-d)},	  \nonumber
\end{align}
where as usual $u,v = \sqrt{\frac1 2 \left(1\pm \frac{\sqrt{\varepsilon^2-\Delta^2}}{\varepsilon}\right)}$. Inside the barrier, the wavefunctions are described by the decaying solutions
\begin{align} 
  \Psi_I &= c_1\left( \begin{array}{c}
		  1\\
		  0
	  \end{array}\right)e^{\kappa_e x} + c_2\left( \begin{array}{c}
		  1\\
		  0
	  \end{array}\right)e^{-\kappa_e x}\\ & \qquad\qquad+c_3\left( \begin{array}{c}
		  0\\
		  1
	  \end{array}\right)e^{\kappa_h x} + c_4\left( \begin{array}{c}
		  0\\
		  1
	  \end{array}\right)e^{-\kappa_h x}\nonumber,
\end{align}
where the subscripts $e$ and $h$ denote the energy dependence
\begin{align}
q_{e,h} &= \left[2m(E_F\pm \sqrt{\varepsilon^2-\Delta^2})\right]^{1/2},\\
\kappa_{e,h} &= \left[2m(U-E_F\mp\varepsilon)\right]^{1/2}, \label{keh}
\end{align}
where to top and bottom signs correspond to electrons (e) and holes (h), respectively. We use the continuity condition at the NS boundary
\begin{equation}
\begin{aligned}
    \Psi_1(0) &= \Psi_I(0), \\
    \Psi'_1(0) &= \Psi'_I(0), \\
    \Psi_2(d) &= \Psi_I(d), \\
    \Psi'_2(d) &= \Psi'_I(d),
\end{aligned} \nonumber
\end{equation}
and solve for the scattering amplitudes
\begin{align}
&\begin{aligned}
  t^{ee}_{21} = \frac{iq_e(u^2-v^2)}{D'}\left[u^2\kappa_e(q_h-i\kappa_h)^2e^{i\varphi/2-\kappa_e d}\right. \quad\qquad&
  \\ -v^2\kappa_h(q_h-i\kappa_e)^2e^{-i\varphi/2-\kappa_h d} \qquad &
  \\ -u^2\kappa_e(q_h+i\kappa_h)^2e^{i\varphi/2-\kappa_e d-2\kappa_h d}\quad
  \\ \left.+v^2\kappa_h(q_h+i\kappa_e)^2e^{-i\varphi/2-\kappa_h d-2\kappa_e d}\right],& 
\end{aligned} \label{tee_square_barrier} \\
&\begin{aligned}
   t^{he}_{21} = \frac{iq_euv(u^2-v^2)}{D'}\left[\kappa_e(q_e+i\kappa_h)(q_h-i\kappa_h)e^{i\varphi/2-\kappa_e d}\right. \quad\,\,&
  \\ -\kappa_h(q_e+i\kappa_e)(q_h-i\kappa_e)e^{-i\varphi/2-\kappa_h d} \quad &
  \\ -\kappa_e(q_e-i\kappa_h)(q_h+i\kappa_h)e^{i\varphi/2-\kappa_e d-2\kappa_h d}\,\,
  \\ \left.+\kappa_h(q_e-i\kappa_e)(q_h+i\kappa_e)e^{-i\varphi/2-\kappa_h d-2\kappa_e d}\right],& 
\end{aligned} \label{the_square_barrier}
\end{align}
where the denominator is
\begin{align*}
& D' = A  + B e^{-2\kappa_e d}+B^\ast e^{-2\kappa_h d} \\
& \qquad\qquad\qquad + C e^{-(\kappa_e+\kappa_h)d} + A^\ast e^{-2(\kappa_e+\kappa_h)d}, \\
&\begin{aligned}
&A = \frac{1}{4}\left[(u^2-v^2)(q_eq_h+\kappa_e\kappa_h)+iu^2(q_h\kappa_e-q_e\kappa_h)\right.\quad \\ 
 &\qquad\left.+iv^2(q_e\kappa_e-q_h\kappa_h)\right]^2,\\
& B = -\frac{1}{4}\left[(u^2-v^2)(q_eq_h-\kappa_e\kappa_h)-iu^2(q_h\kappa_e+q_e\kappa_h)\right. \quad  \\
&\qquad\left.-iv^2(q_e\kappa_e+q_h\kappa_h)\right]^2,\\
&C = -2 u^2v^2(q_e+q_h)^2\kappa_e\kappa_h\cos\varphi. \qquad \qquad&
\end{aligned}
\end{align*}
This is an exact formal solution of Eq.~(\ref{bdgEq}). Next we show how to recover expressions~(\ref{tunnel_lim_Beenakker}) and (\ref{amplitudes_tunHam}) from the solution above.

{\it Limit of weakly-transparent barrier in the scattering formalism [\onlinecite{Beenakker1991}].} We assume that the length of the junction is short enough $d\ll v_F/\Delta$ to be considered a point contact. At the same time, we assume that the junction is long enough $ \kappa_{e,h}\,d\gg 1$, so that it is in the tunneling regime. In a concise form, the condition on the length may be written as  $E_F/\Delta \gg k_Fd\gg \sqrt{E_F/(U-E_F)}$. So, one may retain only the leading order terms $e^{-\kappa_{e,h}d}$ and obtain 
\begin{align*}
&\begin{aligned}
  t^{ee}_{21} = \frac{iq_e(u^2-v^2)}{A}\left[u^2\kappa_e(q_h-i\kappa_h)^2e^{i\varphi/2-\kappa_e d}\right. \quad\qquad&
  \\ \left.-v^2\kappa_h(q_h-i\kappa_e)^2e^{-i\varphi/2-\kappa_h d} \right],& 
\end{aligned}  \\
&\begin{aligned}
   t^{he}_{21} = \frac{iq_euv(u^2-v^2)}{A}\left[\kappa_e(q_e+i\kappa_h)(q_h-i\kappa_h)e^{i\varphi/2-\kappa_e d}\right. \quad\,\,&
  \\ \left. -\kappa_h(q_e+i\kappa_e)(q_h-i\kappa_e)e^{-i\varphi/2-\kappa_h d} \right],& 
\end{aligned}
\end{align*}
where $A$ is defined above. Further, we assume  the following separation of energy scales $E_F \gg U-E_F \gg \varepsilon,\Delta$.  This helpful assumption allows to drop terms $\propto\varepsilon/E_F$ but retain the terms $\propto\varepsilon/(U-E_F)$ which carry information about the particle-hole asymmetry. In other words, we may set $q_e = q_h = k_F$ but retain the energy dependence in $\kappa_{e,h}$. This assumption also allows us to retain only the lowest-order in $\kappa_{e,h}/k_F$ terms, 
\begin{align}
\begin{aligned}
    & t_{21}^{ee} = \frac{4i\varepsilon}{\xi k_F}\left( u^2\kappa_e e^{i\varphi/2-\kappa_e d}-v^2\kappa_h e^{-i\varphi/2-\kappa_h d}\right),\\
    & t_{21}^{he} = \frac{2i\Delta}{\xi k_F}\left( \kappa_e e^{i\varphi/2-\kappa_e d}-\kappa_h e^{-i\varphi/2-\kappa_h d}\right),
\end{aligned}
\end{align}
where the variables $\kappa_{e,h}$ depend on $\varepsilon$ according to Eq.~(\ref{keh}). We may rewrite the scattering amplitudes in the form
\begin{align}
&\begin{aligned}
    & t_{21}^{ee} = \frac{i}{2\xi}\left[ (\varepsilon+\xi)t_{\varepsilon}e^{i\varphi/2}-(\varepsilon-\xi)t_{-\varepsilon}e^{-i\varphi/2}\right],\\
    & t_{21}^{he} = \frac{i\Delta}{2\xi}\left[t_{\varepsilon}e^{i\varphi/2}-t_{-\varepsilon} e^{-i\varphi/2}\right],
\end{aligned}
\label{amplitudes_bdg} 
\end{align}
where
\begin{align}
 \begin{aligned}
    &t_{\varepsilon} = \frac{4\kappa_e}{k_F} e^{-\kappa_e d},  \\
    &t_{-\varepsilon} = \frac{4\kappa_h}{k_F} e^{-\kappa_h d}, \\
\end{aligned} 
\end{align}
with $\kappa_{e,h} = \left[2m(U-E_F\mp\varepsilon)\right]^{1/2}$. The amplitudes~(\ref{amplitudes_bdg}) conform with the corresponding expressions~(\ref{tunnel_lim_Beenakker}) obtained in the scattering formalism. 

{\it The delta-barrier limit.} We introduce a dimensionless parameter $Z$ via the identity $U = k_F Z/2md$ and expand Eqs.~(\ref{tee_square_barrier}) and (\ref{the_square_barrier}) in powers of $k_F d$ to obtain
\begin{align}
    &\begin{aligned}
    & t_{21}^{ee} = {t_{21}^{ee}}^{(0)} +  (k_Fd)\, {t_{21}^{ee}}^{(1)} + \mathcal O(k_F d)^2, \\
    & t_{21}^{he} = {t_{21}^{he}}^{(0)} +  (k_Fd)\, {t_{21}^{he}}^{(1)}  + \mathcal O(k_F d)^2.
\end{aligned}  \label{amplitudes_delta_Fun}
\end{align}
For brevity, we focus on the $\varphi = 0$ case where the amplitudes simplify
\begin{align}
    &\begin{aligned}
    & {t_{21}^{ee}}^{(0)} = \frac{2q_e}{2q_e+ i k_F Z},\\
    & {t_{21}^{he}}^{(0)} = 0.
\end{aligned}  \label{amplitudes_delta_Fun0}
\end{align}
Observe that at $\xi \rightarrow 0$ the amplitudes ${t_{21}^{ee}}^{(0)}$ and ${t_{21}^{he}}^{(0)}$ are regular. The obtained amplitudes (\ref{amplitudes_delta_Fun0}) conform with the corresponding expressions (\ref{amplitudes_tunHam}) obtained from the tunneling Hamiltonian method. 

We give the leading terms in the Laurent series in $\xi$ of the higher-order terms appearing in Eq.~(\ref{amplitudes_delta_Fun})
\begin{align}
&\begin{aligned}
    & {t_{21}^{ee}}^{(1)} = \frac{1}{\xi}\,\frac{\varepsilon^2}{E_F}\, \frac{2ik_Fq_e}{(2q_e+ik_F Z)^2} + \mathcal O (\xi^0)  \\ 
  %  & \qquad \qquad + \xi^0\,\frac{2iq_e\left(k_F^2+q_e^2+ik_Fq_eZ - \frac{1}{6}k_FZ^2\right)}{k_F (2q_e + ik_F Z)^2}+ \mathcal O (\xi),\\
    & {t_{21}^{he}}^{(1)} = \frac{1}{\xi}\,\frac{\varepsilon^2}{E_F}\, \frac{2ik_Fq_e}{(2q_e+ik_F Z)(2q_h-ik_F Z)} + \mathcal O (\xi^0), 
\end{aligned}  \label{amplitudes_delta_Fun1} %\\
%&\begin{aligned}
%    & {t_{RL}^{ee}}^{(2)} = \frac{1}{\xi^2}\frac{\varepsilon^4}{E_F^2} \frac{4k_F^2q_e(q_e+q_h)}{(2q_e+i k_F Z)^3(2q_h-i k_F Z)}  \\ 
%    & \qquad \qquad \qquad  - \frac{1}{\xi}\frac{\varepsilon^2}{E_F} \frac{4k_F(k_Fq_E + \frac{i}{6}q_e^2Z)}{(2q_e + i k_F Z)^3}+ \mathcal O (\xi^0),\\
%    & {t_{RL}^{he}}^{(2)} = -\frac{1}{\xi^2} \frac{\varepsilon^4}{E_F^2} \frac{4k_F^2q_e(q_e+q_h)}{(2q_e+i k_F Z)^2(2q_h-i k_F Z)^2} \\
%    & \qquad\qquad - \frac{1}{\xi}\frac{\varepsilon^2}{E_F} \frac{4 i k_F q_e(k_F^2Z -\frac{1}{3}q_eq_h Z)}{(2q_e + i k_F Z)^2(2q_h - i k_F Z)^2}+ \mathcal O (\xi^0),
%\end{aligned}  \label{amplitudes_delta_Fun2}
\end{align}
where $\mathcal O(\xi^{0})$ and $\mathcal O(\xi)$ denote behavior at $\xi \rightarrow 0$. Note that the amplitudes corresponding to a hole-like quasiparticle may be obtained from equations above by complex conjugation and replacing $\xi \leftrightarrow -\xi$. Observe that ${t_{21}^{ee}}^{(1)}$, ${t_{21}^{he}}^{(1)}$ develop a singularity as $\xi \rightarrow 0$. The singularity in the scattering amplitudes signifies an appearance of the shallow Andreev levels. In order to analyze the Andreev levels, let us expand the denominator of the scattering amplitudes in the studied limit $k_F d \rightarrow 0$. The leading behavior of the denominator is $D' \propto \left[\xi^2 - i\tau\frac{\varepsilon^2 \xi}{E_F}(k_Fd)\right]$, where $\tau = \frac{4}{4+Z^2}$.  This gives the behavior of the Andreev levels 
\begin{equation}
%\varepsilon_A^2 = \Delta^2\left[1- \tau^2 \frac{\Delta^2}{E_F^2} (k_FL)^2\right].
\varepsilon_A = \Delta - \frac{\tau^2}{2} \frac{\Delta^3}{E_F^2} (k_Fd)^2.
\label{andreevEnergy}
\end{equation}
So, we conclude that finite length $d$ generates shallow Andreev levels with energy controlled by length $d$ and the scale of the particle-hole asymmetry $\Delta/E_F$.

{\it Consequence for the particle current.} The appearance of the Andreev levels has consequences for the particle current. In order to evaluate it, we need to retain the $\propto \varepsilon/E_F$ terms, which were dropped in derivation of Eqs.~(\ref{c_cur_one_lead})-(\ref{thermopower1}) as well as (\ref{c_cur_one_lead_ap0})-(\ref{c_cur_final_ap}). For the case of the symmetric junction considered here, we keep the $\varepsilon/E_F$ terms and obtain the thermoelectric coefficient
\begin{align}
    & S^{SS}_I = \label{PH_asymmetry_retained} \\
    & \frac{1}{T^2}\frac{2e}{h} \int_\Delta^\infty d\varepsilon\, \frac{\varepsilon^2}{\xi}   \left[ \left|t^{ee}_{21}\right|^2 - \left|t^{hh}_{21}\right|^2 -\frac{q_h}{q_e} \left|t^{he}_{21}\right|^2 + \frac{q_e}{q_h} \left|t^{eh}_{21}\right|^2\right.\nonumber \\
    & \left.+ \frac{\Delta(q_e-q_h)}{\varepsilon}  {\rm Re}\left(\frac{t^{ee}_{21} {t^{he}_{21}}^\ast}{q_e} + \frac{t^{hh}_{21} {t^{eh}_{21}}^\ast}{q_h}\right)\right][-f'(x)]_{x = \varepsilon/T}. \nonumber
\end{align}
Now we substitute the amplitudes (\ref{amplitudes_delta_Fun})-(\ref{amplitudes_delta_Fun1}) in Eq.~(\ref{PH_asymmetry_retained}) and obtain the correction to the thermoelectric coefficient up to first order in $k_F d$
\begin{align}
   & S^{SS}_I =   \frac{GT}{e}\frac{\partial \ln G}{\partial\mu} 2\int_{\Delta/T}^\infty dx\,x^2[-f'(x)]  \nonumber \\ 
    & \,\,+ (k_Fd)\frac{G \sqrt{\tau(1-\tau)}}{e}\,\frac{ T}{E_F}\,2 \int_{\Delta/T}^\infty dx \frac{x^4}{x^2-(\Delta/T)^2}[-f'(x)]  \nonumber \\ 
       & \qquad \qquad \qquad \qquad \qquad  + \mathcal O(k_Fd)^2 \label{correction_to_thermopower}
\end{align}
where $G = 2e^2 \tau /h$. The first term is regular, whereas the second term has a logarithmic divergence at lower integration limit. Note that the regular (not divergent) terms in the order $\propto k_Fd$ are not displayed. Recalling that there are Andreev levels with energies given by Eq.~(\ref{andreevEnergy}), the logarithmic divergence may be regularized producing for the integral $-\frac{1}{2}\left(\frac{\Delta}{T}\right)^3\, f'\left(\frac{\Delta}{T}\right)\, \ln \frac{T}{\Delta - \varepsilon_A}$.

This resolves the discrepancy between Refs.~[\onlinecite{SmithPRB1980}] and  [\onlinecite{GuttmanPRB1997a}]. Reference [\onlinecite{GuttmanPRB1997a}] used the amplitudes~(\ref{amplitudes_delta_Fun0}) corresponding to the zeroth order in $k_Fd$  (i.e. a delta-barrier limit) and obtained a regular expression for particle current consistent with the first term in Eq.~(\ref{correction_to_thermopower}). However, it completely missed the existence of the Andreev levels, and, thus, missed the logarithmic contribution to the particle current represented by the second term in Eq.~(\ref{correction_to_thermopower}). Given that physical contacts have finite length $k_Fd\gtrsim 1$ (actually $k_Fd\gg 1$ in most cases), the logarithmic term is important, and we favor the approach of Ref. [\onlinecite{SmithPRB1980}].

%%%%%%%%%%%%%%%%%%%%%%%%%%%%%%%%%%%%%%%%%%%%%%%%%%%%%%%%%%%%%%%%%%%%%%%%%%%%%%%%%%%%%
\end{document}